\documentclass[aps,floatfix,notitlepage,nofootinbib,twocolumn,
superscriptaddress]{revtex4-1}

\usepackage{amsmath}
\usepackage{amssymb}
\usepackage{graphicx}
\usepackage[tight]{subfigure}
\usepackage{enumitem}
\usepackage{soul}
\usepackage{pifont}
\usepackage{xspace}
\usepackage{slashed}
\usepackage{changepage}
\usepackage{array}

\usepackage[english]{babel}
\makeatletter
\def\bbl@set@language#1{%
  \edef\languagename{%
    \ifnum\escapechar=\expandafter`\string#1\@empty
    \else\string#1\@empty\fi}%
  \@ifundefined{babel@language@alias@\languagename}{}{%
    \edef\languagename{\@nameuse{babel@language@alias@\languagename}}%
  }%
  \select@language{\languagename}%
  \expandafter\ifx\csname date\languagename\endcsname\relax\else
    \if@filesw
      \protected@write\@auxout{}{\string\select@language{\languagename}}%
      \bbl@for\bbl@tempa\BabelContentsFiles{%
        \addtocontents{\bbl@tempa}{\xstring\select@language{\languagename}}}%
      \bbl@usehooks{write}{}%
    \fi
  \fi}
\newcommand{\DeclareLanguageAlias}[2]{%
  \global\@namedef{babel@language@alias@#1}{#2}%
}
\makeatother
\DeclareLanguageAlias{en}{english}

\usepackage{bbold}
\usepackage[usenames,dvipsnames]{xcolor}
\definecolor{darkblue}{rgb}{0.1,0.2,0.6} \definecolor{darkred}{rgb}{0.8,0.1,0.2}
\usepackage[colorlinks,citecolor=darkblue,linkcolor=darkred,urlcolor=darkblue]{hyperref}
\usepackage{tikz}
\usetikzlibrary{calc,fadings,decorations.pathreplacing,shapes,shapes.multipart,arrows,shapes.misc,intersections,positioning}

\newcommand{\I}{\mathbb{I}}

\usepackage{bm}
\renewcommand{\vec}[1]{\boldsymbol{\mathbf{#1}}}

\newcommand*\circled[1]{\tikz[baseline=(char.base)]{
            \node[shape=circle,draw,inner sep=1pt] (char) {\small #1};}}

\renewcommand{\eqref}{\ref}
\graphicspath{{./}{./images/}}

\newcommand{\bit}{\begin{itemize}}
\newcommand{\eit}{\end{itemize}}

\newcommand{\f}{\frac}
\renewcommand{\>}{\right\rangle}
\newcommand{\<}{\left\langle}
\newcommand{\ba}{\begin{align}}
\newcommand{\ea}{\end{align}}
\newcommand{\be}{\begin{equation}}
\newcommand{\ee}{\end{equation}}
\newcommand{\bi}{\begin{itemize}}
\newcommand{\ei}{\end{itemize}}
\newcommand{\lf}{\left(}
\newcommand{\ri}{\right)}
\newcommand{\dd}{\mathrm{d}}

\newcommand{\Tr}{\operatorname{Tr}}
\newcommand{\tr}{\operatorname{tr}}

\def\braket#1#2{\left\langle #1|#2\right\rangle}

\newcommand{\seq}{s_\text{eq}}

\newcommand{\bra}[1]{\< #1 \right|}
\newcommand{\ket}[1]{\left| #1 \>}

\usepackage{xifthen}
\usepackage{xargs}

\newcommandx{\drawbox}[6][1=0,2=0,3=1,4=1,5=,6=]{
  \ifthenelse{\equal{#5}{}}{
    \draw[line width = 0.5pt] (#1,#2) rectangle (#3,#4);
  }{
    \draw[line width = 0.5pt, fill = #5] (#1,#2) rectangle (#3,#4);
  }
  \node () at (#1*0.5+#3*0.5,#2*0.5+#4*0.5) {#6};
}

\newcommandx{\regbox}[4][1=0,2=0,3=,4=]{
  \begin{scope}[shift={(#1,#2)}]
    \drawbox[0][0][1.5][1][#3][#4];
  \end{scope}
}

\newcommandx{\regboxt}[8][1=0,2=0,3=,4=,5=,6=,7=,8=]{
  \begin{scope}[shift={(#1,#2)}]
    \drawbox[0][0][1.5][1][#3][#4];
    \node[above] () at (0,1) {#5};
    \node[above] () at (1.5,1) {#6};
    \node[below] () at (1.5,0) {#7};
    \node[below] () at (0,0) {#8};
  \end{scope}
}

\newcommandx{\sqzbox}[5][1=0,2=0,3=,4=,5=]{
  \begin{scope}[shift={(#1,#2)}]
    \drawbox[0][0][2.25][1][#3][#4];
    \ifthenelse{\equal{#5}{s}}{
      \draw (0,0.1)--++(-0.1,0)--++(0,1)--++(2.25,0)--++(0,-0.1);
    }{}
  \end{scope}
}

\newcommandx{\ballon}[5][1=0,2=0,3=0.25,4=0.5,5=]{
  \begin{scope}[shift={(#1,#2)}]
    \pgfmathsetmacro{\r}{#4};
    \pgfmathsetmacro{\cent}{#3+#4};
    \draw (0,0)--++(0,#3);
    \draw (0,\cent) circle (\r);
    \draw (0,\cent+\r)--++(0,#3);
    \ifthenelse{\equal{#5}{}}{}{
      \node () at (0,\cent) {#5};
    }
  \end{scope}
}

\newcommandx{\wallebox}[5][1=,2=,3=,4=,5=]
{
  \fineq[-0.8ex][0.8][0.8]{
    \sqzbox[0][0];
    \ifthenelse{\equal{#1}{t}}{
      \ballon[0.5][1][0.25][0.5][$#2$];
      \ballon[1.75][1][0.25][0.5][$#3$];
    }{}
    \ifthenelse{\equal{#1}{b}}{
      \ballon[0.5][-1][0.25][0.5][$#4$];
      \ballon[1.75][-1][0.25][0.5][$#5$];
    }{}
    \ifthenelse{\equal{#1}{tb}}{
      \ballon[0.5][1][0.25][0.5][$#2$];
      \ballon[1.75][1][0.25][0.5][$#3$];
      \ballon[0.5][-1.5][0.25][0.5][$#4$];
      \ballon[1.75][-1.5][0.25][0.5][$#5$];
    }{}
  }
}

\newcommandx{\sufourpart}[5][1=0,2=0,3=0.25,4=0.5,5=]{
  \begin{scope}[shift={(#1,#2)}]
    \pgfmathsetmacro{\r}{#4};
    \pgfmathsetmacro{\cent}{#3+#4};
    \draw (0,0)--++(0,#3);
    \draw (-\r,\cent-\r) rectangle (\r,\cent+\r);
    \draw (0,\cent+\r)--++(0,#3);
    \ifthenelse{\equal{#5}{}}{}{
      \node () at (0,\cent) {#5};
    }
  \end{scope}
}

\newcommandx{\sufour}[5][1=,2=,3=,4=,5=]
{
  \fineq[-0.8ex][0.6][1]{
    \sqzbox[0][0][][$#1$];
    \sufourpart[0.5][1][0.25][0.5][$#2$];
    \sufourpart[1.75][1][0.25][0.5][$#3$];
    \sufourpart[0.5][-1.5][0.25][0.5][$#4$];
    \sufourpart[1.75][-1.5][0.25][0.5][$#5$];
  }
}

\newcommandx{\tisingu}[4][1=,2=,3=,4=]
{
  \fineq[-0.8ex][0.6][1]{
    \sqzbox[0][-1][][$#1$];
    \sqzbox[0][1.5][][$#1$];
    \sufourpart[0.5][0][0.25][0.5][$#2$];
    \sufourpart[1.75][0][0.25][0.5][$#3$];
    \draw (0.5,2.5)--++(0,0.25);
    \draw (1.75,2.5)--++(0,0.25);
    \draw (0.5,-1)--++(0,-0.25);
    \draw (1.75,-1)--++(0,-0.25);
  }
}

\newcommandx{\squbox}[4][1=0,2=0,3=,4=]{
  \begin{scope}[shift={(#1,#2)}]
    \drawbox[0][0][1][1][#3][#4];
  \end{scope}
}

\newcommandx{\uonesite}[7][1=0,2=0,3=,4=,5=,6=,7=]{
  \begin{scope}[shift={(#1,#2)}]
    \draw[line width = 0.5pt] (0.5,0.25)--++(0,-0.25) coordinate (A);
    \draw[line width = 0.5pt] (0.5,1.25)--++(0,0.25) coordinate (B);
    \ifthenelse{\equal{#7}{s}}{
      \squbox[0][0.25][#3][#4];
    }{
      \regbox[-0.25][0.25][#3][#4];      
    }
    \ifthenelse{\equal{#5}{}}{}{
      \node[below] () at (A) {#5};
    }
    \ifthenelse{\equal{#6}{}}{}{
      \node[above] () at (B) {#6};
    }
  \end{scope}
}

\newcommandx{\ugate}[5][1=0,2=0,3=,4=,5=]{
  \begin{scope}[shift={(#1,#2)}]
      \draw[line width = 0.5pt] (0,0)--++(0,0.25);
      \draw[line width = 0.5pt] (0,1.25)--++(0,0.25);
      \draw[line width = 0.5pt] (1,0)--++(0,0.25);
      \draw[line width = 0.5pt] (1,1.25)--++(0,0.25);
      \regbox[-0.25][0.25][#3][#4];
      \ifthenelse{\equal{#5}{p}}{
        \draw[line width = 2pt] (-0.1,0)--(1.1,0);
        \draw (0,0)--++(0,-0.1);
        \draw (1,0)--++(0,-0.1);
      }{}
  \end{scope}
}

\newcommandx{\dualgate}[7][1=0,2=0,3=,4=,5=,6=,7=]{
  \begin{scope}[shift={(#1,#2)}]
    \ifthenelse{\equal{#3}{l} \OR \equal{#3}{}}{
      \draw (0,-0.2)--(0.2,-0.2)--(0.2,0.2)--(0,0.2);
      \draw (0.2,0.2)--(0.5,0.5);
      \node () at (0.6,0.6) {$#5$};
      \draw (0.2,-0.2)--(0.5,-0.5);
      \node () at (0.6,-0.6) {$#7$};
    }{}
    \ifthenelse{\equal{#3}{r} \OR \equal{#3}{}}{
      \draw (0,-0.2)--(-0.2,-0.2)--(-0.2,0.2)--(0,0.2);
      \draw (-0.2,0.2)--(-0.5,0.5);
      \node () at (-0.6,0.6) {$#4$};
      \draw (-0.2,-0.2)--(-0.5,-0.5);
      \node () at (-0.6,-0.6) {$#6$};
    }{}
  \end{scope}
}

\newcommandx{\shortarc}[4][1=0,2=0,3=,4=]{
  \begin{scope}[shift={(#1,#2)}]
    \ifthenelse{\equal{#3}{r}}{
      \draw[line width = 0.5pt] (0,0) to[out=-90,in=0] (-0.1+0.06,0.1-0.18) to[out=180,in=-90] (-0.1,0.1);
      \pgfmathsetmacro{\flag}{-1};
    }{
      \draw[line width = 0.5pt] (0,0) to[out=90,in=0] (-0.06,0.18) to[out=180,in=90] (-0.1,0.1);
      \pgfmathsetmacro{\flag}{1};
    }
    \ifthenelse{\equal{#4}{l}}{
      \draw (0,0)--++(0,-0.2);
      \draw (-0.1,0.1)--++(0,-0.2);
    }{}
  \end{scope}
}

\newcommandx{\longarc}[4][1=0,2=0,3=,4=]{
  \begin{scope}[shift={(#1,#2)}]
    \ifthenelse{\equal{#3}{r}}{
      \draw[line width = 0.5pt] (0,0) to[out=-90,in=0] (-0.1*3+0.06*3,0.1*3-0.14*3) to[out=180,in=-90] (-0.1*3,0.1*3);
      \pgfmathsetmacro{\flag}{-1};
    }{
      \draw[line width = 0.5pt] (0,0) to[out=90,in=0] (-0.06*3,0.14*3) to[out=180,in=90] (-0.1*3,0.1*3);
      \pgfmathsetmacro{\flag}{1};
    }
    \ifthenelse{\equal{#4}{l}}{
      \draw (0,0)--++(0,-0.2);
      \draw (-0.1*3,0.1*3)--++(0,-0.2);
    }{}
  \end{scope}
}

\newcommandx{\idarc}[4][1=0,2=0,3=,4=]{
  \begin{scope}[shift={(#1,#2)}]
    \shortarc[0][0][#3][#4];
    \shortarc[-0.2][0.2][#3][#4];
  \end{scope}
}

\newcommandx{\swaparc}[4][1=0,2=0,3=,4=]{
  \begin{scope}[shift={(#1,#2)}]
    \longarc[0][0][#3][#4];
    \shortarc[-0.1][0.1][#3][#4];
  \end{scope}
}

\newcommandx{\idst}[3][1=0,2=0,3=]{
  \ifthenelse{\equal{#3}{r}}{
    \pgfmathsetmacro{\flag}{-1};
  }{
    \pgfmathsetmacro{\flag}{1};
  }
  \begin{scope}[shift={(#1,#2)}]
    \draw[line width = 0.5pt] (0,0) to[out=\flag*90,in=180] (0.15,\flag*0.2) to[out=0,in=\flag*90] (0.3,0);
    \draw[line width = 0.5pt] (0.4,0) to[out=\flag*90,in=180] (0.4+0.15,\flag*0.2) to[out=0,in=\flag*90] (0.7,0);
  \end{scope}
}

\newcommandx{\swapst}[3][1=0,2=0,3=]{
  \ifthenelse{\equal{#3}{r}}{
    \pgfmathsetmacro{\flag}{-1};
  }{
    \pgfmathsetmacro{\flag}{1};
  }
  \begin{scope}[shift={(#1,#2)}]
      \draw[line width = 0.5pt] (0,0) to[out=\flag*90,in=180] (0.4,\flag*0.25) to[out=0,in=\flag*90] (0.8,0);
      \draw[line width = 0.5pt] (0.3,0) to[out=\flag*90,in=180] (0.4,\flag*0.15) to[out=0,in=\flag*90] (0.5,0);
  \end{scope}
}

\newcommandx{\uonesitestack}[4][1=,2=,3=1,4=]
{
  \begin{scope}[shift={(#1,#2)}]
    \foreach \x in {#3,...,1}{
      \pgfmathsetmacro{\shiftx}{-0.1*(2*\x-2)};
      \pgfmathsetmacro{\shifty}{0.1*(2*\x-2)};
      \uonesite[\shiftx-0.1][\shifty+0.1][blue!50];
      \uonesite[\shiftx][\shifty][red!50];
    }
    \ifthenelse{\equal{#4}{id}}{
      \idarc[0.5][1.5];
    }{}
    \ifthenelse{\equal{#4}{swap}}{
      \swaparc[0.5][1.5];
    }{}
  \end{scope}
}
\newcommandx{\ugatestack}[6][1=0,2=0,3=1,4=,5=,6=]
{
  \begin{scope}[shift={(#1,#2)}]
    \foreach \x in {#3,...,1}{
      \pgfmathsetmacro{\shiftx}{-0.1*(2*\x-2)};
      \pgfmathsetmacro{\shifty}{0.1*(2*\x-2)};
      \ifthenelse{\equal{#6}{l}}{
        \ugatel[\shiftx-0.1][\shifty+0.1][][blue!50];
        \ugatel[\shiftx][\shifty][][red!50];
      }{}
      \ifthenelse{\equal{#6}{r}}{
        \ugater[\shiftx-0.1][\shifty+0.1][][blue!50];
        \ugater[\shiftx][\shifty][][red!50];
      }{}
      \ifthenelse{\equal{#6}{}}{
        \ugate[\shiftx-0.1][\shifty+0.1][blue!50];
        \ugate[\shiftx][\shifty][red!50];
      }{}
    }
    \ifthenelse{\equal{#4}{id}}{
      \idarc[0][1.5];
    }{}
    \ifthenelse{\equal{#4}{swap}}{
      \swaparc[0][1.5];
    }{}
    \ifthenelse{\equal{#5}{id}}{
      \idarc[1][1.5];
    }{}
    \ifthenelse{\equal{#5}{swap}}{
      \swaparc[1][1.5];
    }{}
  \end{scope}
}

\newcommandx{\dptri}[7][1=0,2=0,3=,4=,5=,6=,7=]{
  \begin{scope}[shift={(#1,#2)}]
    \ifthenelse{\equal{#3}{}}{
      \draw (0,0)--(-1,1.732)--(1,1.732)--cycle;
    }{
      \draw[fill=#3] (0,0)--(-1,1.732)--(1,1.732)--cycle;
    }
    \node () at (0,1.15) {#4};
    \node[below] () at (0,0) {#5};
    \node[left] () at (-1,1.732) {#6};
    \node[right] () at (1,1.732) {#7};
  \end{scope}
}

\newcommandx{\dptridash}[7][1=0,2=0,3=,4=,5=,6=,7=]{
  \begin{scope}[shift={(#1,#2)}]
    \dptri[0][0][#3][#4][#5][#6][#7]
    \draw[dashed] (0,0)--++(0,1.732*2/3)--(0-1,0+1.732);
    \draw[dashed] (0,1.732*2/3)--(0+1,0+1.732);
  \end{scope}
}

\newcommandx{\dtptri}[4][1=0,2=0,3=,4=]
{
  \begin{scope}[shift={(#1,#2)}]
    \draw (0,0)--(-1,1.732)--(1,1.732)--cycle;
    \draw (0,0)--(-1,-1.732)--(1,-1.732)--cycle;
  \end{scope}
}

\newcommandx{\fineq}[4][1=-.8ex,2=1,3=1]{
  \begin{tikzpicture}[baseline={([yshift=#1]current  bounding  box.center)}, scale = #2, every node/.style={scale = #3}]
    #4
  \end{tikzpicture}
}

\newcommandx{\ideq}[1][1=]{
  \fineq{
    \idst[0][0][#1]
  }
}
\newcommandx{\swapeq}[1][1=]{
  \fineq{
    \swapst[0][0][#1]
  }
}

\newcommandx{\idket}[0]{
  |
  \fineq[-0.6ex]{
    \idst[0][0][r]
  } \rangle 
}

\newcommandx{\swapket}[0]{
  |
  \fineq[-0.4ex]{
    \swapst[0][0][r]
  } \rangle 
}

\newcommandx{\idbra}[0]{
  \langle 
  \fineq{
    \idst[0][0][]
  } |
}

\newcommandx{\swapbra}[0]{
  \langle 
  \fineq{
    \swapst[0][0][]
  } |
}

\newcommandx{\uonesiteeq}[3][1=,2=,3=]
{
  \fineq{
    \uonesite[0][0][][#1][#2][#3];
    \node () at (-0.5,0.75) {};
    \node () at (1.5,0.75) {};
  }
}

\newcommandx{\ugateeq}[3][1=,2=,3=]
{
  \fineq{
    \ugate[0][0][][#1][#2][#3];
    \node () at (-0.5,0.75) {};
    \node () at (1.5,0.75) {};
  }
}

\newcommandx{\uonesitetwoeq}[4][1=,2=,3=,4=]
{
  \fineq{
    \uonesite[0][0][][#1][#3];
    \uonesite[0][1.5][][#2][#4];
    \node () at (-0.5,0.75) {};
    \node () at (1.5,0.75) {};
  }
}

\newcommandx{\hfbox}[5][1=0,2=0,3=,4=l,5=]{
  \ifthenelse{\equal{#4}{r}}{
    \pgfmathsetmacro{\flag}{-1};
  }{
    \pgfmathsetmacro{\flag}{1};
  }
  \begin{scope}[shift={(#1,#2)}]
    \draw (0,0) --++ (\flag*1,0) --++ (0,1) --++(-\flag*1,0);
    \node () at (\flag*0.5,0.5) {#3};
    \ifthenelse{\equal{#5}{s}}{
      \ifthenelse{\equal{#4}{l}}{
        \draw (0.9,1)--++(0,0.1) --++ (-1,0);
      }{}
      \ifthenelse{\equal{#4}{r}}{
        \draw (-1,0.1)--++(-0.1,0)--++(0,1) --++ (1,0);
      }{}
    }{}
  \end{scope}
}

\newcommandx{\tribox}[4][1=,2=,3=,4=]{
  \fineq[-0.8ex][0.35][0.8]{
    \hfbox[0][1.25][$#1$][l][#4];
    \hfbox[2.25][1.25][$#2$][r][#4];
    \sqzbox[0][0][][$#3$][#4];
    \node () at (-0.25,1.25) {};
    \node () at (2.5,1.25) {};
  }
}

\newcommandx{\lshapebox}[4][1=,2=,3=l,4=]{
  \fineq[-0.8ex][0.35][0.55]{
    \ifthenelse{\equal{#3}{l}}{
      \hfbox[0][1.25][$#1$][l][#4];
    }{
      \hfbox[2.25][1.25][$#1$][r][#4];
    }
    \sqzbox[0][0][][$#2$][#4];
    \node () at (-0.25,1.25) {};
    \node () at (2.5,1.25) {};
  }
}

\newcommandx{\hexbox}[8][1=,2=,3=,4=,5=,6=,7=,8=]{
  \fineq[-0.8ex][0.3][0.48]{
    \sqzbox[0][2.5][][$#1$][#8];
    \sqzbox[2.5][2.5][][$#2$][#8];
    \hfbox[0][1.25][$#3$][l][#8];
    \sqzbox[1.25][1.25][][$#4$][#8];
    \hfbox[2.5+2.25][1.25][$#5$][r][#8];
    \sqzbox[0][0][][$#6$][#8];
    \sqzbox[2.5][0][][$#7$][#8];
    \node () at (-0.25,1.25) {};
    \node () at (5,1.25) {};
  }
}

\newcommandx{\regboxteq}[5][1=,2=,3=,4=,5=]{
  \fineq[-0.8ex][0.5][0.7]{
    \regboxt[0][0][][#1][$#2$][$#3$][$#4$][$#5$];
  }
}

\newcommandx{\enttworandqubit}[2][1=,2=]{
  \fineq[-0.8ex][0.6][0.5]{
    \sqzbox[0][0][][4U];
    \draw (0.5,1)--++(0,0.5) node[above] () {$+$};
    \draw (2.25-0.5,1)--++(0,0.5) node[above] () {$-$};
    \draw (0.5,0)--++(0,-0.5);
    \draw (2.25-0.5,0)--++(0,-0.5);
    \draw (0.5,-1.5)--++(0,-0.5) node[below] () {$e$};
    \draw (2.25-0.5,-1.5)--++(0,-0.5) node[below] () {$e$};
    \draw (0,-1.5) rectangle (1,-0.5);
    \draw (2.25-1,-1.5) rectangle (2.25,-0.5);
    \node () at (0.5,-1) {#1};
    \node () at (2.25-0.5,-1) {#2};
  }
}

\newcommandx{\ugatetext}[7][1=0,2=0,3=,4=,5=,6=,7=]
{
  \begin{scope}[shift={(#1,#2)}]
    \ugate[0][0][][#3];
    \node[above] () at (0,1.5) {$#4$};
    \node[above] () at (1,1.5) {$#5$};
    \node[below] () at (0,0) {$#6$};
    \node[below] () at (1,0) {$#7$};
  \end{scope}
}

\newcommandx{\stairgateone}[1][1=]{
  \fineq[-0.8ex][0.75][0.75]{
    \ugatetext[0][0][#1][+][-][\frac{+}{q^2}][\frac{-}{q^2}];
  }
}

\newcommandx{\stairgatebottom}[1][1=]{
  \fineq[-0.8ex][0.75][0.75]{
    \ugatetext[0][0][#1][+][][\frac{+}{q^2}][\frac{-}{q^2}];
  }
}

\newcommandx{\stairgatetwo}[2][1=,2=]{
  \fineq[-0.8ex][0.75][0.75]{
    \ugatetext[0][0][#1][+][][\frac{+}{q^2}][\frac{-}{q^2}];
    \ugatetext[1][1.5][#2][+][-][][\frac{-}{q^2}];
    \node () at (-0.5,1) {};
    \node () at (2.5,1) {};
  }
}

\newcommandx{\sqzboxwithantennae}[7][1=0,2=0,3=,4=,5=,6=,7=]
{
  \begin{scope}[shift={(#1,#2)}]
    \sqzbox[0][0][][#3];
    \ifthenelse{\equal{#4}{}}{
    }{
      \draw (0.625,1)--++(0,0.5) node[above] {#4};
    }
    \ifthenelse{\equal{#5}{}}{
    }{
      \draw (2.25-0.625,1)--++(0,0.5) node[above] {#5};
    }
    \ifthenelse{\equal{#6}{}}{
    }{
      \draw (0.625,0)--++(0,-0.5) node[below] {#6};
    }
    \ifthenelse{\equal{#7}{}}{
    }{
      \draw (2.25-0.625,0)--++(0,-0.5) node[below] {#7};
    }
  \end{scope}
  
}

\newcommandx{\pyramidone}[1][1=]{
  \fineq[-0.8ex][0.5][0.8]{
    \sqzboxwithantennae[0][0][#1][$+$][$-$];
    \node () at (-0.25,1.25) {};
    \node () at (2.5,1.25) {};
  }
}

\newcommandx{\pyramidtwo}[3][1=,2=,3=]{
  \fineq[-0.8ex][0.5][0.8]{
    \sqzboxwithantennae[1.25][1.25][#1][$+$][$-$];
    \sqzboxwithantennae[0][0][#2][$+$][];
    \sqzboxwithantennae[2.5][0][#3][][$-$];
    \node () at (-0.25,1.25) {};
    \node () at (5,1.25) {};
  }
}

\newcommandx{\pyramidthree}[6][1=,2=,3=,4=,5=,6=]{
  \fineq[-0.8ex][0.5][0.8]{
    \sqzboxwithantennae[2.5][2.5][#1][$+$][$-$];
    \sqzboxwithantennae[1.25][1.25][#2][$+$][];
    \sqzboxwithantennae[3.75][1.25][#3][][$-$];
    \sqzboxwithantennae[0][0][#4][$+$][];
    \sqzboxwithantennae[2.5][0][#5];
    \sqzboxwithantennae[5][0][#6][][$-$];
    \node () at (-0.25,1.25) {};
    \node () at (7.5,1.25) {};
  }
}

\newcommandx{\pyramidZone}[2][1=,2=]{
  \fineq[-0.8ex][0.4][0.64]{
    \sqzboxwithantennae[0][0][][$+$][$-$][$#1$][$#2$];
    \node () at (-0.25,1.25) {};
    \node () at (2.5,1.25) {};
  }
}

\newcommandx{\pyramidZtwo}[4][1=,2=,3=,4=,]{
  \fineq[-0.8ex][0.4][0.64]{
    \ifthenelse{\equal{#1}{l}}{
      \sqzboxwithantennae[1.25][1.25][][$+$][$-$][][$#4$];
      \sqzboxwithantennae[0][0][][$+$][][$#2$][$#3$];
    }{
      \ifthenelse{\equal{#1}{r}}{
        \sqzboxwithantennae[1.25][1.25][][$+$][$-$][$#2$];
        \sqzboxwithantennae[2.5][0][][][$-$][$#3$][$#4$];
      }{
        \sqzboxwithantennae[1.25][1.25][][$+$][$-$];
        \sqzboxwithantennae[0][0][][$+$][][$#1$][$#2$];
        \sqzboxwithantennae[2.5][0][][][$-$][$#3$][$#4$];
      }
    }
    \node () at (-0.25,1.25) {};
    \node () at (5,1.25) {};
  }
}

\newcommandx{\pyramidZttwo}[3][1=,2=,3=]{
  \fineq[-0.8ex][0.4][0.64]{
    \sqzboxwithantennae[2.5][2.5][][$+$][$-$];
    \sqzboxwithantennae[1.25][1.25][][$+$][];
    \sqzboxwithantennae[3.75][1.25][][][$-$];
    \hfbox[0][1.25][$+$][l];
    \sqzbox[0][0][][$#1$];
    \sqzboxwithantennae[2.5][0][$#2$];
    \hfbox[2.25+5][1.25][$-$][r];    
    \sqzbox[5][0][][$#3$];
    \node () at (-0.25,1.25) {};
    \node () at (7.5,1.25) {};
  }
}

\newcommandx{\stairgateany}[2][1=,2=]{
  \fineq[-0.8ex][0.75][0.75]{
    \ugatetext[0][0][#1][+][][\frac{+}{q^2}][\frac{-}{q^2}];
    \node () at (1.5,2.4) {$\iddots$};
    \draw (0.75,1.5)--++(0,0.5);
    \draw (0.75,1.5)--++(0.5,0);
    \draw (2.25,3)--++(0,-0.5);
    \draw (2.25,3)--++(-0.5,0);
    \ugatetext[2][3][#2][+][-][][\frac{-}{q^2}];
    \node () at (-0.5,1) {};
    \node () at (3.5,1) {};
  }
}

\newcommandx{\stairgatethree}[3][1=,2=,3=]{
  \fineq[-0.8ex][0.75][0.75]{
    \ugatetext[0][0][#1][+][][\frac{+}{q^2}][\frac{-}{q^2}];
    \ugatetext[1][1.5][#2][+][][][\frac{-}{q^2}];
    \ugatetext[2][3][#3][+][-][][\frac{-}{q^2}];
    \node () at (-0.5,1) {};
    \node () at (3.5,1) {};
  }
}

\newcommandx{\pingbox}[3][1=,2=,3=]{
  \fineq[-0.8ex][0.5][0.5]{
    \ugate[1][1.5][][#1];
    \ugate[0][0][][#2];
    \ugate[2][0][][#3];
    \foreach \x in {0,...,3}{
      \node[below] () at (\x,0) {$e$};
    }
    \draw[line width = 0.5pt] (0,1.5)--++(0,1.5);
    \draw[line width = 0.5pt] (3,1.5)--++(0,1.5);
    \node[above] () at (0,3) {$+$};
    \node[above] () at (1,3) {$+$};
    \node[above] () at (2,3) {$-$};
    \node[above] () at (3,3) {$-$};
  }
}

\newcommandx{\opsqu}[1][1=1]{
  \fineq[-0.8ex][0.2]{
    \ifthenelse{\equal{#1}{1}}{
      \draw[line width = 1pt] (0.125,0)--++(0,1);
      \draw[line width = 1pt] (0.875,0)--++(0,1);
    }{}
    \ifthenelse{\equal{#1}{2}}{
      \draw[line width = 1pt] (0,0)--++(1,1);
      \draw[line width = 1pt] (1,0)--++(-1,1);
    }{}
  }
}

\newcommandx{\bottomtri}[6][1=,2=,3=,4=,5=,6=]{
  \fineq[-0.8ex][0.4][0.6]{
    \dptri[0][0][][$#6$][][$#2$][$#3$];
    \node[right] () at (0,0) {$#1$};
    \draw (0,0)--++(-120:1) node[below left] () {$#4$};
    \draw (0,0)--++(-60:1) node[below right] () {$#5$};
  }
}

\newcommandx{\jabc}[4][1=,2=,3=,4=]
{
  \fineq[-0.8ex][0.4][0.8]{
    \dptri[0][0][][$#4$][$#1$][$#2$][$#3$];
  }
}

\newcommandx{\joneperp}[8][1=,2=,3=,4=,5=,6=,7=,8=]
{
  \fineq[-0.8ex][0.4][0.8]{
    \dptri[0][0][][$#4$][$#1$][$#2$][$#3$];
    \dptri[-1][-1.732][][$#6$][][$#5$][];
    \dptri[1][-1.732][][$#8$][][][$#7$];
  }
}

\newcommandx{\sdw}[1][1=l]
{
  \fineq[-0.4ex][0.5][1]{
    \dptri;
    \ifthenelse{\equal{#1}{l}}{
      \draw (0,1.155) -- +(90:1.077);
      \draw (0,1.155) -- +(210:1.077);
    }{}
    \ifthenelse{\equal{#1}{r}}{
      \draw (0,1.155) -- +(90:1.077);
      \draw (0,1.155) -- +(-30:1.077);
    }{}
    \ifthenelse{\equal{#1}{lp}}{
      \draw (0,1.2) -- +(90:1.077);
      \draw (0,1.2) -- +(210:1.077);
      \draw (0,1.1) -- +(-30:1.02);
      \draw (0,1.1) -- +(210:1.02);
    }{}
    \ifthenelse{\equal{#1}{lr}}{
      \draw (0,0)--(-1,-1.732)--(1,-1.732)--cycle;
      \draw (-0.05,1.732+0.5)--++(0,-0.5-1/3*1.732)--++(-1,-1/3*1.732)--++(0,-2/3*1.732)--++(1,-1/3*1.732)--++(0,-0.5-1/3*1.732);
    }{}
    \ifthenelse{\equal{#1}{rl}}{
      \draw (0,0)--(-1,-1.732)--(1,-1.732)--cycle;
      \draw (0.05,1.732+0.5)--++(0,-0.5-1/3*1.732)--++(1,-1/3*1.732)--++(0,-2/3*1.732)--++(-1,-1/3*1.732)--++(0,-0.5-1/3*1.732);
    }{}
    \node () at (-1.1,0) {};
    \node () at (1.1,0) {};
  }
}

\newcommandx{\sdwi}[1][1=]
{
  \fineq[-0.8ex][0.5][1]{
    \dptridash;
    \ifthenelse{\equal{#1}{l}}{
      \draw (0,1.732+0.5)--(0,1.732)--(-0.5,0.5*1.732)--++(-0.25*1.732,-0.25);
    }{}
    \ifthenelse{\equal{#1}{t}}{
      \draw (0,1.732+0.5)--(0,1.732)--(+0.5,0.5*1.732);
      \draw[red] (0.5, 0.5*1.732)--(-0.5,0.5*1.732);
      \draw (-0.5, 0.5*1.732)--++(-0.25*1.732,-0.25);
    }{}
    \ifthenelse{\equal{#1}{tri}}{
      \draw (0,1.732+0.5)--(0,1.732)--(-0.5,0.5*1.732)--++(-0.25*1.732,-0.25);
      \draw[red] (-0.4,0.5*1.732)--++(0.85,0)--++(120:0.85)--cycle;
    }{}
    \ifthenelse{\equal{#1}{lpd}}{
      \draw (0,1.732+0.5)--(0,1.732)--(-0.5,0.5*1.732)--++(-0.25*1.732,-0.25);
      \draw (-0.45,0.45*1.732)--++(60:0.9)--++(-60:0.9);
      \draw (-0.45,0.45*1.732)--++(-0.25*1.732,-0.25);
      \draw (0.45,0.45*1.732)--++(0.25*1.732,-0.25);
    }{}
    \ifthenelse{\equal{#1}{lpw}}{
      \draw (0,1.732+0.5)--(0,1.732)--(-0.5,0.5*1.732)--++(-0.25*1.732,-0.25);
      \draw (-0.45,0.45*1.732)--++(-0.25*1.732,-0.25);
      \draw (0.45,0.45*1.732)--++(0.25*1.732,-0.25);
      \draw[red] (-0.45,0.45*1.732)--(0.45,0.45*1.732);
    }{}
    \ifthenelse{\equal{#1}{lpt}}{
      \draw (0,1.732+0.5)--(0,1.732)--(+0.5,0.5*1.732);
      \draw[red] (0.5, 0.5*1.732)--(-0.5,0.5*1.732);
      \draw (-0.5, 0.5*1.732)--++(-0.25*1.732,-0.25);
      \draw (-0.45,0.45*1.732)--++(60:0.9)--++(-60:0.9);
      \draw (-0.45,0.45*1.732)--++(-0.25*1.732,-0.25);
      \draw (0.45,0.45*1.732)--++(0.25*1.732,-0.25);
    }{}
    \ifthenelse{\equal{#1}{lpwt}}{
      \draw (0,1.732+0.5)--(0,1.732)--(+0.5,0.5*1.732);
      \draw[red] (0.5, 0.5*1.732)--(-0.5,0.5*1.732);
      \draw (-0.5, 0.5*1.732)--++(-0.25*1.732,-0.25);
      \draw (-0.45,0.45*1.732)--++(-0.25*1.732,-0.25);
      \draw (0.45,0.45*1.732)--++(0.25*1.732,-0.25);
      \draw[red] (-0.45,0.45*1.732)--(0.45,0.45*1.732);
    }{}
    \node () at (-1.1,0) {};
    \node () at (1.1,0) {};
  }
}

\newcommandx{\ddw}[1][1=]
{
  \fineq[-0.4ex][0.5][1]{
    \dptri;
    \ifthenelse{\equal{#1}{l}}{
      \draw (0,1.155) -- +(90:1.077);
      \draw (0,1.155) -- +(210:1.077);
      \draw (-0.1,1.23) -- +(90:1.008);
      \draw (-0.1,1.23) -- +(210:1.02);  
    }{}
    \ifthenelse{\equal{#1}{lr}}{
      \draw (0.06,1.155) -- +(90:1.077);
      \draw (0.06,1.155) -- +(-30:1.077);
      \draw (-0.06,1.155) -- +(90:1.077);
      \draw (-0.06,1.155) -- +(210:1.077);
    }{}
    \ifthenelse{\equal{#1}{r}}{
      \draw (0,1.155) -- +(90:1.077);
      \draw (0,1.155) -- +(-30:1.077);
      \draw (0.1,1.23) -- +(90:1.008);
      \draw (0.1,1.23) -- +(-30:1.02);  
    }{}
    \ifthenelse{\equal{#1}{lp}}{
      \draw (0,1.04) -- +(-30:1.02);
      \draw (0,1.04) -- +(210:1.02);
      \draw (-0.1,1.732+0.5)--++(0,-0.45-1/3*1.732)--++(210:1.0);
      \draw (-0.0,1.732+0.5)--++(0,-0.5-1/3*1.732)--++(210:1.077);
    }{}
    \ifthenelse{\equal{#1}{lrp}}{
      \draw (0.05,1.732+0.5)--++(0,-0.5-1/3*1.732)--++(1,-1/3*1.732);
      \draw (-0.05,1.732+0.5)--++(0,-0.5-1/3*1.732)--++(-1,-1/3*1.732);
      \draw (0,0.616*1.732)--++(210:2/3*1.76);
      \draw (0,0.616*1.732)--++(-30:2/3*1.76);
    }{}
    \ifthenelse{\equal{#1}{lrpp}}{
      \draw (0.05,1.732+0.5)--++(0,-0.5-1/3*1.732)--++(1,-1/3*1.732);
      \draw (-0.05,1.732+0.5)--++(0,-0.5-1/3*1.732)--++(-1,-1/3*1.732);
      \draw (0,0.616*1.732)--++(210:2/3*1.76);
      \draw (0,0.616*1.732)--++(-30:2/3*1.76);
      \draw (0,0.54*1.732)--++(210:2/3*1.66);
      \draw (0,0.54*1.732)--++(-30:2/3*1.66);
    }{}
    \ifthenelse{\equal{#1}{regvert}}{
      \draw (0.05,1.732+0.5)--++(0,-0.5-1/3*1.732)--++(1,-1/3*1.732);
      \draw (-0.05,1.732+0.5)--++(0,-0.5-1/3*1.732)--++(-1,-1/3*1.732);
      \draw (0,0.616*1.732)--++(210:2/3*1.76) coordinate (A);
      \node[left] () at (A) {\footnotesize $(12)$};
      \draw (0,0.616*1.732)--++(-30:2/3*1.76) coordinate (B);
      \node[right] () at (B) {\footnotesize$(34)$};
      \node[above] () at (0,1.732+0.5) {\footnotesize $(12)(34)$};
      \node () at (-1-0.3,-1/1.732) {\footnotesize $\alpha$};
      \node () at (1+0.65,-1/1.732+0.1) {\footnotesize $\alpha^{-1}$};
      \draw[dashed] (A)--++(0,-2/3*1.55)--(0,-0.616*1.732);
      \draw[dashed] (B)--++(0,-2/3*1.55)--(0,-0.616*1.732);
    }{}
    \ifthenelse{\equal{#1}{specvert}}{
      \draw (0.05,1.732+0.5)--++(0,-0.5-1/3*1.732)--++(1,-1/3*1.732);
      \draw (-0.05,1.732+0.5)--++(0,-0.5-1/3*1.732)--++(-1,-1/3*1.732);
      \draw (0,0.616*1.732)--++(210:2/3*1.65);
      \draw (0,0.616*1.732)--++(-30:2/3*1.65);
      \fill (0,2/3*1.732-0.03) circle (0.1);
    }{}
    \node () at (-1.1,0) {};
    \node () at (1.1,0) {};
  }
}

\newcommandx{\ddwi}[1][1=]
{
  \fineq[-0.4ex][0.5][1]{
    \dptridash;
    \ifthenelse{\equal{#1}{l}}{
      \draw (0,1.732+0.5)--(0,1.732)--(-0.5,0.5*1.732)--++(-0.25*1.732,-0.25);
      \draw (-0.1,1.732+0.5)--(-0.1,1.732)--(-0.55,0.55*1.732)--++(-0.25*1.732,-0.25);
    }{}
    \ifthenelse{\equal{#1}{lr}}{
      \draw (-0.05,1.732+0.5)--++(0,-0.5)--(-0.55,0.55*1.732)--++(-0.25*1.732,-0.25);
      \draw (0.05,1.732+0.5)--++(0,-0.5)--(0.55,0.55*1.732)--++(0.25*1.732,-0.25);
    }{}
    \ifthenelse{\equal{#1}{r}}{
      \draw (0,1.732+0.5)--(0,1.732)--(0.5,0.5*1.732)--++(0.25*1.732,-0.25);
      \draw (0.1,1.732+0.5)--(0.1,1.732)--(0.55,0.55*1.732)--++(0.25*1.732,-0.25);
    }{}
    \ifthenelse{\equal{#1}{lrt}}{
      \draw (-0.05,1.732+0.5)--++(0,-0.5)--(-0.55,0.55*1.732)--++(-0.25*1.732,-0.25);
      \draw (0.05,1.732+0.5)--++(0,-0.5)--(-0.5,0.5*1.732);
      \draw[red] (-0.5,0.5*1.732)--(0.5,0.5*1.732);
      \draw (0.5,0.5*1.732)--++(0.25*1.732,-0.25);
    }{}
    \ifthenelse{\equal{#1}{lrtt}}{
      \draw (-0.05,1.732+0.5)--++(0,-0.5)--(0.55,0.55*1.732);
      \draw[red] (0.55,0.55*1.732)--(-0.55,0.55*1.732);
      \draw (-0.55,0.55*1.732)--++(-0.25*1.732,-0.25);
      \draw (0.05,1.732+0.5)--++(0,-0.5)--(-0.5,0.5*1.732);
      \draw[red] (-0.5,0.5*1.732)--(0.5,0.5*1.732);
      \draw (0.5,0.5*1.732)--++(0.25*1.732,-0.25);
    }{}
    \ifthenelse{\equal{#1}{lrtri}}{
      \draw (-0.05,1.732+0.5)--++(0,-0.5)--(-0.55,0.55*1.732)--++(-0.25*1.732,-0.25);
      \draw (0.05,1.732+0.5)--++(0,-0.5)--(0.55,0.55*1.732)--++(0.25*1.732,-0.25);
      \draw[red] (-0.45,0.5*1.732)--++(0.9,0)--++(120:0.9)--cycle;
    }{}
    \ifthenelse{\equal{#1}{lt}}{
      \draw (0,1.732+0.5)--(0,1.732)--(+0.5,0.5*1.732);
      \draw[red] (0.5, 0.5*1.732)--(-0.5,0.5*1.732);
      \draw (-0.1,1.732+0.5)--(-0.1,1.732)--(-0.55,0.55*1.732)--++(-0.25*1.732,-0.25);
      \draw (-0.5, 0.5*1.732)--++(-0.25*1.732,-0.25);
    }{}
    \ifthenelse{\equal{#1}{ltt}}{
      \draw (0,1.732+0.5)--(0,1.732)--(+0.5,0.5*1.732);
      \draw[red] (0.5, 0.5*1.732)--(-0.5,0.5*1.732);
      \draw (-0.5, 0.5*1.732)--++(-0.25*1.732,-0.25);
      \draw (-0.1,1.732+0.5)--(-0.1,1.732)--(0.35,0.55*1.732);
      \draw[red] (0.35,0.55*1.732)--(-0.55,0.55*1.732);
      \draw (-0.55,0.55*1.732)--++(-0.25*1.732,-0.25);
    }{}
    \ifthenelse{\equal{#1}{ltri}}{
      \draw (0,1.732+0.5)--(0,1.732)--(-0.5,0.5*1.732)--++(-0.25*1.732,-0.25);
      \draw[red] (-0.4,0.5*1.732)--++(0.85,0)--++(120:0.85)--cycle;
      \draw (-0.1,1.732+0.5)--(-0.1,1.732)--(-0.55,0.55*1.732)--++(-0.25*1.732,-0.25);
    }{}
    \ifthenelse{\equal{#1}{lp}}{
      \draw (0,1.732+0.5)--(0,1.732)--(-0.5,0.5*1.732)--++(-0.25*1.732,-0.25);
      \draw (-0.45,0.45*1.732)--++(60:0.9)--++(-60:0.9);
      \draw (-0.45,0.45*1.732)--++(-0.25*1.732,-0.25);
      \draw (0.45,0.45*1.732)--++(0.25*1.732,-0.25);
      \draw (-0.1,1.732+0.5)--(-0.1,1.732)--(-0.55,0.55*1.732)--++(-0.25*1.732,-0.25);
    }{}
    \ifthenelse{\equal{#1}{lpw}}{
      \draw (0,1.732+0.5)--(0,1.732)--(-0.5,0.5*1.732)--++(-0.25*1.732,-0.25);
      \draw (-0.45,0.45*1.732)--++(-0.25*1.732,-0.25);
      \draw (0.45,0.45*1.732)--++(0.25*1.732,-0.25);
      \draw[red] (-0.45,0.45*1.732)--(0.45,0.45*1.732);
      \draw (-0.1,1.732+0.5)--(-0.1,1.732)--(-0.55,0.55*1.732)--++(-0.25*1.732,-0.25);
    }{}
    \ifthenelse{\equal{#1}{lpt}}{
      \draw (0,1.732+0.5)--(0,1.732)--(+0.5,0.5*1.732);
      \draw[red] (0.5, 0.5*1.732)--(-0.5,0.5*1.732);
      \draw (-0.5, 0.5*1.732)--++(-0.25*1.732,-0.25);
      \draw (-0.45,0.45*1.732)--++(60:0.9)--++(-60:0.9);
      \draw (-0.45,0.45*1.732)--++(-0.25*1.732,-0.25);
      \draw (0.45,0.45*1.732)--++(0.25*1.732,-0.25);
      \draw (-0.1,1.732+0.5)--(-0.1,1.732)--(-0.55,0.55*1.732)--++(-0.25*1.732,-0.25);
    }{}
    \ifthenelse{\equal{#1}{lpwt}}{
      \draw (0,1.732+0.5)--(0,1.732)--(+0.5,0.5*1.732);
      \draw[red] (0.5, 0.5*1.732)--(-0.5,0.5*1.732);
      \draw (-0.5, 0.5*1.732)--++(-0.25*1.732,-0.25);
      \draw (-0.45,0.45*1.732)--++(-0.25*1.732,-0.25);
      \draw (0.45,0.45*1.732)--++(0.25*1.732,-0.25);
      \draw[red] (-0.45,0.45*1.732)--(0.45,0.45*1.732);
      \draw (-0.1,1.732+0.5)--(-0.1,1.732)--(-0.55,0.55*1.732)--++(-0.25*1.732,-0.25);
    }{}
    \node () at (-1.1,0) {};
    \node () at (1.1,0) {};
  }
}

\newcommandx{\ddwhex}[0]
{
  \draw (0.05,1.732+0.5)--++(0,-0.5-1/3*1.732)--++(1,-1/3*1.732)--++(0,-2/3*1.732)--++(-1,-1/3*1.732)--++(0,-0.5-1/3*1.732);
  \draw (-0.05,1.732+0.5)--++(0,-0.5-1/3*1.732)--++(-1,-1/3*1.732)--++(0,-2/3*1.732)--++(1,-1/3*1.732)--++(0,-0.5-1/3*1.732);
  \draw (0,0.616*1.732)--++(210:2/3*1.65)--++(0,-2/3*1.55)--(0,-0.616*1.732);
  \draw (0,0.616*1.732)--++(-30:2/3*1.65)--++(0,-2/3*1.55)--(0,-0.616*1.732);
  \fill (0,2/3*1.732-0.03) circle (0.1);
  \fill (0,-2/3*1.732+0.03) circle (0.1);
}

\newcommandx{\ddwtwostep}[1][1=]{
  \fineq[-0.4ex][0.5][1]{
    \dtptri;
    \ifthenelse{\equal{#1}{}}{
      \draw (0.05,1.732+0.5)--++(0,-0.5-1/3*1.732);
      \draw (-0.05,1.732+0.5)--++(0,-0.5-1/3*1.732);
      \draw (0.05,-1.732-0.5)--++(0,0.5+1/3*1.732);
      \draw (-0.05,-1.732-0.5)--++(0,0.5+1/3*1.732);
      \node () at (0,2/3*1.732-0.2) {$\vdots$};
      \node () at (0,-2/3*1.732+0.5) {$\vdots$};
    }{}
    \ifthenelse{\equal{#1}{ll}}{
      \draw (-0.05,1.732+0.5)--++(0,-0.5-1/3*1.732)--++(-1,-1/3*1.732)--++(0,-2/3*1.732)--++(1,-1/3*1.732)--++(0,-0.5-1/3*1.732) coordinate (A);
      \draw (0.05,1.732+0.5)--++(0,-0.5-0.05-1/3*1.732)--(0,0.616*1.732);
      \draw (0,0.616*1.732)--++(210:2/3*1.65)--++(0,-2/3*1.55)--(0,-0.616*1.732)--++(0.05,-0.05/1.732)--([shift=({0.1,0})]A);
    }{}
    \ifthenelse{\equal{#1}{lr}}{
      \draw (-0.05,1.732+0.5)--++(0,-0.5-1/3*1.732)--++(-1,-1/3*1.732)--++(0,-2/3*1.732)--++(1,-1/3*1.732)--++(0,-0.5-1/3*1.732);
      \draw (0.05,1.732+0.5)--++(0,-0.5-1/3*1.732)--++(1,-1/3*1.732)--++(0,-2/3*1.732)--++(-1,-1/3*1.732)--++(0,-0.5-1/3*1.732);
    }{}
    \ifthenelse{\equal{#1}{rl}}{
      \draw (-0.05,1.732+0.5)--++(0,-0.5-1/3*1.732)--++(1,-1/3*1.732)--++(0,-2/3*1.732)--++(-1,-1/3*1.732)--++(0,-0.5-1/3*1.732);
      \draw (0.05,1.732+0.5)--++(0,-0.5-1/3*1.732)--++(-1,-1/3*1.732)--++(0,-2/3*1.732)--++(1,-1/3*1.732)--++(0,-0.5-1/3*1.732);
    }{}
    \ifthenelse{\equal{#1}{rr}}{
      \draw (0.05,1.732+0.5)--++(0,-0.5-1/3*1.732)--++(1,-1/3*1.732)--++(0,-2/3*1.732)--++(-1,-1/3*1.732)--++(0,-0.5-1/3*1.732) coordinate (A);
      \draw (-0.05,1.732+0.5)--++(0,-0.5-0.05-1/3*1.732)--(0,0.616*1.732);
      \draw (0,0.616*1.732)--++(-30:2/3*1.65)--++(0,-2/3*1.55)--(0,-0.616*1.732)--++(-0.05,-0.05/1.732)--([shift=({-0.1,0})]A);
    }{}
    \ifthenelse{\equal{#1}{hex}}{
      \ddwhex;
    }{}
    \ifthenelse{\equal{#1}{hexa}}{
      \ddwhex;
      \node[above] () at (0,1.732+0.5) {\footnotesize $(12)(34)$};
      \node[below] () at (0,-1.732-0.5) {\footnotesize $(12)(34)$};
      \node[left] () at (-1,0) {\footnotesize $(14)(23)$};
      \node[right] () at (1,0) {\footnotesize $(24)(13)$};
    }{}
    \ifthenelse{\equal{#1}{hexb}}{
      \ddwhex;
      \node[above] () at (0,1.732+0.5) {\footnotesize $(12)(34)$};
      \node[below] () at (0,-1.732-0.5) {\footnotesize $(12)(34)$};
      \node[left] () at (-1,0) {\footnotesize $(24)(13)$};
      \node[right] () at (1,0) {\footnotesize $(14)(23)$};
    }{}
    \node () at (-1.1,0) {};
    \node () at (1.1,0) {};
  }
}

\newcommandx{\tdw}[1][1=]
{
  \fineq[-0.4ex][0.5][1]{
    \dptri;
    \ifthenelse{\equal{#1}{l}}{
      \draw (-0.1,1.732+0.5)--++(0,-0.45-1/3*1.732)--++(210:1.0);
      \draw (0,1.732+0.5)--++(0,-0.5-1/3*1.732)--++(210:1.077);
      \draw (0.1,1.732+0.5)--++(0,-0.55-1/3*1.732)--++(210:1.154);
    }{}
    \ifthenelse{\equal{#1}{r}}{
      \draw (-0.1,1.732+0.5)--++(0,-0.55-1/3*1.732)--++(-30:1.154);
      \draw (0,1.732+0.5)--++(0,-0.5-1/3*1.732)--++(-30:1.077);
      \draw (0.1,1.732+0.5)--++(0,-0.45-1/3*1.732)--++(-30:1.0);
    }{}
    \ifthenelse{\equal{#1}{llr}}{
      \draw (-0.1,1.732+0.5)--++(0,-0.45-1/3*1.732)--++(210:1.0);
      \draw (0,1.732+0.5)--++(0,-0.5-1/3*1.732)--++(210:1.077);
      \draw (0.1,1.732+0.5)--++(0,-0.45-1/3*1.732)--++(-30:1.0);
    }{}
    \ifthenelse{\equal{#1}{lrr}}{
      \draw (-0.1,1.732+0.5)--++(0,-0.45-1/3*1.732)--++(210:1.0);
      \draw (0,1.732+0.5)--++(0,-0.5-1/3*1.732)--++(-30:1.077);
      \draw (0.1,1.732+0.5)--++(0,-0.45-1/3*1.732)--++(-30:1.0);
    }{}
    \node () at (-1.1,0) {};
    \node () at (1.1,0) {};
  }
}

\newcommandx{\ugater}[4][1=0,2=0,3=,4=]{
  \begin{scope}[shift={(#1,#2)}]
      \ifthenelse{\equal{#4}{}}{
      }{
        \fill[#4] (0.75,0.25)--++(0.5,0)--++(0,1)--++(-0.5,0)--cycle;
      }
      \ifthenelse{\equal{#3}{p}}{
        \draw[line width = 2pt] (0.75,0)--(1.1,0);
        \draw (1,0)--++(0,-0.1);
      }{}
      \draw[line width = 0.5pt] (1,0)--++(0,0.25);
      \draw[line width = 0.5pt] (1,1.25)--++(0,0.25);
      \draw[line width = 0.5pt] (0.75,0.25)--++(0.5,0)--++(0,1)--++(-0.5,0);
  \end{scope}
}

\newcommandx{\ugatel}[4][1=0,2=0,3=,4=]{
  \begin{scope}[shift={(#1,#2)}]
      \ifthenelse{\equal{#4}{}}{
      }{
        \fill[#4] (0.25,0.25)--++(-0.5,0)--++(0,1)--++(0.5,0)--cycle;
      }
      \ifthenelse{\equal{#3}{p}}{
        \draw[line width = 2pt] (-0.1,0)--(0.25,0);
        \draw (0,0)--++(0,-0.1);
      }{}
      \draw[line width = 0.5pt] (0,0)--++(0,0.25);
      \draw[line width = 0.5pt] (0,1.25)--++(0,0.25);
      \draw[line width = 0.5pt] (0.25,0.25)--++(-0.5,0)--++(0,1)--++(0.5,0);
  \end{scope}
}

\newcommandx{\partitionZ}[1][1=]
{
  \foreach \x in {0,2,4}{
  	\ugate[\x][0][][][#1];
    \ugate[\x][3][][][#1];
  	\ugate[\x][6][][][#1];
  }
  \foreach \x in {1,3}{
    \ugate[\x][4.5][][][#1];
    \ugate[\x][1.5][][][#1];
  }
  \foreach \y in {1.5,4.5}{
  	\ugater[-1][\y][#1];
    \ugatel[5][\y][#1];
  }
}

\newcommandx{\ugateonlyp}[5][1=0,2=0,3=,4=,5=]{
  \begin{scope}[shift={(#1,#2)}]
      \draw[line width = 0.5pt] (0,0)--++(0,0.25);
      \draw[line width = 0.5pt] (0,1.25)--++(0,0.25);
      \draw[line width = 0.5pt] (1,0)--++(0,0.25);
      \draw[line width = 0.5pt] (1,1.25)--++(0,0.25);
      \draw[line width = 2pt] (-0.1,0)--(1.1,0);
      \draw (0,0)--++(0,-0.1);
      \draw (1,0)--++(0,-0.1);
  \end{scope}
}

\newcommandx{\ugateronlyp}[3][1=0,2=0,3=]{
  \begin{scope}[shift={(#1,#2)}]
    \draw[line width = 2pt] (0.75,0)--(1.1,0);
    \draw (1,0)--++(0,-0.1);
  \end{scope}
}

\newcommandx{\ugatelonlyp}[3][1=0,2=0,3=]{
  \begin{scope}[shift={(#1,#2)}]
    \draw[line width = 2pt] (-0.1,0)--(0.25,0);
    \draw (0,0)--++(0,-0.1);
  \end{scope}
}

\newcommandx{\partitionZc}[1][1=]
{
  \foreach \x in {0,...,5}{
  	\draw (\x, 0)--++(0,7.5);
  }
  \foreach \y in {0,...,4}{
  	\fill[black!40] (-0.25,1.25+\y*1.5)--++(5.5,0)--++(0,-1)--++(-5.5,0)--cycle;
    \draw (-0.25,1.25+\y*1.5)--++(5.5,0);
    \draw (-0.25,0.25+\y*1.5)--++(5.5,0);
  }
}

\newcommandx{\partitionZtwosite}[1][1=]
{
  \partitionZc
  \foreach \x in {0,2,4}{
  	\ugateonlyp[\x][0][][];
    \ugateonlyp[\x][3][][];
  	\ugateonlyp[\x][6][][];
  }
  \foreach \x in {1,3}{
    \ugateonlyp[\x][4.5][][];
    \ugateonlyp[\x][1.5][][];
  }
  \foreach \y in {1.5,4.5}{
  	\ugateronlyp[-1][\y];
    \ugatelonlyp[5][\y];
  }
}

\newcommandx{\partitionZonesite}[1][1=]
{
  \partitionZc
  \foreach \x in {0,...,5}{
   	\foreach \y in {0,1.5,3,4.5,6}{
      \begin{scope}[shift={(\x,\y)}]
        \draw[line width = 2pt] (-0.25,0)--(0.25,0);
      \end{scope}
    }
  }
}

\newcommandx{\threegate}[1][1=]
{\fineq[-0.8ex][0.4][0.8]{
  \ugate[-1][1.5][][$\sigma_b$];
  \ugate[1][1.5][][$\sigma_c$];
  \ugate[0][0][][$\sigma_a$];
}
}

\newcommandx{\ptensor}[2][1=,2=]
{
  \fineq[-0.8ex][0.35][0.8]{
    \sqzbox[0][0][#2][$#1$];
    \draw (0.5,0)--++(0,-0.25);
    \draw (1.75,0)--++(0,-0.25);
    \draw (0.5,1)--++(0,0.25);
    \draw (1.75,1)--++(0,0.25);
  }
}

\newcommandx{\pdecomp}[8][1=0,2=0,3=,4=,5=,6=,7=,8=]
{
  \begin{scope}[shift={(#1,#2)}]
    \drawbox[0][0][1][0.4][#6][$#5$];
    \drawbox[1.25][0][2.25][0.4][#8][$#7$];
    \drawbox[0][0.6][2.25][1][#4][$#3$];
    \draw (0.5,0)--++(0,-0.25);
    \draw (1.75,0)--++(0,-0.25);
    \draw (0.5,1)--++(0,0.25);
    \draw (1.75,1)--++(0,0.25);
  \end{scope}
}

\newcommandx{\pdecompeq}[8][1=0,2=0,3=,4=,5=,6=,7=,8=]
{
   \fineq[-0.8ex][0.5][0.55]{
     \pdecomp[#1][#2][#3][#4][#5][#6][#7][#8]
  }
}

\newcommandx{\hfpdecomp}[5][1=0,2=0,3=,4=,5=]
{
  \begin{scope}[shift={(#1,#2)}]
    \drawbox[0][0][1][0.4][#5][$#4$];
    \draw (0.5,0)--++(0,-0.25);
    \draw (0.5,1)--++(0,0.25);
    \ifthenelse{\equal{#3}{l}}{
      \draw (0,0.6)--++(1,0)--++(0,0.4)--++(-1,0);
    }{}
    \ifthenelse{\equal{#3}{r}}{
      \draw (1,0.6)--++(-1,0)--++(0,0.4)--++(1,0);
    }{}
  \end{scope}
}

\newcommandx{\lshapetensor}[4][1=,2=,3=l,4=]{
  \fineq[-0.8ex][0.5][0.55]{
    \drawbox[0][-0.65][2.25][-0.25][#4][$#2$];
    \draw (0.5,0)--++(0,-0.25);
    \draw (2.25-0.5,0)--++(0,-0.25);
    \ifthenelse{\equal{#3}{l}}{
      \drawbox[0][0][1][0.4][#4][$#1$];
    }{
      \drawbox[1.25][0][2.25][0.4][#4][$#1$];
    }
  }
}

\newcommandx{\hfpdecompeq}[6][1=0,2=0,3=,4=,5=,6=]
{
   \fineq[-0.8ex][0.35][0.8]{
     \hfpdecomp[#1][#2][#3][#4][#5][#6];
  }
}

\newcommandx{\hexboxdecomp}[8][1=,2=,3=,4=,5=,6=,7=,8=]
{
  \fineq[-0.8ex][0.5][0.55]{
    \pdecomp[0][2.5][][][][][+][blue!50];
    \pdecomp[2.5][2.5][][][-][blue!50];
    \hfpdecomp[0][1.25][l][+][green!50];
    \sqzbox[1.25][1.25][red!50][\Huge $\perp$][#8];
    \hfpdecomp[1.5+2.25][1.25][r][-][green!50];
    \pdecomp[0][0][\sigma_1][green!50];
    \pdecomp[2.5][0][\sigma_2][green!50];
  }
}

\newcommandx{\triboxdecomp}[8][1=,2=,3=,4=,5=,6=,7=,8=]
{
  \fineq[-0.8ex][0.5][0.55]{
    \hfpdecomp[0][1.25][l][#1][red!50];
    \hfpdecomp[1.25][1.25][r][#2][red!50];
    \pdecomp[0][0][#3][red!50];
  }
}

\newcommandx{\partitionZs}[1][1=]
{
  \ugate[0][0][][$s_{ x-2,t-2}$];
  \ugate[1][1.5][][$s_{ x-1,t-1}$];
  \ugate[0][3][][$s_{ x-2,t}$];
  \ugate[1][4.5][][$s_{ x-1,t+1}$];
  \ugate[0][6][][$s_{ x-2,t+2}$];

  \ugate[2][0][][$s_{ x,t-2}$];
  \ugate[4][0][][$s_{ x+2,t-2}$];
  \ugate[3][1.5][][$s_{ x+1,t-1}$];
  \ugate[2][3][][$s_{ x,t}$];
  \ugate[4][3][][$s_{ x+2,t}$];
  \ugate[3][4.5][][$s_{ x+1,t+1}$];
  \ugate[2][6][][$s_{ x,t+2}$];
  \ugate[4][6][][$s_{ x+2,t+2}$];
  \ugater[-1][1.5][];
  \ugater[-1][4.5][];
  \ugatel[5][1.5][];
  \ugatel[5][4.5][];
}

\newcommand{\sdensity}{s}

\newcommandx{\uc}[3][1=U,2=N,3=]
{
  \ifthenelse{\equal{#3}{}}{
    #1^{(#2)}
  }{
    #1^{(#2)}_{#3}
  }
}

\begin{document}
\date{\today}

\title{The entanglement membrane in chaotic many-body systems}

\begin{abstract}
In certain analytically-tractable quantum chaotic systems, the calculation of out-of-time-order correlation functions,  entanglement entropies after a quench, and other related dynamical observables, 
reduces to an effective theory of an ``entanglement membrane'' in spacetime. 
These tractable systems involve an average over random local unitaries defining the dynamical evolution.
We show here how to make sense of this membrane in more realistic models, which do not involve an average over random unitaries.
Our approach relies on introducing effective pairing degrees of freedom in spacetime, describing a pairing of forward and backward Feynman trajectories, inspired by the structure emerging in random unitary circuits.
This provides a framework for applying ideas of coarse-graining to dynamical quantities in chaotic systems.
We apply the approach to some translationally invariant Floquet spin chains studied in the literature. We show that a consistent line tension may be defined for the entanglement membrane, 
and that there are qualitative differences in this tension between generic models and ``dual-unitary'' circuits.
These results allow scaling pictures for out-of-time-order correlators and for entanglement to be taken over from random circuits to non-random Floquet models.
We also provide an efficient numerical algorithm for determining the entanglement line tension in 1+1D. 
\end{abstract}
\author{Tianci Zhou}
\affiliation{Kavli Institute for Theoretical Physics, University of California, Santa Barbara, CA 93106, USA
}
\author{Adam Nahum}
\affiliation{
Theoretical Physics, University of Oxford, Parks Road, Oxford OX1 3PU, United Kingdom}
\maketitle

\section{Introduction}

This paper is about universality in the dynamics of chaotic
many-body systems.
One familiar type of universality is encapsulated in hydrodynamics for conserved quantities and other slow modes \cite{kadanoff_hydrodynamic_1963, hohenberg1977theory}.
But random circuits \cite{oliveira2007generic,
hamma2012quantum,brandao2016local,nahum_quantum_2017,
nahum_operator_2017,
von_keyserlingk_operator_2017,
rakovszky_diffusive_2017,
khemani_operator_2017,
chan_solution_2018,
chan_spectral_2018,
rowlands2018noisy,
zhou2019emergent,
rakovszky_sub-ballistic_2019,
friedman_spectral_2019,
hunter-jones_unitary_2019}, a family of tractable many-body systems, have suggested new aspects of universality \cite{nahum_quantum_2017,
nahum_operator_2017,
von_keyserlingk_operator_2017,
rakovszky_diffusive_2017,
khemani_operator_2017,
chan_solution_2018,zhou2019emergent}.
In particular they have revealed a generic structure associated with an emergent ``membrane'' in spacetime \cite{nahum_quantum_2017,jonay_coarse-grained_2018}. The effective statistical mechanics of this membrane determine the production of entanglement after a quantum quench, the spreading of quantum operators, and other aspects of the ``scrambling'' \cite{lieb_robinson, calabrese2005evolution, kim2013ballistic,kitaev2014hidden, liu2014entanglement, shenker_black_2014, roberts2015localized, maldacena_bound_2015,roberts_lieb-robinson_2016, kaufman2016quantum, aleiner2016microscopic, patel2017quantum, hosur_chaos_2016, mezei2017entanglement2, nahum_operator_2017, von_keyserlingk_operator_2017, chan_solution_2018,kos2017many,mezei_exploring_2019} of quantum information. 

One way to motivate this membrane, which is relevant to the approach we take here, is via the multi-layer structure of the quantum circuit (or the multi-sheet structure of the path integral, in a continuum language) representing the observables in question.
Algebraically, conventional correlation functions, such as an expectation value ${\<\mathcal{O}(t)\>}$ following a quench, involve a single copy of the unitary time evolution operator $U(t)$, and a single copy of its Hermitian conjugate $U^\dag(t)$. 
Formally, we can represent them in terms of matrix elements of the operator ${U(t)\otimes U^*(t)}$.
But the quantities mentioned above 
(for example the $N$th R\'enyi entropy, obtained by tracing the $N$th power of the reduced density matrix) 
require matrix elements of the multiply ``replicated'' operator
\begin{equation}
\label{eq:replicatedunitaryintro}
\uc[U][N] \equiv {U\otimes U^* \cdots  U\otimes U^*}
\end{equation}
with $N>1$ copies of $U$, and $N$ copies of $U^*$.
In a path integral language, we require multiple forward and backward paths \cite{calabrese2005evolution, maldacena_bound_2015}.

The structure arising from this is most easily understood in random unitary circuits, which are chaotic models built from random unitary gates \cite{nahum_quantum_2017,nahum_operator_2017,von_keyserlingk_operator_2017,chan_solution_2018,zhou2019emergent}. Averaging $\uc[U][N]$ in Eq.~\ref{eq:replicatedunitaryintro} over the random unitaries introduces a degree of freedom $\sigma(x,t)$, at each location in spacetime, 
that labels a pairing between the set of forward replicas and the set of backward replicas, i.e. a pairing of trajectories \cite{zhou2019emergent}. 
Formally this pairing is a permutation, $\sigma\in S_N$, 
and quantities like the R\'enyi entropies and the out-of-time ordered correlator (OTOC) map to effective partition functions for $\sigma(x,t)$.
In this setting, the membrane can be understood as a domain wall between different values of the pairing field $\sigma(x,t)$ \cite{nahum_operator_2017, chan_solution_2018, zhou2019emergent}.

The goal of this work is to show how to make sense of the pairing field $\sigma(x,t)$ and the membrane in nonrandom systems. This allows the language of the renormalization group to be applied to chaotic dynamics.

\begin{figure}[b]
\centering
\subfigure[]{
  \label{fig:ev_intro}	
  \includegraphics[width=0.46\columnwidth]{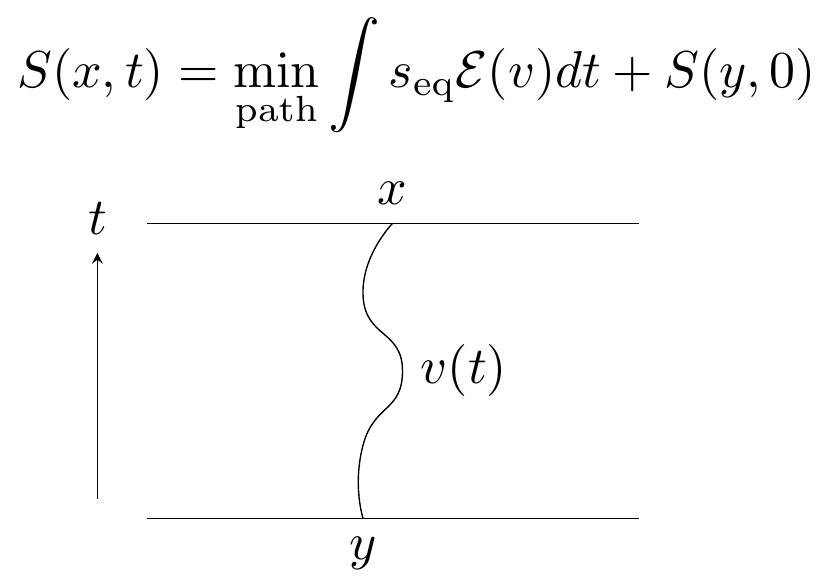}
}
\subfigure[]{
  \label{fig:ev_typical}	
  \includegraphics[width=0.4\columnwidth]{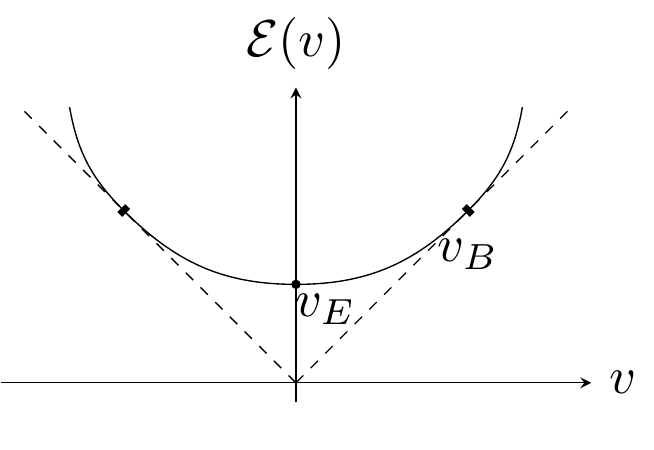}
}
\caption{The membrane picture of entanglement and other dynamical quantities. (a) Example of  membrane for evaluating R\'enyi entropy growth in a 1D quench. Minimizing the membrane tension gives the entanglement. The tension is a function of local velocity ${v=\dot x}$. (b) Schematic line tension function for a chaotic system with parity symmetry. $\mathcal{E}(v)$ is convex and tangential to $|v|$ at $(\pm v_B, v_B)$.}
\label{fig:membrane}
\end{figure}

The basic quantity characterizing the membrane is its coarse-grained ``line tension'' $\mathcal{E}(v)$  \cite{jonay_coarse-grained_2018}. This line tension determines the system's butterfly velocity or operator spreading speed and its entanglement growth rate.
It can be thought of loosely as associating an entanglement cost with a given curve in spacetime (in general this tension depends on $N$). 
See Fig.~\ref{fig:membrane} for a schematic picture.

We expect that the pairings, and the membrane description, continue to make sense in systems more general and more realistic than random circuits, for the heuristic reason described in the next paragraph.
In support of this, there is numerical evidence that a consistent membrane tension can be defined in a generic spin chain \cite{jonay_coarse-grained_2018}, 
and in a dual-unitary circuit \cite{bertini_entanglement_2018,piroli_exact_2019}. 
There are analytic computations in random Floquet circuits in the large Hilbert space dimension (large $q$) limit  \cite{chan_solution_2018}, showing that  the  domain wall structure makes sense in that limit.
Finally, there is an analytical derivation of the  form of the membrane tension in  holographic systems \cite{mezei2018membrane,mezei_exploring_2019}, at leading order in the number of degrees of freedom;
this also begins to make a concrete connection between the entanglement membrane, and the Ryu-Takayanagi/Hubeny-Rangamani-Takayanagi geometrical pictures for entanglement in AdS space \cite{ryu2006holographic,hubeny2007covariant,agon2019bit,freedman_bit_2017}. (Lattice toy models for the AdS-CFT correspondence, using random tensor networks, also exhibit domain walls between permutation degrees of freedom  \cite{hayden2016holographic, vasseur2018entanglement}.)
But at present there is no formalism for explaining how the pairing degrees of freedom emerge when there is no average over randomness, or how to compute the properties of the membrane in a generic system without randomness and away from any large-$N$-like limit.
As a more general point, random circuits and related stochastic models have shed light on various aspects of chaotic dynamics \cite{nahum_quantum_2017,
nahum_operator_2017,
von_keyserlingk_operator_2017,
rakovszky_diffusive_2017,
khemani_operator_2017,
chan_solution_2018,
rowlands2018noisy,
zhou2019emergent,
rakovszky_sub-ballistic_2019,
chan_spectral_2018,
friedman_spectral_2019,
hunter-jones_unitary_2019}, and a natural question is how to generalize these calculations to non-random systems. Our aim here is to provide a formalism for doing this.

Heuristically, the paired configurations mentioned above can be thought of as saddle points of the path integral for the replicated evolution operator (\ref{eq:replicatedunitaryintro}). 
This path integral contains multiple forward and backward paths. 
In a continuum language, these are  weighted by $e^{-iS}$ and $e^{+iS}$ respectively, where $S$ is the action; 
in the discrete setting, $e^{-iS}$ is replaced by a product of matrix elements of local unitaries. 
By pairing each forward path with a backward path we can cancel the phases in the exponent,
so we might guess that paired configurations predominate.
In the cases of interest to us, however, the boundary conditions force the pairing to differ in different regions of spacetime. 
This leads to saddle-point solutions with a nontrivial
membrane structure.

In this work we show explicitly, for a large class of systems without conservation laws, how the pairing field $\sigma(x,t)$ and the entanglement membrane emerge, making the above heuristic picture precise. We emphasize that we do \textit{not} require either a limit of large local Hilbert space dimension, or any kind of randomness. 
To demonstrate the utility of our approach, we compute the entanglement line tension (for $N=2$, the case relevant to the OTOC  and to the second R\'enyi entropy) for a variety of translationally-invariant Floquet spin-1/2 chains.

Formally, our approach is to work in the replicated Hilbert space (acted on by Eq.~\ref{eq:replicatedunitaryintro})  and to insert, in each time slice, 
an exact local resolution of identity. This includes states  $\ket{\sigma}$ associated with pairings, but also a projector onto the complementary part of the Hilbert space, which we denote by $\perp$, containing  non-pairing states.
Traces of $\uc[U][N]$, yielding for example correlation functions or R\'enyi entropies, can then be written as partition functions for an effective ``spin'' $s(x,t)$ which can take the values either $s=\sigma$ (a permutation) or $s=\perp$.

\begin{figure}[t]
\centering
\subfigure[]{
  \label{fig:ev_renorm}	
  \includegraphics[width=0.8\columnwidth]{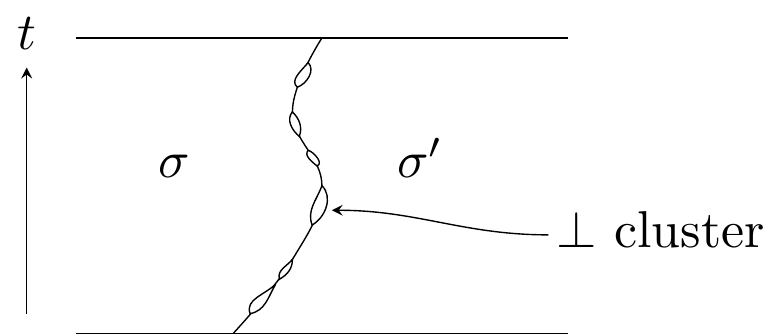}
}
\caption{Thick membrane (schematic). 
Except in the special case of Haar-averaged unitaries, the membrane (a domain wall between permutations $\sigma$ and $\sigma'$) is thickened by clusters where the effective spin takes the value ``$\perp$''.  In the multilayer tensor network for $U^{(N)}$, this represents the propagation of states orthogonal to paired states.  (In general, the interior of the thick  membrane also contains paired states. For $N>2$ these may be distinct from $\sigma$ or $\sigma'$ \cite{chan_solution_2018, zhou2019emergent}.)
}
\label{fig:membrane2}
\end{figure}

Including these non-pairing states allows the structures that appear in the random circuit to be generalized to models without any random average.
The Haar-averaged random circuit can be thought of as a special case: there, the contribution from  any spin configuration that includes a $\perp$ vanishes after averaging, leaving a spin model for permutations only, and the membrane is a domain wall between permutations  \cite{nahum_operator_2017, zhou2019emergent}.
In, say, a translationally-invariant system, 
the domain wall structure becomes more complex, with the appearance of the spin value $s=\perp$ along the domain wall.
(In fact, this is also true in a particular realization of the random circuit, \textit{before} we average over the random unitaries.)
This leads to a thickened domain wall with a more complex structure.  However we argue that in chaotic systems, the domain wall's width remains of order 1 in terms of microscopic lengthscales. 
Therefore, on large scales, the membrane is still well-defined, but with a nontrivial, renormalized membrane tension function.
We illustrate this situation in  Fig.~\ref{fig:membrane2}.

Our initial introduction of the pairing field is exact but formal, because the microscopic weight for a configuration $s(x,t)$ is in general complicated.
To make progress we first argue that configurations with large unpaired regions (large regions of $\perp$) are exponentially suppressed. 
Then we show how to resum an infinite number of configurations, or ``Feynman diagrams'', exactly, to give a simpler description of the entanglement membrane, in terms of a reduced set of parameters. 

This is loosely analogous to the renormalization of the mass of a quasi-particle due to interactions --- here the $\perp$ states dress the structure of the membrane, giving it a larger width
(than in the averaged random circuit, where $\perp$ states are suppressed) 
and a modified line tension. This is an application of the renormalization group idea to find the coarse grained quantities characterizing scrambling.

Before tackling translation-invariant systems, we analyze the case of a  \textit{fixed realization} of a random unitary circuit. This illustrates the basic mechanism that makes the approach possible --- suppression of $\perp$ states by phase cancellation --- in a tractable setting.
Since the disorder realization is fixed, we cannot introduce the pairing degrees of freedom by Haar-averaging.
Instead we apply the new method described here.
Working at large but finite Hilbert local space dimension, we show that the membrane remains well-defined, and is thickened slightly by occasional appearances of $\perp$ states.
This leads to a membrane that inhabits a disordered ``potential'' in spacetime.
This picture reproduces our previous results on Kardar-Parisi-Zhang (KPZ) fluctuations in the R\'enyi entropy $S_2$ between different realizations of the random circuit \cite{zhou2019emergent}. But previously these results required the replica trick, which we are able to dispense with here. 

Next we turn to Floquet dynamics of spin-$\frac{1}{2}$ chains, without any kind of randomness. In order to control the calculations, we introduce a systematic expansion in the maximal temporal extent of connected $\perp$ clusters in the effective spin model.
A priori, this expansion does not have a small parameter. 
However we conjecture (based on the tractable case of the random circuit), and give numerical evidence, that it is a convergent expansion in chaotic Floquet systems.

Using this observation, we construct an efficient numerical scheme for computing the membrane line tension $\mathcal{E}(v)$ using the expansion above: we truncate the maximum size of a connected $\perp$ cluster (but we do not require that these clusters are dilute), and we examine convergence as a function of the order at which we truncate. We first discuss  Floquet models with a local unitary circuit structure, since these are the simplest case to visualize. However the circuit structure is not required: we also apply the method to completely generic Floquet models, involving Hamiltonian evolution in continuous time without any circuit structure.

The numerical results show good convergence for a wide range of chaotic models, including prototypical spin models for quantum chaotic systems, such as Floquet Ising models with longitudinal and transverse fields (kicked Ising models) \cite{kim2013ballistic,kim_testing_2014,prosen_chaos_2007}. 

The line tension $\mathcal{E}(v)$ satisfies some general constraints \cite{jonay_coarse-grained_2018} which provide a highly nontrivial check on our results.
Recall that $\int \dd t \, \mathcal{E}(\dot x(t))$ is proportional to the ``free energy'', in the scaling limit,  of a membrane whose shape is parameterized by $x(t)$  (Fig.~\ref{fig:membrane}).
$\mathcal{E}(v)$  is a convex function of velocity, which is tangential to the line $\mathcal{E}=\pm v$ at the point $( \pm v_B, v_B)$, where $v_B$ is the quantum butterfly velocity (for parity symmetric systems): see the schematic picture in Fig.~\ref{fig:ev_typical}. 
For the chaotic systems we study, our numerical approximations to $\mathcal{E}(v)$ indeed appear to  converge a form satisfying the constraints above.

We also investigate a special case of the kicked Ising model that has maximal entanglement growth \cite{bertini_entanglement_2018,piroli_exact_2019} and the  property of ``dual unitarity'' which, remarkably, allows a range of exact computations \cite{gopalakrishnan_unitary_2019,bertini2019exact, akila_particle-time_2016-2,  bertini2019operator,piroli_exact_2019}.  

In addition to studying particular models, we discuss the general continuum theory for the membrane in 1+1D. We argue that the entanglement membrane has special properties in dual-unitary circuits: its equation of motion becomes a wave equation, rather than being diffusive as it is in generic models.

\subsection{Organization of the paper}

Sec.~\ref{sec:def_spin_model} introduces    pairing degrees of freedom in the multi-layer unitary circuit, leading to an effective spin model (Sec.~\ref{subsec:eff_spin}--\ref{sec:symmetry}) in which a domain wall structure emerges (Sec.~\ref{sec:domainwallstructure}).

We first test our formalism in an analytically tractable setting, a fixed realization of a random circuit, in Sec.~\ref{sec:tractable_limit}. We show that large clusters of $\perp$ in the effective spin model are exponentially suppressed (Sec.~\ref{sec:expsup}) and use this to obtain Kardar-Parisi-Zhang scaling of the R\'enyi entropy  (Sec.~\ref{sec:kpz}). 

We then develop a systematic formalism for resumming the effect of $\perp$ clusters for translationally invariant Floquet circuits. Sec.~\ref{sec:floq_pre} describes  technical preliminaries, and Sec.~\ref{sec:floq_app} applies the method to various models from the literature. Sec.~\ref{sec:domainwalldualunitary} discusses the coarse-graining of the membrane, showing that two distinct universality classes arise, for a generic circuit and a self-dual circuit. 

We discuss the application of this formalism to operator spreading in Sec.~\ref{sec:opspread}, and to general  Floquet models in continuous time, without any circuit structure, in Sec.~\ref{sec:beyond_circuit}. 

Sec.~\ref{sec:outlook} describes some open questions. Appendices contain technical details about the spin model and further numerical results.

\tableofcontents


\section{Defining the spin model}
\label{sec:def_spin_model}

\begin{figure}[h]
\centering
\includegraphics[width=0.5\columnwidth]{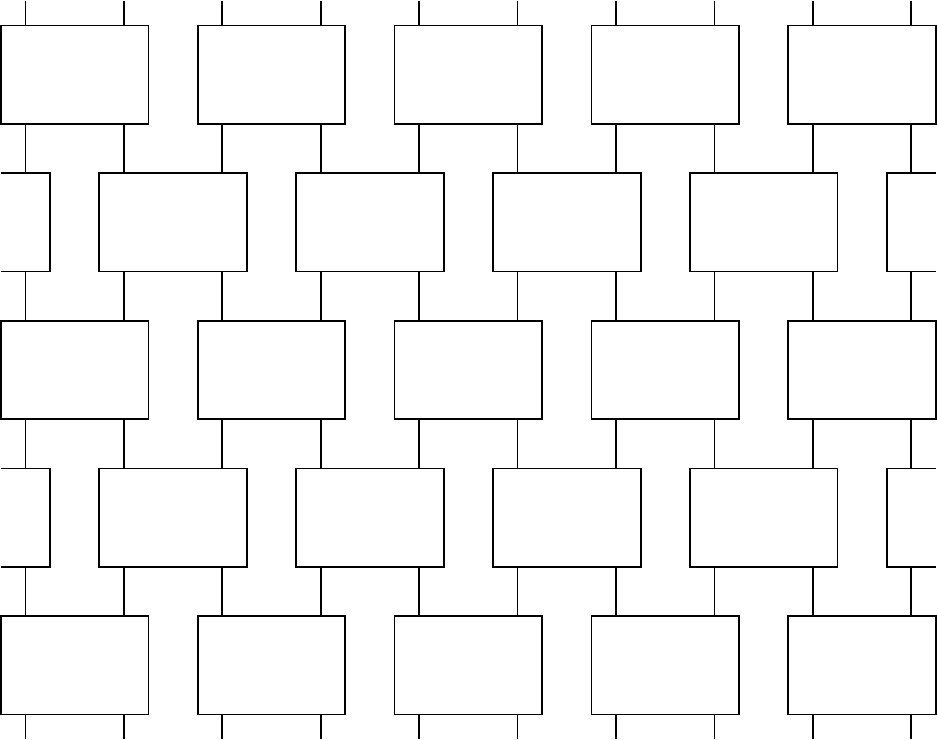}
\caption{The structure of the circuit with gates acting on nearest-neighbor sites. In the Floquet models we study, each gate is identical. In the random unitary case, each gate is independently sampled  from the Haar ensemble.}
\label{fig:ruc_struct}
\end{figure}

For concreteness, let us focus on time evolution with a unitary circuit made of two-site gates, with the structure in Fig.~\ref{fig:ruc_struct}.
(The constructions below generalize straightforwardly to evolution without a circuit structure: we describe this  explicitly in Sec.~\ref{sec:beyond_circuit}.)
We will use $U(t)$ or simply $U$ to denote the full many-body unitary represented by the entire circuit, and the lower-case $u$ to stand for a local two-site gate.

At this point the local gates $u$ are left arbitrary.
In some sections we will restrict to Floquet models with both space and time translation invariance, but this is not necessary. 
The circuit geometry guarantees that there is no propagation of information outside a strict lightcone with speed $v=1$; the actual butterfly speed in these circuits is in general strictly smaller than 1.

Conventional time ordered correlation functions such as ${\Tr \mathcal{O}(x_n, t_n)\ldots \mathcal{O}(x_1, t_1)}$ involve a `two-layer' quantum circuit made of $U$ and $U^*$, in the sense that they can be obtained from the contraction of ${U\otimes U^*}$ by inserting operators in the intermediate time-steps (here $U^*$ is the complex conjugate of $U$, taken in an arbitrary local basis, and not the Hermitian conjugate).
In contrast, the quantities of interest to us here require traces of the $2N$-layer unitary circuit 
\begin{equation}\label{eq:replicatedunitary}
\uc[U][N] \equiv \underbrace{U\otimes U^* \cdots  U\otimes U^*}_{\text{$N$ $U$s and $N$ $U^*$s}},
\end{equation}
where $N > 1$. For example, the out-of-time order correlation function
\begin{equation}
C(x, t ) = \langle \mathcal{O}(x, t) \mathcal{O}'( 0, 0 ) \mathcal{O}(x, t) \mathcal{O}'( 0, 0 )    \rangle 
\end{equation}
requires $N = 2$. The R\'enyi entanglement entropy for subsystem $A$
\begin{equation}
S_n = -\frac{1}{n - 1} \ln \tr( \rho_A^n (t)) 
\end{equation}
requires $N = n$ to write the $n$-th power of the time-evolved reduced density matrix. 

Taking $N>1$ leads to new structure. 
One consequence is that it multiplies the number of conservation laws, since formally each `copy' of the system undergoes independent unitary evolution. This will not concern us here, however, since we study models without conservation laws. 
The additional structure we examine is instead associated with pairings between spacetime histories in the `worlds' associated with different layers \cite{zhou2019emergent}. 

The multi-layer circuit has $N$ layers of $U$ and of $U^*$, and can be thought of as describing evolution of a `replicated', multi-copy system. 
For a heuristic picture, we may think in terms of Feynman histories of this replicated system in our chosen local basis. The amplitude for a given Feynman history is given by a product of matrix elements of the local unitaries. 
Feynman histories that are `paired', so that the $N$ histories occuring in the $U$ replicas are pairwise equal to the $N$ histories occuring in the $U^*$ replicas, avoid phase cancellation \cite{zhou2019emergent}.  
(This pairing can happen in $N!$ different ways.)
We will argue that in quantum chaotic systems, the observables mentioned above are dominated by Feynman histories that have a well-defined pairing structure after coarse-graining. 
We will show how a corresponding local `pairing field' $\sigma(x,t)$ can be defined by coarse-graining over microscopic length and time scales.

\subsection{Introducing the effective spins}
\label{subsec:eff_spin}

To motivate the structures we will consider, we first recall what happens when the local unitaries are averaged over the Haar ensemble instead of being fixed.

For a given two-site gate $u$, acting on sites $i$ and ${i+1}$, the replicated gate $\uc[u]$ is a unitary operator on $2N$ copies ($N$ layers of $u$, $N$ layers of $u^*$) of the two-site Hilbert space, which has dimension $d = q^2$. 
Its average (which is no longer unitary) is the operator 
\begin{equation}
\label{eq:u_haar_aver}
\overline{u_{\text{Haar}}^{(N)}} =  \sum_{ \tau, \sigma \in S_N}  \text{Wg}( \tau^{-1} \sigma; d  ) | \tau \tau \rangle \langle \sigma \sigma | 
\end{equation}
where $\text{Wg}$ is the unitary Weingarten function \cite{collins_integration_2006,gu_moments_2013,weingarten1978asymptotic}. Here $\tau$ and $\sigma$ are elements of the permutation group $S_N$.
These denote a pairing of the $u$ layers, labelled ${1,2,\ldots,N}$, with the $u^*$ layers, labelled ${\bar 1,\bar 2,\ldots, \bar N}$. For example $\sigma$ corresponds to pairing  $1$  with $\overline{\sigma(1)}$, etc. At a given spatial site, the state $\ket{\sigma}_i$ is a product of maximally entangled states between paired copies of the site, and the states above are for a pair of spatial sites: ${\ket{\sigma\sigma}= \ket{\sigma}_i\otimes\ket{\sigma}}_{i+1}$. 

For example, for $N = 2$, there are two possible permutations: the identity, $\I$, and the transposition, denoted $(12)$ in cycle notation. 
In our local basis the corresponding states at a site are
\begin{equation}
\begin{aligned}
\langle a_1,\overline{a_1}, a_2, \overline{a_2} | \I  \rangle  &= \delta_{a_1 \overline{a_1} } \delta_{a_2 \overline{a_2} },   \\
\langle a_1,\overline{a_1}, a_2, \overline{a_2} | (12)  \rangle  &= \delta_{a_1 \overline{a_2} } \delta_{a_2 \overline{a_1} }
\end{aligned}
\end{equation}
(note that we do not normalize these states).

In Ref.~\onlinecite{zhou2019emergent} we showed how, starting with the above expression, the expression for $\uc$ could be reduced to a lattice magnet, with local interactions, for the permutations $\sigma$ (the permutations $\tau$ were integrated out in this mapping). At the end of this mapping, there is a single $\sigma$ spin associated with each local block in the circuit.

Now let us consider a general circuit with the brickwork geometry in Fig.~\ref{fig:ruc_struct}.
The basic idea will be to separate out the pairing (permutation) states, $\ket{\sigma\sigma}$, in the multi-copy Hilbert space for a pair of sites, from states in orthogonal complement of this space. 


Regardless of the choice of unitary gate, we will be able to decompose the multi-layer  unitary as
\begin{equation}
    \uc[u][N][] =
    P_\parallel
    +  ``\perp",
\end{equation}
where $P_\parallel$ is a projection operator onto paired states, which as discussed below is also equal to $\overline{u_{\text{Haar}}^{(N)}}$, and ``$\perp$" represents the projection of $\uc[u][N][]$ to the complement of the space of paired states.
Eq.~\ref{eq:u_haar_aver} shows that in the Haar-averaged case the ``$\perp$''  contributions vanish identically.
More generally they do not vanish, and their operator norm is not negligible.
However we will argue that the permutation states dominate in a certain sense, and that states in the orthogonal complement can be taken into account via a systematic procedure.

The expression for the Haar average of a local unitary in Eq.~\ref{eq:u_haar_aver} can be regarded simply as the  projection operator onto the subspace of permutation states. By the invariance of the Haar measure under (say) left multiplications,
\begin{equation}
\overline{ \uc[u][N][\text{Haar}]} 
\times
\overline{ \uc[u][N][\text{Haar}]} 
=
\overline{ \uc[u][N][\text{Haar}]}, 
\end{equation}
showing that this is a projector. The state $\ket{\sigma\sigma}$ satisfies $\uc[u] | \sigma \sigma \rangle = |\sigma \sigma \rangle$ for any choice of two-site unitary, so after Haar averaging we also have
\begin{equation}
\overline{ \uc[u][N][\text{Haar}]} \,
| \sigma \sigma \rangle = |\sigma  \sigma \rangle
\end{equation}
for any permutation $\sigma$. 
This shows that the Haar averaged object, which is a Hermitian operator on the replicated space, is equal to $P_{\parallel}$, the orthogonal projector onto the subspace spanned by permutation states \cite{harrow_church_2013}.

To separate out contributions from permutation states for a general circuit, we will insert the resolution of the identity (on two physical sites)
\begin{equation}
\label{eq:paraandperp}
\uc[I] = P_{\parallel} + P_{\perp} 
\end{equation}
immediately before each two-site gate in the circuit. Here $P_\perp$ is the projector onto the subspace complementary to that spanned by pairing states.
This insertion is shown in (\ref{eq:def_Z}) below as a bar below each gate.

We will view the quantity to be computed, for example the purity, as a partition function which we denote $Z$. (In the purity example, the  second R\'enyi entropy is the free energy associated with this partition function.) 
After using \ref{eq:paraandperp}, this partition function is a sum of terms with all possible choices of projection operator inserted below each gate. Graphically,
\begin{equation}
\label{eq:def_Z}
\begin{aligned}
Z &= 
\fineq[0][0.4][1]{
  \partitionZ;
} = 
\sum
\,\,
\fineq[0][0.4][1]{
  \partitionZ[p];
} 
\end{aligned}
\end{equation}
where the sum is over all possible assignments of ${\parallel, \perp}$ to the projection operators, which are represented by horizontal bars. We leave the boundary conditions at the initial and final times unspecified at this point, since they depend on the observable to be computed.

Note that the operator $P_\parallel$ can ``absorb'' an arbitrary two-site gate.
This can be seen from the invariance of the Haar measure. For any 2-site unitary $u$,
\begin{equation}
\uc[u] \times P_{\parallel} =\uc[u] \times \overline{\uc[u][N][\text{Haar}]} = \overline{\uc[u][N][\text{Haar}]} = P_{\parallel} .
\end{equation}
As a result, the term in the partition function where all the insertions are $P_{\parallel}$ is identical to the expression where all gates are replaced with Haar-averaged gates. Therefore the decomposition in Eq.~\ref{eq:def_Z} gives us a well-understood starting point, which we can go beyond by taking into account domains of $\perp$ insertions. Roughly speaking, this approach will be useful if these domains do not form a connected cluster `percolating' from the top to the bottom of the spacetime slab, but instead form finite  clusters.
Their effect can then be taken into account as a renormalization of the interactions for the permutation degrees of freedom that are present in the random circuit.

To make these permutations explicit we decompose $P_{\parallel}$ using  Eq.~\ref{eq:u_haar_aver}.
In the random circuit it was useful to perform the sum over the $\tau$ variables (cf. Eq.~\ref{eq:u_haar_aver}) explicitly, leaving a partition sum solely for the $\sigma$ variables. 
Algebraically, that corresponds to splitting $P_\parallel$ into  projection operators $P_\sigma$,
\begin{equation}
P_{\parallel}  = \sum_{ \sigma \in S_N} P_{\sigma },
\end{equation}
which are given by grouping together terms in Eq.~\ref{eq:u_haar_aver} with a given $\sigma$:
\begin{equation}
\label{eq:p_sigma} 
P_{\sigma } = \left( \sum_{\tau} \text{Wg}( \tau^{-1} \sigma; { q^2} ) | { \tau \tau} \rangle \right)  \langle  {
 \sigma \sigma }|. 
\end{equation}
Using the properties of the Weingarten function, one can check that the states appearing in the brackets
\begin{equation}
\label{eq:dual_basis}
| (\sigma \sigma)^* \rangle  = \sum_{\tau} \text{Wg}( \tau^{-1} \sigma; { q^2} ) | { \tau \tau} \rangle
\end{equation}
form the dual basis of $| \sigma \sigma \rangle $,
i.e. $\langle (\sigma \sigma)^* | \tau \tau \rangle = \delta_{ \tau, \sigma}  $, so that the non-Hermitian operators $P_\sigma = | (\sigma \sigma)^* \rangle \langle \sigma \sigma |  $ are projectors satisfying (App.~\ref{app:P_sigma}):
\begin{equation}
\label{eq:proj_prop}
P_{\sigma} P_{\sigma'} = \delta_{\sigma \sigma'} P_{\sigma}.
\end{equation}
Here we assume that $q$ is generic, to avoid divergences in the Weingarten function, which can arise for small values of $q$  when $N$ is sufficiently large (for $q=2$, such divergences can arise for $N>4$).
This is not a fundamental obstacle, as we could always define a ``spin'' with a larger local Hilbert space dimension by grouping sites.
 
Our resolution of the identity, inserted below each block in the multi-layer circuit, is now refined to
\begin{equation}
I^{(N)} = \sum_{ \sigma \in S_N} P_{\sigma } + P_\perp.
\end{equation}
Let us represent the possibilities graphically as:
\begin{equation}
\label{eq:p_box}
\fineq[0.6ex][0.4][0.8]{
  \ugate[0][0][][][p];
  \node () at (0.5,-0.5) {$P_{\sigma}$};
}
=
\fineq[0][0.4][0.8]{
  \ugate[0][0][][$\sigma$];
},
\qquad
\fineq[0.6ex][0.4][0.8]{
  \ugate[0][0][][][p];
  \node () at (0.5,-0.5) {$P_{\perp}$};
}
=
\fineq[0][0.4][0.8]{
  \ugate[0][0][][$\perp$];
}
\end{equation}
Importantly, the blocks on the left, labelled by permutations, are \textit{independent} of the choice of local unitary gate, because $P_\sigma$ again ``absorbs'' any unitary. For any gate $u$, we have 
$\uc[u][N] |\sigma \sigma \rangle = |\sigma \sigma \rangle$, so 
\begin{equation}
\label{eq:un_sigma}
\uc[u] P_{\sigma} = P_{\sigma}.
\end{equation}
However the $\perp$ blocks, which represent the tensor $\uc[u] P_\perp$, depend on the local unitaries defining the dynamics. 

We have now written $Z$ as a sum over spins $s$, associated with each block in the circuit, which take values in $S_N \cup \{\perp\}$:
\begin{equation}
\label{eq:Zs}
 Z = \, \sum_{\{s\}} \,\,
\fineq[0][0.6][0.75]{
  \partitionZs
}
\end{equation}
Throughout this paper, we will use $x,t$ to label the bond between two lattice sites. The gates are located on half of the bonds.

Note that ``$\perp$'' is simply a choice of label for one of the values that $s$ can take; we could also have denoted this state by ``$0$'' or anything else. At each location, $s$ runs over ${N!+1}$ different values.

The weight of a spin configuration $\{s\}$ in this partition sum is given by contracting the network. In general, this weight is not simply a product of  local terms. 
However we will argue that for the boundary conditions of interest, and for generic models,
there is an effective notion of locality. We first describe some basic properties of the weights (Secs.~\ref{sec:spinmodelgeneralities},~\ref{sec:symmetry}), and then specify to the case of interest, where the boundary conditions induce a domain wall structure in the spin configuration (Sec.~\ref{sec:domainwallstructure}).

\subsection{General properties of the spin model}
\label{sec:spinmodelgeneralities}

First consider the sub-partition-function without any $P_{\perp}$ insertions (i.e. with $s\neq \perp$ everywhere), which we denote $Z_{0\perp}$. 

$Z_{0\perp}$ is a statistical model whose ``spins'' $\sigma$ are permutations \cite{zhou2019emergent}. 
The spins have local three-body interactions on ``down-pointing triangles'' (and no interactions on up-pointing triangles).  To be specific, each down-pointing triangle such as
\begin{equation}
\threegate
\end{equation}
has an associated weight
\begin{align}
\label{eq:jabc}
\jabc[\sigma_a][\sigma_b][\sigma_c][] &\equiv \tribox[\sigma_b][\sigma_c][\sigma_a]  \equiv J(\sigma_b, \sigma_c; \sigma_a ).
\end{align}
From Eq.~\ref{eq:p_sigma}, the interaction  triangle is (see App.~\ref{app:P_sigma})
\begin{equation}
\label{eq:J_in_weingarten}
J( \sigma_b, \sigma_c; \sigma_a ) = \sum_{\tau} \langle \sigma_b  \sigma_c  | \tau \tau \rangle  \text{Wg}( \tau^{-1}\sigma_a; q^2 ).
\end{equation}
Equivalently, the triangle is the coefficient in the expansion
\begin{equation}\label{eq:triangleasexpansioncoeff}
\bra{\sigma_b \,\sigma_c} P_\parallel^{(N)}
=\sum_{\sigma_a\in S_N} 
\jabc[\sigma_a][\sigma_b][\sigma_c] \bra{\sigma_a \,\sigma_a},
\end{equation}
where the two arguments in the bras refer to the pair of spatial states.

This spin model was described in Ref.~\onlinecite{zhou2019emergent} for the random circuit. In that context, the spins $\sigma$ arose from the Haar average over the physical random unitary in Eq.~\ref{eq:u_haar_aver}. The three-body interaction appears after integrating out the ``$\tau$'' spins. Here the sum over the $\tau$ spins appeared at the state level in defining the projection operator in Eq.~\ref{eq:p_sigma}.

The simplest case is $N=2$ \cite{nahum_operator_2017}.
There are then two possible permutations, $\I$ and $(12)$, which we denote $+$ and $-$. The weights are symmetric under exchange of $+$ and $-$ and under spatial reflection.  They are
\begin{align}
\tribox[+][+][-] & = 0, & 
\tribox[+][+][+] &= 1,   & 
\tribox[+][-][+] &= K, \label{eq:tribox_3} 
\end{align}
with
\begin{equation}
\label{eq:Kdef}
K = \frac{q}{q^2 + 1},
\end{equation}
which follows from Eq.~\ref{eq:p_sigma} with the explicit expression 
\begin{equation}
\label{eq:p_pm}
\begin{aligned}
P_{+} &=
 |(++)^* \rangle \langle +\!+ | \\
&= 
\frac{1}{q^4 - 1} \left( |+\!+ \rangle \langle +\!+ | - \frac{1}{q^2} |-\!-\rangle \langle +\!+ |    \right) 
\end{aligned}
\end{equation}
and symmetrically for $P_-$. 

Note the vanishing of the first weight, which amounts to a hard constraint on the spin configurations. 
The  spin configuration $\{ s\}$ is in fact highly constrained for any $N$: the underlying unitarity means that $J$ vanishes for many configurations of the triangle.  In other words, many terms in the sum over $s\in S_N$ defining $Z_{0\perp}$ are zero  (see Ref.~\cite{hunter-jones_unitary_2019} and App.~\ref{app:N_indep} for a further simplification). 
For example, as a generalization of the first equality in Eq.~\ref{eq:tribox_3} above,
\begin{equation}
\label{eq:j_abb}
\begin{aligned}
\jabc[\sigma_a][\sigma_b][\sigma_b][] \equiv
 \tribox[\sigma_b][\sigma_b][\sigma_a]  &  = \delta_{\sigma_a, \sigma_b }.
\end{aligned}
\end{equation}
This enforces a light cone structure. For example, for the  purity calculation which we discuss below in Sec.~\ref{sec:domainwallstructure},
the boundary spins on the top are $s_{x,t} = \I $ for $x < 0$ and $s_{x,t} = (12)$ for $x \ge 0$. This means that the spin configuration is only nontrivial within a backwards lightcone emanating from the entanglement cut at the top boundary [i.e. we must have $s_{x' <  x - t + t' , t'< t } = \I $ and $s_{x' \ge x + t - t' , t'< t} = (12)$].

This light-cone structure is preserved when we include configurations with $s=\perp$. If the two spins at the top of a triangle are the same permutation $\sigma_b$, the lower spin cannot be $\perp$, because by definition the projector onto $\perp$ is orthogonal to $\ket{\sigma_b\sigma_b}$:
\begin{equation}
\label{eq:boxabc}
\begin{aligned}
  \tribox[\sigma_b][\sigma_b][\perp]  &= 0.
\end{aligned}
\end{equation}
So again in the purity example above $\perp$ can only be inserted in the backward lightcone emanating from the entanglement cut.

Let us discuss the locality of the spin interactions. 
We have seen that in the absence of $\perp$ the Boltzmann weight factorizes into three-body terms representing interactions between three adjacent spins. 
Once we have large  clusters of $\perp$ this is no longer true. 
However, the weight \textit{does} factorize into a product of separate weights for each cluster of $\perp$ (together with the local weights mentioned above): there is no interaction between disconnected clusters of $\perp$. This is because surrounding a cluster of $\perp$ with permutation states dictates a definite way of contracting up the $\perp$ blocks, yielding a c-number. (We will give some explicit formulas below.)
Therefore, if the partition function is dominated by configurations in which clusters of $\perp$ have a finite typical size, locality will be regained after coarse-graining beyond this scale.

\subsection{Symmetry of the spin model}
\label{sec:symmetry}

Finally, we note that the model we have defined has a global symmetry, with the symmetry group
\begin{equation}\label{eq:symmgp}
G_N = (S_N\times S_N) \rtimes \mathbb{Z}_2.
\end{equation}
This is a consequence of the fact that Eq.~\ref{eq:replicatedunitary} involves multiple copies of the same unitary \cite{zhou2019emergent,vasseur2018entanglement}.
The importance of ${S_N\times S_N}$ symmetry has been emphasized in random tensor networks \cite{vasseur2018entanglement},
where permutations labelling pairings also appear \cite{hayden2016holographic}.

Graphically, the symmetry arises from the the possibility of permuting the layers of the original unitary circuit without changing Eq.~\ref{eq:replicatedunitary}. 
We can permute the $U$ layers among themselves with a permutation $g_L\in S_N$, and we can also separately permute the $U^*$ layers among themselves with another permutation $g_R$. This gives the $S_N\times S_N$ subgroup of $G_N$, which acts on the spins via
\begin{align}
\sigma & \rightarrow g_L \sigma g_R^{-1},
& 
\perp & \rightarrow \perp.
\end{align}
The $\mathbb{Z}_2$ subgroup of $G_N$ arises from the fact that the multilayer circuit is invariant if we exchange \textit{all} the $U$ layers with all the $U^*$ layers, and also complex conjugate the circuit. The resulting symmetry acts by
\begin{align}
\label{eq:inversionsymm}
\sigma & \rightarrow \sigma^{-1},
& 
\perp & \rightarrow \perp.
\end{align}
For both types of operation, the spin state $\perp$ is invariant because these exchanges of layers preserve the $\parallel$ and $\perp$ subspaces of the two-site Hilbert space.

The simplest case  is $N=2$, when  $G_N$  reduces to an Ising-like $\mathbb{Z}_2$ symmetry relating ${+=\I}$ and ${-=(12)}$. In this case the symmetry in Eq.~\ref{eq:inversionsymm} becomes trivial, since $\sigma=\sigma^{-1}$.

In the sub-partition function $Z_{0\perp}$ the weights $J$ are invariant under $G$ because the Weingarten function and the overlaps  in Eq.~\ref{eq:p_sigma} depend only on the cycle structure of products of the form $\sigma \tau^{-1}$. This cycle structure is invariant under the above operations.

The $G_N$ symmetry is a symmetry of the bulk interactions for the spin model. It will however be strongly broken by the boundary conditions we require.

\subsection{Domain wall structure}
\label{sec:domainwallstructure}

As a useful illustrative example, let us consider the purity of a region $A$, ${e^{-S_2(A)}=\tr( \rho_A^2 )}$.
We take $A$ be a semi-infinite half-system to the left of the origin in an infinite chain. 
This quantity requires $N=2$.
If we take the physical system to start off in the state ${\ket{\psi}}$,
and if we denote the replicated state in the four-copy Hilbert space by ${|\psi^{(2)}\rangle}$, then
\begin{equation}
\exp \lf {-S_2} \ri = \bra{\ldots ++--\ldots} \uc[U][2](t)  {|\psi^{(2)}\rangle},
\end{equation}
where we have labelled the two permutation states at a site,  $\I$ and $(12)$, by $+$ and $-$.
For simplicity we take the initial state to be a product state, ${\ket{\psi}=\otimes_x \ket{e_x}}$.

\begin{figure}[t]
\centering
\subfigure[]{
  \label{fig:dw_width_1}	
  \includegraphics[width=0.9\columnwidth]{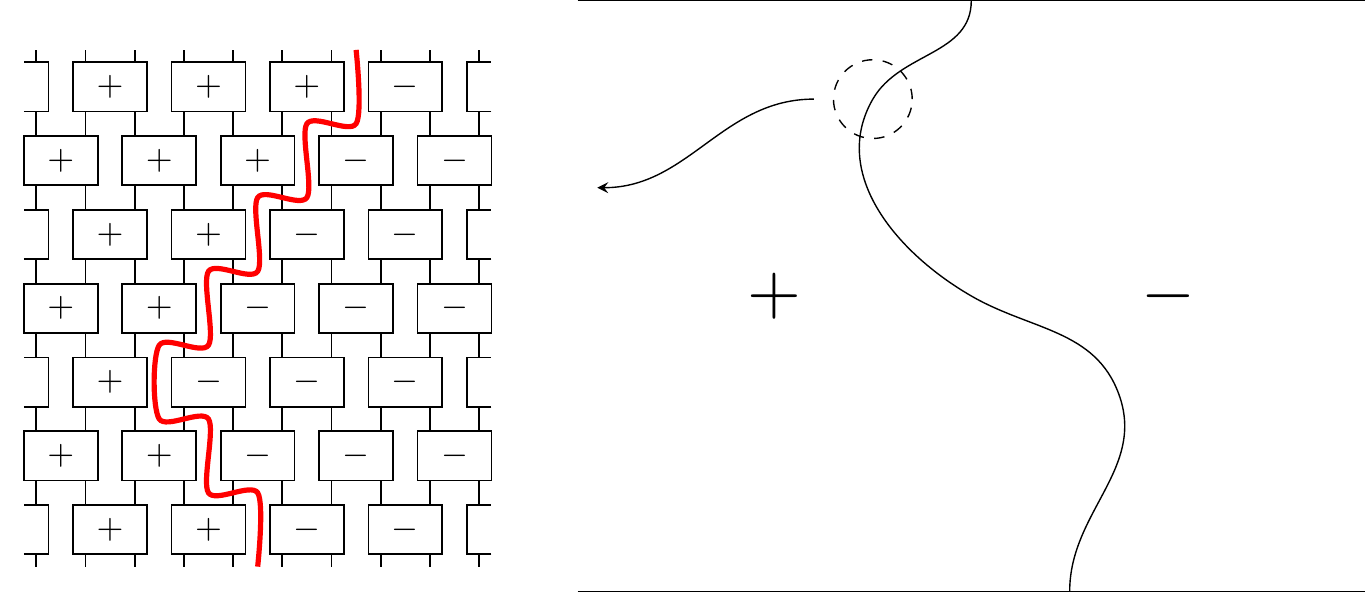}
} \\ 
\subfigure[]{
  \label{fig:dw_width_order_1}	
  \includegraphics[width=0.9\columnwidth]{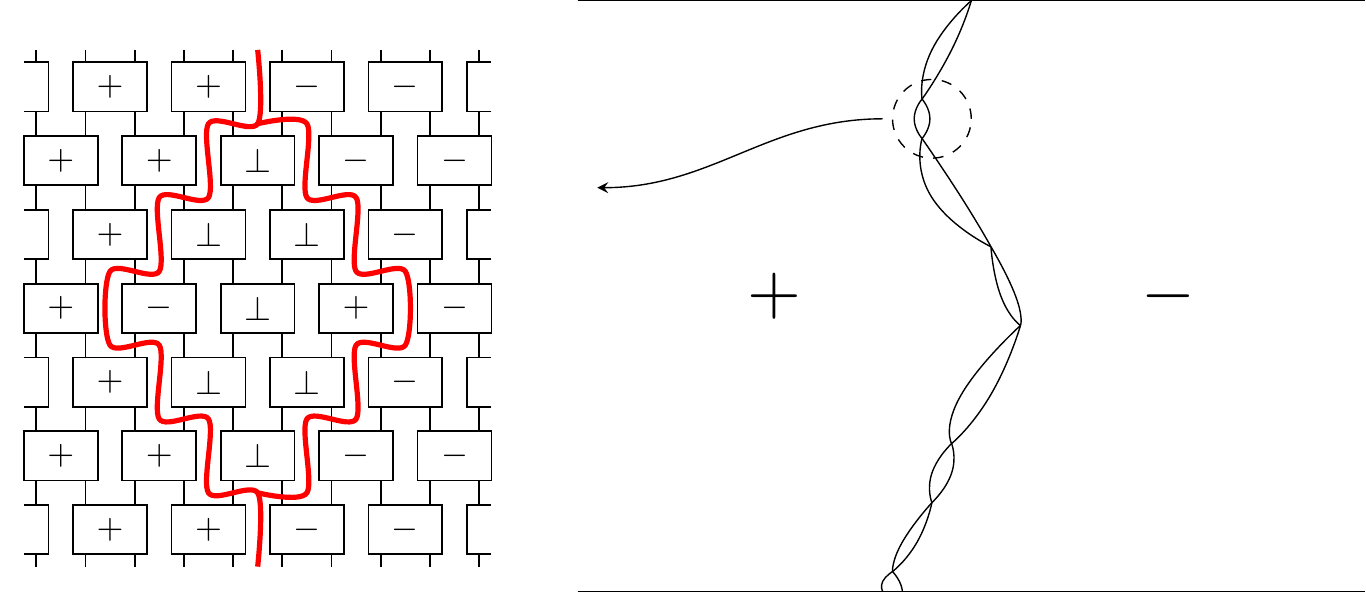}
}
\caption{ Domain walls in (a) $Z_{0 \perp}$ (or in the Haar-averaged circuit) (b) $Z$, taking into account $\perp$ states. In (a), the width of the domain wall is $1$. The width of the domain wall in (b) is argued to be of order $1$.}
\label{fig:dw_width}
\end{figure}

Again let us first consider the sub-partition-function $Z_{0 \perp}$, which is equivalent to that for Haar-random unitaries treated in Ref.~\cite{nahum_operator_2017}.
The spins  have fixed boundary conditions at the top, enforcing a domain wall between $+$ and $-$, and free boundary conditions at the bottom, because ${\langle + | e^{(2)}\rangle}  ={\langle - | e^{(2)}\rangle} =1$.
As a result of the fact that $J(+,+;-)=J(-,-;+)=0$
(the first diagram in Eq.~\ref{eq:tribox_3})
the only nonvanishing configurations in 
$Z_{0 \perp}$ contain a \textit{single} directed domain wall emanating from the entanglement cut, separating an infinite domain of $+$ on the left and an infinite domain of $-$ on the right. This is illustrated in Fig.~\ref{fig:dw_width}, top. Formally
\begin{equation}
Z_{0\perp} = \sum_{\substack{\text{directed}\\ \text{paths}}} K^t = 2^t K^t,
\end{equation}
where $K$, defined in Eq.~\ref{eq:Kdef}, is the weight for a single step in the directed walk.

Now we consider including the terms in the partition function with $s=\perp$, in order to address models without a Haar average.

We have shown that the spins outside the backward light cone are equal and fixed by the final-time boundary condition  (a consequence of causality) and that $\perp$ can only appear within the backward light cone. 
Hence, we can define a `thick' domain wall separating the connected infinite domain of $+$ on the left from the connected infinite domain of $-$ on the right. In principle, this thick domain wall could fill the entire backward lightcone: for example the $\perp$ blocks could form a large percolating cluster of $O(t^2)$ size. If such configurations are dominant, then the representation of the partition function in Eq.~\ref{eq:Zs} is not useful. 

However, we will argue that this does not happen for a typical chaotic choice of the circuit $U$, because of a phase cancellation effect which suppresses large $\perp$ clusters.
Instead the typical thickness of the domain wall remains of order $1$ in $t$ as $t\rightarrow \infty$, so that in a typical configuration the domain wall has the schematic structure in shown in Fig.~\ref{fig:dw_width}. 
The domain wall has a nontrivial structure made up of clusters that can contain $\perp$, $+$ and $-$; these clusters are separated by the places in spacetime where the domain wall becomes ``thin'', which we will take to mean ``of minimal width''.
After coarse graining we recover a membrane picture similar to that in the random circuit, and we can define a renormalized line tension $\mathcal{E}(v)$ for this membrane.

We will give analytical and numerical evidence for this picture in Secs.~\ref{sec:tractable_limit},~\ref{sec:floq_pre}.
In some cases, there may be a small parameter which makes the analytical calculation of renormalized line tension simple. We will argue in Sec.~\ref{sec:tractable_limit} that this is the case for typical choices of $u$ at large local Hilbert space dimension $q$. But even if there is no small parameter, the renormalized line tension can be well defined. We will see below how to extract it from simulations.

We now rewrite the partition function of the domain wall in a way which will be useful when the width of the thick domain wall does not grow with $t$.\footnote{As we will discuss later, this property depends on the boundary conditions.}
As noted above, we then expect that  a typical configuration has order $t$ locations where the domain wall is thin.
We define a directed path that is made of steps connecting these locations: i.e. steps of length $t\geq 1$ connecting the ``pinch points'' in Fig.~\ref{fig:dw_width_order_1}.
The definition of one of these ``irreducible steps'' is that domain wall is thin at the beginning and end of an irreducible step, but nowhere in between.

The weight for a given domain wall configuration can then be written as a product of weights $W(\Delta x, \Delta t)$ for each step of extent $\Delta t$ in the time direction and $\Delta x$ in the space direction. 
Factorization into such a product follows from the locality property mentioned in the previous subsection (in the paragraph following Eq.~\ref{eq:boxabc}).

To simplify the formulas, we restrict for now to systems with space and time translation symmetries such that all the local gates are identical, but the generalization to other cases is direct (Sec.~\ref{sec:tractable_limit}).

Define a partition function $Z(x,y;t)$ with a modified bottom boundary condition, such that there is a  domain wall between $+$ and $-$ at position  $x$ at the top boundary, and at $y$ at the bottom boundary. This can be enforced using the dual states $|(\pm\pm)^* \rangle$ defined in Eq.~\ref{eq:dual_basis}:
\begin{equation}
\label{eq:Z_x_y_t}
Z(x,y;t) = \langle \dots + + -- \dots | U^{(2)}(t) | \dots (++)^* (--)^* \dots. \rangle 
\end{equation}
($x$ labels bonds of the lattice as in Eq.~\ref{eq:Zs}). 
In a translationally invariant system, 
 the domain wall connecting  $(x,t)$ and $(y,0)$
(cf. the schematic picture in Fig.~\ref{fig:ev_intro}) 
forms a path that is straight on scales of order ${t\gg 1}$, with coarse-grained velocity ${v=(x-y)/t}$. 
The ``free energy'' of the domain wall  is proportional to the line tension ${\mathcal{E}(v)}$ for a path with this velocity: at leading order in $t$,
\begin{align}
\label{eq:ZasymptoticsDWsec}
- \ln Z(x, y; t) & \sim s_{\rm eq} \,  \mathcal{E}( v  ) \, t , &  v & \equiv \f{x-y}{t}.
\end{align}  
This asymptotic scaling is one way to define ${\mathcal{E}(v)}$.
However, it is more efficient to extract $\mathcal{E}(v)$ directly from $W$, as we will discuss.

Let the number of timesteps where the domain wall is thin be ${M+1}$, and let their spacetime coordinates be:
\begin{align}
&(y,0), &
&(x_1, t_1), &
&\ldots, &
&{(x_{M-1}, t_{M-1})}, &
&(x, t).
\end{align}
The steps are of time duration and spatial extent
\begin{align}
\Delta t_i &= t_i - t_{i-1} \geq 1,\\
\Delta x_i &= x_i-x_{i-1} \in \mathbb{Z}.
\end{align}
For our choice of lattice geometry, $\Delta x + \Delta t$ is necessarily even.
Defining the weight for an irreducible  step to be $W(\Delta x, \Delta t)$, the total partition function is given by summing over paths of all possible lengths $M$:
\begin{equation}
\label{eq:w_z_recur}
Z(x,y;t ) =  { \sum_{M\ge 1}} \,\,\,
\sum_{
\substack{
 \{\Delta x\}, \,\, \{\Delta t\}
 \\ 
\sum \hspace{-0.5mm} \Delta x_i = x-y
\\
\sum \hspace{-0.5mm} \Delta t_i = t
}
}
\,\,\,
{ \prod_{i=1}^M }
\,\,
W(\Delta x_i, \Delta t_i ).
\end{equation}
Because of translational invariance $Z$ only depends on one spatial argument, so we will also write ${Z(x,t)\equiv Z(x,0;t)}$.

The weight for a step of duration 1 is
\begin{align}\label{eq:minimalstepweight}
W (\pm 1, 1) = K.
\end{align}
The simplest case is for $Z_{0\perp}$, when these are the only steps allowed:
\begin{align}
W_{0\perp} (x,t> 1) = 0.
\end{align}
The $\perp$ insertion produces non-zero values for ${W(x, t>1)}$, which may be either positive or negative. The presence of negative steps is not necessarily an obstacle to defining a coarse-grained line tension. If positive steps predominate then these negative weights will disappear under coarse-graining. However we will suggest that in some fine-tuned cases  negative weights are important  (Sec.~\ref{sec:domainwalldualunitary}).

The rewriting in Eq.~\ref{eq:w_z_recur} is exact for the boundary conditions we have chosen, but this rewriting will only be useful if two conditions are satisfied. First, that the irreducible step weights decay sufficiently fast for large $\Delta t$: 
at a minimum we require that the ratio ${W(x,t) / Z(x,t)}$ tends to zero at large $t$.
Second we require that the original partition function of interest, where the lower boundary condition is free, is simply related to $Z(x,y;t)$. 

To be more precise, we should distinguish between different kinds of usefulness.
First, we conjecture that the above is useful for generic chaotic models for deriving a coarse-grained picture that is in the right universality class. 
Second we will argue that for some strongly chaotic models the above representation is also practically useful for numerical determination of quantities like the line tension $\mathcal{E}(v)$ and the butterfly speed $v_B$.  
Third, in a more restricted class of models with a large parameter, the above can be used to obtain these quantities analytically.

We will describe the numerical algorithm that follows naturally from the above representation in Sec.~\ref{sec:floq_pre}.
Its starting point is to use the recursive form of Eq.~\ref{eq:w_z_recur} to compute $W(x,t)$:
\begin{equation}
\label{eq:Wbyrecursion}
W(x,t) = Z(x,t) -\sum_{t'<t} \sum_{y}  W(x-y, t-t')  Z(y,t').
\end{equation}
This expression follows simply from splitting the partition function $Z(x,t)$ for the directed path into the term $W(x,t)$ from the single-step path with $N=1$, and from paths with $N>1$ whose final irreducible step has weight $W(x-y, t-t')$. Since $Z(x,t)$ can be computed numerically by taking appropriate traces of $U(t)^{(N)}$, we can compute the irreducible step weights $W(x,t)$ by starting with the weight for $t=1$ in Eq.~\ref{eq:minimalstepweight} and employing Eq.~\ref{eq:Wbyrecursion} recursively to get the irreducible weights for larger $t$. The weights can then be used to extract $\mathcal{E}(v)$ and $v_B$.

In practise we will be limited to $t\leq t_\text{max}$ for some $t_\text{max}$, so the usefulness of the algorithm will depend on how rapidly the weights $W$ get smaller at larger $t$.
We apply the algorithm to some non-random but chaotic  Floquet models in Sec.~\ref{sec:floq_app}, with encouraging results. The results support the universal picture described in this section, with a domain wall that is not microscopically thin but has an order 1 width.

The approach can be modified in many ways which do not change the basic structure but which might improve the practical usefulness of the algorithm for a given choice of model. For example, we can  insert projection operators on single sites instead of on double sites, yielding a slightly modified spin model. This has the advantage that the weight of a step of duration $\Delta t = 1$ is no longer independent of the gate $u$ (which it is in the present setup, see Eq.~\ref{eq:minimalstepweight}). It is clear that at least for some choices of $u$ this will give better approximations for small $t_\text{max}$. 
Nevertheless here we will stick with the geometry above, since it is convenient for making contact with the random circuit case, and is sufficient to illustrate the basic ideas.

\subsection{Weights for  smallest nontrivial $\perp$ cluster}
\label{subsec:perp_cluster}

In some limits it is a good approximation to consider only the smallest possible $\perp$ cluster, namely a single isolated $\perp$ block (Sec.~\ref{sec:tractable_limit}).
We can easily write down the weights for such a minimal cluster. They depend on the singular value structure of the local unitary gate $u$ where $\perp$ is inserted. 

Recall that 
\begin{equation}
\label{eq:perpcausality2}
\begin{aligned}
  \tribox[+][+][\perp]  &=
  \tribox[-][-][\perp] = 0.
\end{aligned}
\end{equation}
as a consequence of unitarity (Eq.~\ref{eq:boxabc}).
Therefore there are four possible local configurations for one isolated $\perp$, up to $+\leftrightarrow -$ symmetry (assuming the row above contains a single domain wall):
\begin{equation}
\label{eq:hexbox}
\underbrace{\hexbox[+][-][+][\perp][-][+][-]}_{\circled{1}}
\quad
\underbrace{\hexbox[+][-][+][\perp][-][+][+]}_{\circled{2}}
\quad 
\underbrace{\hexbox[+][-][+][\perp][-][-][-]}_{\circled{3}}
\quad
\underbrace{\hexbox[+][-][+][\perp][-][-][+]}_{\circled{4}}
\end{equation}
In fact only three of these are independent, as $\circled{2}$ and $\circled{3}$ are equal for any $u$, even if it is not reflection-symmetric. Explicit formulas are given below in Eq.~\ref{eq:124weights}.

For the translationally invariant case, the first three diagrams above give the irreducible domain wall weights $W(x,t)$ defined in the previous subsection for $t=2$:
\begin{align}
W(0, 2) & = \circled{1}, 
&
W(\pm 2, 2) = \circled{2}.
\end{align}
By contrast $\circled{4}$ does not contribute to $W(x,2)$, as it does not yield a thin domain wall at the bottom. This diagram will appear only in configurations contributing to $W(x,t>2)$.
In the limit discussed in Sec.~\ref{sec:tractable_limit}, namely typical unitaries at large $q$, only $\circled{1}$ is required at leading nontrivial order in $1/q$.

In order to express the values of $\circled{1}$ to $\circled{4}$,
recall that we can regard the gate $u$ as a quantum state for \text{four} $q$-state spins (``vectorisation'' of the operator). 
In a tensor network language, this simply means that we regard all four of the legs sticking out of $u$ as physical spin indices.
Let us label these legs $A$, $B$, $C$, $D$ as 
\begin{equation}
\label{eq:u_tensor}
u = 
\fineq[-0.8ex][0.6][0.8]{
   \ugatetext[0][0][][A][B][C][D]
}.
\end{equation}
Regarding $u$ as a state makes it clear that (after normalizing this state) we can define the entanglement between any subset of $\{ A, B, C, D\}$ and the complement following the usual prescription for states.
These are referred to as operator entanglements \cite{zanardi_entanglement_2001,prosen_operator_2007,pizorn_operator_2009,dubail_entanglement_2016,zhou_operator_2017},
and have been studied in the context of quantifying the ``entangling power'' of unitary gates \cite{zanardi_entangling_2000,zanardi_entanglement_2001}.
Unitarity implies that $\{A,B\}$ is maximally entangled with $\{C,D\}$, but other entanglements depend on the gate $u$. For example, if $u$ is close to the identity then $\{A,C\}$ is weakly entangled with $\{B, D\}$.

For $N=2$ we require the purities $\mathcal{P} = e^{-S_2}$
\begin{align}
    &\mathcal{P}_{\opsqu[1]} \equiv \mathcal{P}_{AC},
    &
    &\mathcal{P}_{\opsqu[2]} \equiv \mathcal{P}_{AD}.
\end{align}
All others are equivalent (e.g. $\mathcal{P}_{BD}=\mathcal{P}_{AC}$) or independent of $u$ (e.g. $\mathcal{P}_A=q^{-1}$, $\mathcal{P}_{AB}=q^{-2}$).
In terms of these, the weights in Eq.~\ref{eq:hexbox} are (recall $ {\circled{3}} = {\circled{2}}$):
\begin{align}\label{eq:124weights}
   {\circled{1}}& = \phantom{+}
   \left( \frac{  q^4}{q^4 - 1} \right)^2  \left(  \mathcal{P}_{\opsqu[1]} + \frac{1}{q^6} \mathcal{P}_{\opsqu[2]} - \frac{2K}{q} - \frac{2K}{q^7} \right)\\
   {\circled{2}}& = -  \lf \frac{ q^3}{q^4 - 1}\ri^2\left( \mathcal{P}_{\opsqu[1]} + \frac{1}{q^2}\mathcal{P}_{\opsqu[2]} - \frac{2K}{q} - \frac{2K}{q^3} \right)\\
   {\circled{4}}& =  \phantom{+}\lf \frac{q^3}{q^4 - 1}\ri^2 \left( \mathcal{P}_{\opsqu[2]} + \frac{1}{q^2} \mathcal{P}_{\opsqu[1]}  - \frac{2K}{q} - \frac{2K}{q^3} \right) 
\end{align}
See App.~\ref{app:perp_op_purity} for computations.

We now describe an analytically tractable case where it is sufficient to consider dilute $\perp$ insertions of the above form
(Sec.~\ref{sec:tractable_limit}). Then in Sec.~\ref{sec:floq_pre} we consider Floquet models for spin-1/2s, where it is important to allow for larger clusters of $\perp$.


\section{A tractable limit:  large local Hilbert space dimension}
\label{sec:tractable_limit}

We now apply the above formalism to a typical realization of a random unitary circuit at large $q$. This is a useful test ground for our approach, because we can quantify the suppression of large clusters of $\perp$ blocks analytically. 
This section provides analytical support for the thin domain wall conjecture: readers keen to see visual evidence for the success of the method can skip ahead to the numerics in Sec.~\ref{sec:floq_app}.

It is important here to distinguish between the \textit{Haar average} of $e^{-S_2}$ or of $U(t)^{(2)}$, and and a typical  \textit{individual realization} drawn from the random circuit ensemble. The latter represents a specific chaotic time evolution, with much more structure than is captured by simply averaging $U(t)^{(2)}$ over the ensemble.

If we average $U(t)^{(2)}$, the weight of any configuration with a $\perp$ is set to zero, leading to $Z_{0\perp}$. 
But in a given realization of the circuit this is \textit{not} the case, and the $\perp$ insertions have an important effect. 
For example, they modify the growth rate of averaged entropy $\overline{S_2}$, because this quantity is not equal to the simpler ``annealed average'' $- \ln \overline{e^{-S_2}}$.

Consider a particular spin configuration in the partition sum $Z$ (Eq.~\ref{eq:Zs}). As noted in Sec.~\ref{sec:spinmodelgeneralities}, each cluster of $\perp$ blocks in the configuration contributes separately to the Bolzmann weight of the spin configuration. We denote the weight of a given  cluster $C$ by $\Omega_C$. It depends on the geometry of the cluster, the spins $\pm$ on its boundary, and the local unitaries in the spacetime region inside the cluster. 

If we average $\Omega_C$ over the random unitaries inside the cluster region, then we obtain $\overline{\Omega_C}=0$ because of cancellation between positive and negative values. This is just the statement that, after Haar averaging, $Z$ becomes equal to $Z_{0\perp}$, where no $\perp$s appear. 

However, we can ask what the magnitude of $\Omega_C$ is for a typical realization of the circuit. We can quantify this by computing $\overline{\Omega_C^2}$, which is not zero.
We show below that (at least for large $q$, where the calculation is controlled) $\overline{\Omega_C^2}$ is exponentially suppressed in the temporal duration of the cluster, with the exponential suppression getting stronger and stronger as $q$ is increased.

Therefore the Haar circuit at large $q$ is a setting where we can put the phase cancellation conjecture on a quantitative footing. We do this next (Sec.~\ref{sec:expsup}). 
The present approach to the random circuit also allows us to compute statistical fluctuations in the second R\'enyi entropy  due to circuit randomness in Sec.~\ref{sec:kpz}. We recover the result that these are in the  KPZ/DPRM universality class. Previously this result required the replica trick \cite{zhou2019emergent}. Here however we can make the mapping to a  directed polymer at the level of an \textit{individual}  circuit, so the correspondence with known classical problems can be made without use of the replica trick (Sec.~\ref{sec:kpz}).

\subsection{Exponential suppression of large $\perp$ clusters}
\label{sec:expsup}

\begin{figure}[t]
\centering
\subfigure[]{
  \label{fig:perp_cluster}	
  \includegraphics[width=0.46\columnwidth]{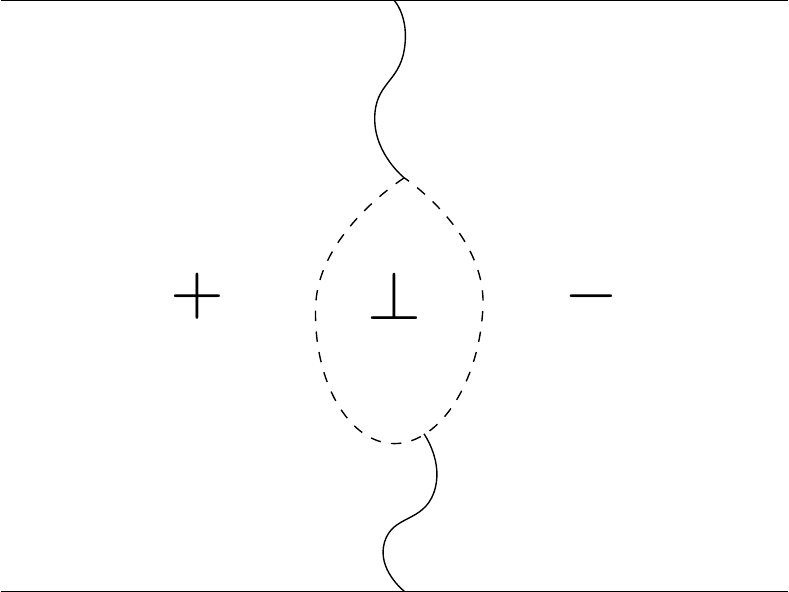}
}
\subfigure[]{
  \label{fig:supp_perp}	
  \includegraphics[width=0.46\columnwidth]{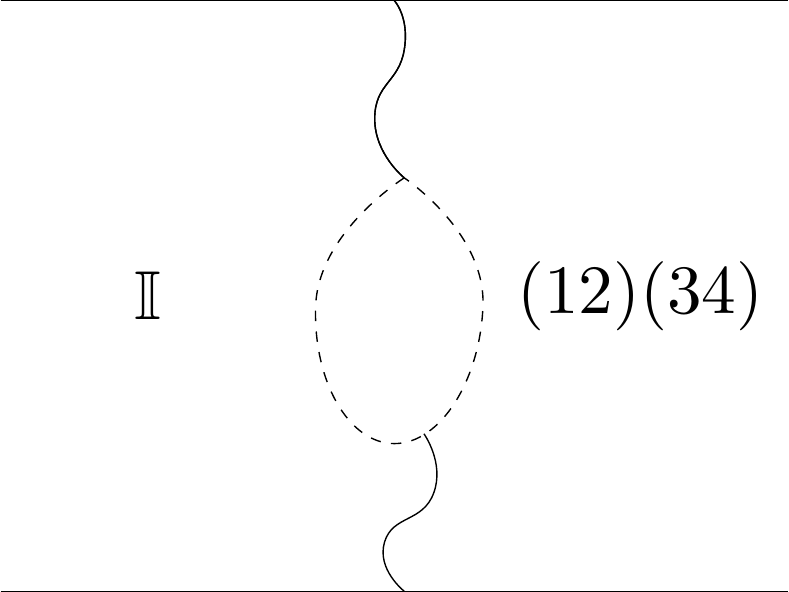}
}
\caption{(a) A domain wall configuration with a $\perp$ cluster. The weight is nonzero only when the domain wall passes through the $\perp$ cluster. (b) The average of the square of (a) in terms of the $S_4$ spin model. The spin states outside the cluster are fixed.}
\label{fig:supp_perp_cluster}
\end{figure}

Let us consider an isolated cluster $C$ of $\perp$ blocks. For simplicity we consider a cluster with trivial topology like that shown in Fig.~\ref{fig:perp_cluster}, but a general cluster with holes inside can be treated similarly. Eq.~\ref{eq:perpcausality2} shows that the weight of the spin configuration vanishes if any $\perp$ block has two spins of the same sign ($++$ or $--$) above it, so the $\perp$ cluster must lie on a domain wall between $+$ and $-$.
We wish to compute the mean square weight of the cluster,  $\overline{\Omega_C^2}$.

Haar-averaging maps 
$\overline{\Omega_C^2}$ to a partition sum for spins ${\sigma\in S_4}$.
The permutations are now in $S_4$ rather than $S_2$ because  squaring the diagram doubles the number of layers in the replicated circuit. 
The spins outside the cluster are fixed, so simply provide a boundary condition for those inside.

This partition sum can be computed relatively simply when $q$ is large.
We summarize the main features here and give a detailed analysis in  App.~\ref{app:supp_perp}. The domain wall labelling convention we use is described in  Ref.~\cite{zhou2019emergent}:  a domain wall with a domain of  $\sigma_L$ to its left, and a domain of $\sigma_R$ to its right, is labelled by the permutation $\sigma_L^{-1}\sigma_R^{\phantom{1}}$.

If we Haar-average the circuit $u^{(4)}$ we obtain spins $\sigma\in S_4$ with the interactions $J$ specified in Eq.~\ref{eq:p_sigma}.
Here the interactions are modified, because we fixed the configurations of spins $s\in \{+,-,\perp\}$ before squaring and averaging. This involves insertions of $P_+$, $P_-$ and $P_\perp$ that modify the triangle weights discussed in Sec.~\ref{sec:spinmodelgeneralities}.

In the doubled problem there is a doubled $(12)(34)$ domain wall incoming at the entanglement cut, as illustrated in Fig.~\ref{fig:supp_perp_cluster}.
In a $+$ domain the $S_4$ spins are fixed to $\mathbb{I}$, and in  a $-$ domain they are fixed to $(12)(34)$.
Inside the $\perp$ cluster the spins must be summed over. However, they are very restricted.
Because of the modified triangle weights, spin values in the subgroup  $S_2\times S_2 \subset S_4$ are suppressed in the interior of the cluster: a triangle whose vertices are all equal and take one of the values  $\mathbb{I}$, $(12)$, $(34)$ or  $(12)(34)$ has weight zero. In fact we have
\begin{align}\label{eq:perptriangleallequal}
\jabc[\sigma][\sigma][\sigma][\perp]  & =
\left\{
\begin{array}{ccc}
 1 & \text{for} &  \sigma\notin S_2\times S_2,     \\
 0 &  \text{for} &  \sigma \in S_2\times S_2,
\end{array}
\right.
\end{align}
where the symbol on the left denotes the weight for a triangle whose lower spin is associated with a $\perp$ block.

In the large $q$ limit, the leading term involves the incoming doubled $(12)(34)$ domain wall splitting at the top of the cluster into a pair of doubled domain walls, according to the allowed multiplication (cf. Sec.~VI B of \cite{zhou2019emergent}):
\begin{align}\label{eq:splitting}
(12)(34) & = (14)(23) \times (24)(13).
\end{align}
One doubled domain wall, say $(14)(23)$, travels along the left boundary of the cluster, and the other along the right. 
[Since $(14)(23)$ commutes with $(24)(13)$, either of the two doubled domain walls can be on the left.]
In the interior of the cluster, the spin value is then $(14)(23)$  [or $(24)(13)$]. This does not incur any ``bulk'' cost, because a triangle all of whose vertices are equal to $(14)(23)$ has weight unity by Eq.~\ref{eq:perptriangleallequal}.

The two doubled domain walls along the two boundaries of the cluster incur a weight $q^{-4}$ per time step.\footnote{At large $q$, an ``elementary'' domain wall, i.e. a transposition, costs $1/q$ per time step. Here we have a total of 4 elementary domain walls (the right-hand side of Eq.~\ref{eq:splitting}).} Consequently the mean squared cluster weight is of order  $\overline{\Omega^2} \sim q^{- 4 t_{\rm cluster}}$ or even smaller, where $t_{\rm cluster}$ is the time duration of the cluster. (We have not included the cost of the ``thin'' domain wall sections above and below the cluster.) 
This indicates a typical magnitude
\begin{equation}
\Omega \sim q^{- 2 t_{\rm cluster}}
\end{equation}
for a $\perp$ cluster of a given size in a typical large $q$ circuit.
Note that this weight is much \textit{smaller} than the weight $\sim q^{-t_{\rm cluster}}$ of a section of thin domain wall of the same time  duration.
Similar considerations apply to clusters with more complex topology, i.e. with holes inside.

Therefore, any configuration with a $\perp$ cluster lasting for $t_{\rm cluster}$ steps has a weight that is suppressed by \textit{at least} $q^{-t_{\rm cluster}}$ compared to the leading configurations in the same spacetime region. 

It should be noted that {$\perp$ clusters}
are suppressed by a cost in the exponent that scales not with their area in spacetime, but instead with the  length {of their boundary.}
This is reminiscent of a domain wall between two different spin states in the ordered phase of the Potts model with $Q>2$ states, where bubbles of other spin states can appear on the domain wall \cite{cardy2000renormalisation}. Here this scaling is a necessary consequence of unitarity.
As a result of it, the effect of boundary conditions can be subtle. For example, a boundary condition at $t=0$ that favours $\perp$ may be able to induce a large domain of $\perp$, because a boundary free energy of order $t$ can compensate the additional domain wall cost of order $t$ in the bulk. This is not the case for the quantities considered here, however. Indeed, for entanglement growth starting from a product state, the $t=0$ boundary condition favours $\pm$ over $\perp$.

The results above strongly support the conjectures of the previous section.
For example, there we defined weights $W$ for ``steps'' of variable temporal duration that connect locations where a domain wall between $+$ and $-$ is ``thin''. 
We conjectured that the contribution of long steps to the partition function was suppressed. 
The calculation above confirms this conjecture explicitly for a typical circuit realization at large $q$. (A realization of a random circuit is not translation-invariant, but $W$ generalizes directly to this case.)
We find that the contribution of long steps to the line tension is  suppressed exponentially (${\sim q^{-\Delta t}}$) in the step duration ${\Delta t}$.
Note that, according to our definition in Sec.~\ref{sec:domainwallstructure}, a step of duration $\Delta t \gg 1$ does not need to include $\mathcal{O}(\Delta t)$ $\perp$ blocks: for example, we can have step with a single $\perp$ block at the top, which allows the domain wall to branch into three domain walls between $+$ and $-$, which  merge again at the bottom of the step.  Such configurations are also exponentially suppressed, simply because they contain extra domain walls, each contributing an $\mathcal{O}(q^{-\Delta t})$ factor to the cost \cite{zhou2019emergent}.

Our results in Sec.~\ref{sec:floq_app} for translation-invariant Floquet {spin-1/2} circuits will also be compatible with exponential suppression of large steps (App.~\ref{subsec:W_decay}).

The results in this Section also mean that, at large $q$, we can obtain the nontrivial fluctuations of the entanglement entropy in a random circuit by considering only single-$\perp$ clusters. We discuss this next, before moving on to deterministic models.

\subsection{KPZ/DPRM scaling in the random circuit}
\label{sec:kpz}

Recall that $e^{-S_2} = Z$ with the appropriate (free) lower boundary condition.
Averaging over Haar-random gates gives $\overline{Z} = Z_{0 \perp}$ (Sec.~\ref{sec:def_spin_model}).
However, the averaged R\'enyi entanglement entropy $\overline{S_2 }$ cannot be obtained from $-\log( \overline{Z} )$, because of fluctuations.

Let us first recall the replica approach used previously. This involves computing the replicated partition function, $\overline{Z^k}$, and extracting quantities like $\overline{S_2}$ from the $k \rightarrow 0$ limit. At large $q$ there is a simple picture for the replicated partition function $\overline{Z^k}$. It describes $k$ domain walls that interact via a mutual attraction. In the microscopic $S_{2k}$ spin model, the attraction is due to an additional local spin configuration that is possible when two domain walls come into contact with each other \cite{zhou2019emergent}.

This description can be identified with the replica description of a well-known classical problem, the directed polymer in a random medium \cite{HuseHenleyFisherRespond,kardar_dynamic_1986}. In that context,
the free energy of the polymer can be calculated by replicating the system  $k$ times and integrating out the disorder. For an appropriate choice of lattice model, the resulting $k$ polymers will have exactly the same mutually attractive interactions as the domain walls here.
Therefore, we have indirectly mapped the entanglement to the free energy of a single domain wall in a random potential that modifies the weights of steps.
This leads to  KPZ scaling \cite{HuseHenleyFisherRespond,kardar_dynamic_1986} for the fluctuations of the entanglement. 

In the present approach we consider a single domain wall, without replicas. 
However, to go beyond $Z_{0\perp}$, we must ``dress'' the domain wall with insertions of $\perp$. We have shown above that clusters of $\perp$ are exponentially suppressed in their size. At large $q$ it suffices to consider only dilute, isolated $\perp$ clusters made up of a single block. Further, at large $q$ the local configuration labelled $\circled{1}$ in Eq.~\ref{eq:hexbox} dominates over the other configurations in Eq.~\ref{eq:hexbox} (App.~\ref{app:size_fluct}).
 
\begin{figure}[t]
\centering
\includegraphics[width=0.9\columnwidth]{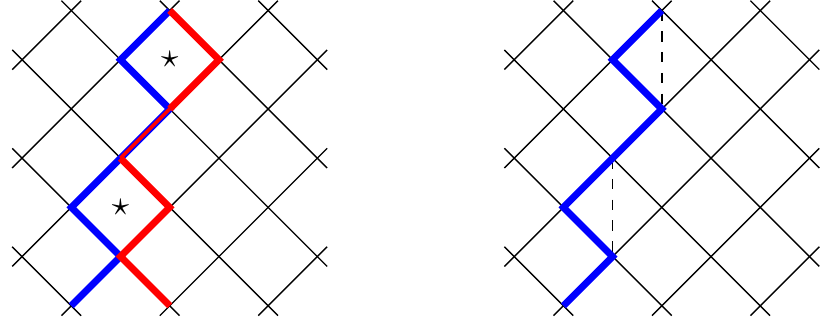}
\caption{Domain walls on a tilted square lattice. Left: In the replica approach, two domain walls have an attractive interaction when they both visit two sides of a plaquette (starred). Right: This replica treatment describes a single domain wall subject to a random potential on the vertical bonds. The domain wall in the figure is affected by the potentials on the two dashed bonds.}
\label{fig:walk_on_lattice}
\end{figure}

By Eq.~\ref{eq:perpcausality2}, these $\perp$ insertions are restricted to lie \textit{on} the domain wall. 
In the notation of Sec.~\ref{sec:domainwallstructure}, this gives us a nonzero weight
\begin{equation}
W(0, 2; u)
\end{equation}
for a step with $\Delta x = 0$ and $\Delta t = 2$. The notation indicates that this weight depends on the local gate $u$, unlike the deterministic weights $W(\pm 1, 1)$. Since all the $u$s are independent, this yields an uncorrelated random potential for our ``polymer'' in spacetime.\footnote{In more detail: since the weight $W(0,2;u)$ can be either positive or negative, it is convenient to absorb it as a correction to the weight of a pair of length-1 steps. This yields a directed path made of length-1 steps, with a random potential associated with pairs of sites vertically above each other: see  Fig.~\ref{fig:walk_on_lattice}, Right. The Boltzmann weight is positive in this representation \cite{zhou2019emergent}.}
In an equivalent stochastic differential equation representation for $S_2(x,t)$, which is the KPZ equation,  $W(0,2;u)$ determines spatiotemporal noise.

The above picture, with random weights assigned to vertical steps, agrees with the replica approach: we just need to check the  strength of the noise matches.

The average $\overline{W(0,2;u)}$ vanishes. Its variance is given  by the squared average of  $\circled{1}$ in Eq.~\ref{eq:hexbox} (App.~\ref{app:size_fluct}):
\begin{equation}
\overline{\hexbox[+][-][+][\perp][-][+][-][s]} \simeq  \frac{2 K^4}{q^4}
\end{equation}
where ${K = {q}/({q^2 + 1})}$ was defined in (\ref{eq:Kdef}). This matches the result for the variance in Ref.~\cite{zhou2019emergent}.

We therefore confirm that a thin domain wall with dilute $\perp$ insertions reproduces the result of the replica calculation.


\section{Floquet spin-1/2 chains: preliminaries}
\label{sec:floq_pre}

We now apply the formalism of Sec.~\ref{sec:def_spin_model} to models that are  invariant under both space and time translations. In the brickwork circuit geometry, such a model is defined by a single 2-site unitary $u$.

In Sec.~\ref{sec:domainwallstructure} we defined the ``thick'' domain wall that appears in the calculation of $Z(x,t)$.
This reduces this partition function to one for a directed path with irreducible step weights $W(\Delta x, \Delta t)$ that take into account successively larger clusters of $\perp$.
The line tension of this path can be extracted easily from the weights $W$.

In Sec.~\ref{sec:floq_app} we will present numerical data for several models using this scheme. 
In preparation for this we next describe how to obtain systematic approximations to $\mathcal{E}(v)= \mathcal{E}_2(v)$ and the butterfly velocity $v_B$ from the weights $W(\Delta x, \Delta t)$ (Sec.~\ref{sec:generatingfunctions}). 

The numerical method for obtaining the weights $W$ themselves is described in App.~\ref{subsec:protocol_WZ}. We use exact diagonalization, which allows us to treat $\Delta t$ up to $t_\text{max}=8$. Together with the analytical results below, this yields a straightforward algorithm which can be applied to any circuit. Sec.~\ref{sec:beyond_circuit} gives an extension to dynamics that are not of circuit form.

\subsection{Extracting $\mathcal{E}(v)$ from $W$}
\label{sec:generatingfunctions}

Recall that the line tension function is encoded in the asymptotics of the partition function $Z(x, t)$ defined in Eq.~\ref{eq:w_z_recur}:
\begin{align}
\label{eq:Zasymptotics}
Z(x, t) & \sim \exp( - s_{\rm eq} \,  \mathcal{E}( v  ) \, t ), &  v & = x/t.
\end{align} 
We use generating functions to determine $\mathcal{E}( v )$. Define generating functions for the partition function $Z$ and for the irreducible weights $W$, 
\begin{align}
\label{eq:zetadef}
\zeta ( \sdensity, b ) &= \sum_{x, t } Z(x, t) e^{\sdensity x} b^t,  \\
\omega(\sdensity, b ) &= \sum_{x, t } W(x, t) e^{\sdensity x} b^t.
\label{eq:omegadef}
\end{align}
(The variable $s$ in this section should not be confused with the spin variables elsewhere in the paper.) We can relate these generating functions using Eq.~\ref{eq:w_z_recur}, which yields a geometric sum for $\zeta$:
\begin{equation}\label{eq:geometricsummed}
\begin{aligned}
  \zeta(\sdensity, b) &= 
  \frac{\omega( \sdensity, b) }{ 1 - \omega( \sdensity, b) }.
\end{aligned}
\end{equation}

By truncating the sum in Eq.~\ref{eq:omegadef} at order $b^{t_\text{max}}$ we will obtain a polynomial approximation to $\omega(s,b)$. 
This approximation is physically motivated: it amounts to setting a maximum step length $t_\text{max}$ in the directed walk of Eq.~\ref{eq:w_z_recur}. We expect this approximation to improve systematically with $t_\text{max}$ for chaotic models.

We then need the prescription for obtaining  the asymptotics of $Z$, or in other words the line tension  $\mathcal{E}(v)$, from the numerically accessible quantity $\omega(s,b)$.
This can be done with the pole method. For a given $s$, let $b_0(s)$ be the smallest root (in absolute value)  of the denominator ${\omega(s, b)-1}$ in  Eq.~\ref{eq:geometricsummed}. Then the desired relation is
\begin{equation}
\label{eq:ev_legendre}
\mathcal{E}(v) = \frac{1}{s_{\rm eq}} { \max_\sdensity   \big(   \ln |b_0 (\sdensity) | + v \sdensity \big). }
\end{equation}
This relation is explained in Appendix~\ref{app:legendre}.
The root $b_0$ appearing here has a physical meaning (although we will not need it here): if we write
\begin{equation}
s_{\rm eq} \Gamma(\sdensity ) =  \ln |b_0(\sdensity) |,
\end{equation}
then $s_{\rm eq} \Gamma(s)$, 
which is related to ${\seq \mathcal{E}(v)}$ by a  Legendre transformation, 
Eq.~\ref{eq:ev_legendre},  is the R\'enyi entropy growth rate in a state with $\partial S/\partial x= s$ \cite{jonay_coarse-grained_2018}.

As an example, consider the lowest-order approximation where we truncate $W$ at $t_\text{max}=1$, 
corresponding to the partition function without any $\perp$ blocks. We then need only
\begin{align}\label{eq:lowestorderW}
W( x, 0 ) & = 0, & 
W( \pm 1,  1 )  &= K. 
\end{align}
At this order, $\omega( \sdensity, b ) = 2K b \cosh \sdensity$, giving the root
\begin{equation}
b_0(\sdensity) = \frac{1}{2K \cosh \sdensity }.
\end{equation}
The quantity $\ln b_0(\sdensity)+v \sdensity$ in  Eq.~\ref{eq:ev_legendre} is maximized by $\sdensity = \tanh^{-1} v$. Plugging this in gives
\begin{equation}
\label{eq:ruc_ev}
s_{\rm eq} \mathcal{E}(v) = \ln K^{-1} + \frac{1 - v}{2} \ln \f{1 - v }{2} + \frac{1 + v}{2} \ln \frac{1 + v}{2}.
\end{equation}
This is indeed the correct result for $Z_{0\perp}$ \cite{jonay_coarse-grained_2018,zhou2019emergent} (and also the leading order result in the random circuit at large $q$).

In the numerical algorithm  we compute $W( x,  t)$ up to ${t = t_\text{max}}$ and neglect $W$ for larger times. 
Given a constant $s$, we then  solve numerically for the smallest zero $b_0(s)$ of
\begin{equation}
\label{eq:polytmax}
1- \sum_{x, t\le t_\text{max}} W(x, t) e^{ s x} b_0^t = 0.
\end{equation}
If we define\footnote{$W$ here is real by definition. In principle the smallest root $b_0$ may be complex (in which case it has a complex conjugate partner $b_0^*$). We have allowed for this above. However from the physical interpretation of the quantities above we expect that $b_0$ is real and positive at least for large $t_\text{max}$.}
\begin{align}
v_s = \operatorname{Re}\,
 \frac{\sum_{x, t\le t_\text{max}} W(x, t) e^{ s x}  b_0(s)^t \times x}{\sum_{x, t\le t_\text{max}} W(x, t) e^{ s x}  b_0(s)^t \times t },
\end{align}
then $\mathcal{E}(v_s)$ is given by  ${s_\text{eq}^{-1} (\ln |b_0(s)|+v_s s)}$. (We see this by  setting the $s$-derivative of 
the argument in  Eq.~\ref{eq:ev_legendre} to zero, and extracting the derivative of $\ln |b_0(s)|$ from Eq.~\ref{eq:polytmax}.)
We iterate over $s$ to construct the entire curve.

This approach also allows us to extract the butterfly speed $v_B$, as the point\footnote{
If $u$ is not reflection-symmetric, there are separate left and right butterfly velocities $v_B^L<0$ and $v_B^R>0$ \cite{jonay_coarse-grained_2018, stahl2018asymmetric, zhang2019asymmetric}.  In this situation $v_E= \min_v \mathcal{E}(v)$,
and ${\mathcal{E}(|v_B^L|)=|v_B^L|}$, and 
${\mathcal{E}(-|v_B^R|)= |v_B^R|}$.
Our convention here is that the support of the operator $\mathcal{O}(t)$ grows, with $t$, to the left at speed $|v^L_B|$ and to the right at $|v^R_B|$.
}
where ${\mathcal{E}(v_B)=v_B}$. The ``entanglement speed'' $v_E$ for the second R\'enyi entropy, for a quench from the product state, is simply $\mathcal{E}(0)$.

It is worth noting that this generating function approach sums up an infinite number of domain wall configurations directly in the thermodynamic limit, eliminating a significant source of finite-$t$ effects.
As an illustration, if (hypothetically) steps of size greater than $t_\text{max}$ had  zero weight, the present approach would give the \textit{exact} result for $\mathcal{E}(v)$ already at time $t_\text{max}$. In reality the weights are not zero for large $t$, but the analytical results in Sec.~\ref{sec:expsup} suggest that the convergence in $t_\text{max}$ is typically exponential.

By contrast, attempting to directly extract the exponential decay rate of $Z(vt,t)$ by a fit for $t\leq t_\text{max}$ is subject to  finite-$t$ effects that are generically only polynomially small in $t_\text{max}$ (as one can see in the case where the domain wall is a simple random walk).
Therefore understanding the domain wall structure improves the computation of the line tension.

Ref.~\cite{jonay_coarse-grained_2018} extracted the line tension associated with the von Neumann entropy for a chaotic Ising model directly from the operator entanglement of $U(t)$. That study indeed noted larger finite--$t$ effects than those found here. However, we do not address the von Neumann entropy (as opposed to the higher R\'enyi entropies) here.

\subsection{General parameterization}
\label{sec:uparameterization}

In our discussion of the numerical results, it will sometimes be useful to use the following parameterization of the two-site gate $u$, which is an arbitrary ${\rm SU}(4)$ matrix
\cite{kraus2001optimal,bertini2019exact,khaneja_cartan_2000,piroli_exact_2019}:
\begin{equation}
\label{eq:sufour}
u=
\sufour[u_{\rm sym}][u_1][u_2][u_3][u_4]
\end{equation}
where
\begin{align}
u_{i}  &= \exp\left( - i \sum_{\alpha = x, y, z} h^{(i)}_\alpha \sigma^\alpha  \right), 
\\
u_{\rm sym}  & = \exp \lf -i \sum_{\alpha = x, y,z} J_{\alpha}\, \sigma^\alpha \otimes \sigma^\alpha  \ri.
\label{eq:u_sym}
\end{align}
The $u_i$s scramble the individual sites and the two-site unitary $u_{\rm sym}$ entangles the two sites. 

We will mostly restrict to the reflection-symmetric case such that $u_{1}  = u_2 $ and $u_3 = u_4$. Additionally,  $Z(x,t)$ is unchanged by the transformation 
${u\rightarrow (v  \otimes v )^{-1} u (v \otimes v)}$, where $v$ is any single site unitary. This amounts to a trivial redefinition of the circuit, and one may check that the conjugate operators $v$ and $v^{-1}$ all cancel out in the weights in the partition function. Exploiting this symmetry, we can take
\begin{equation}\label{eq:hequal}
\vec{h}^{(i)} = \vec{h} \quad \text{ for } i = 1, 2, 3, 4,
\end{equation}
so that the general symmetric gate is parameterized only by the two vectors $\vec{h}$ and $\vec{J}$.


\section{Application to Floquet circuits}
\label{sec:floq_app}

\begin{figure}[t]
\subfigure[]{
  \label{fig:sym_gate}	
  \includegraphics[width=\linewidth]{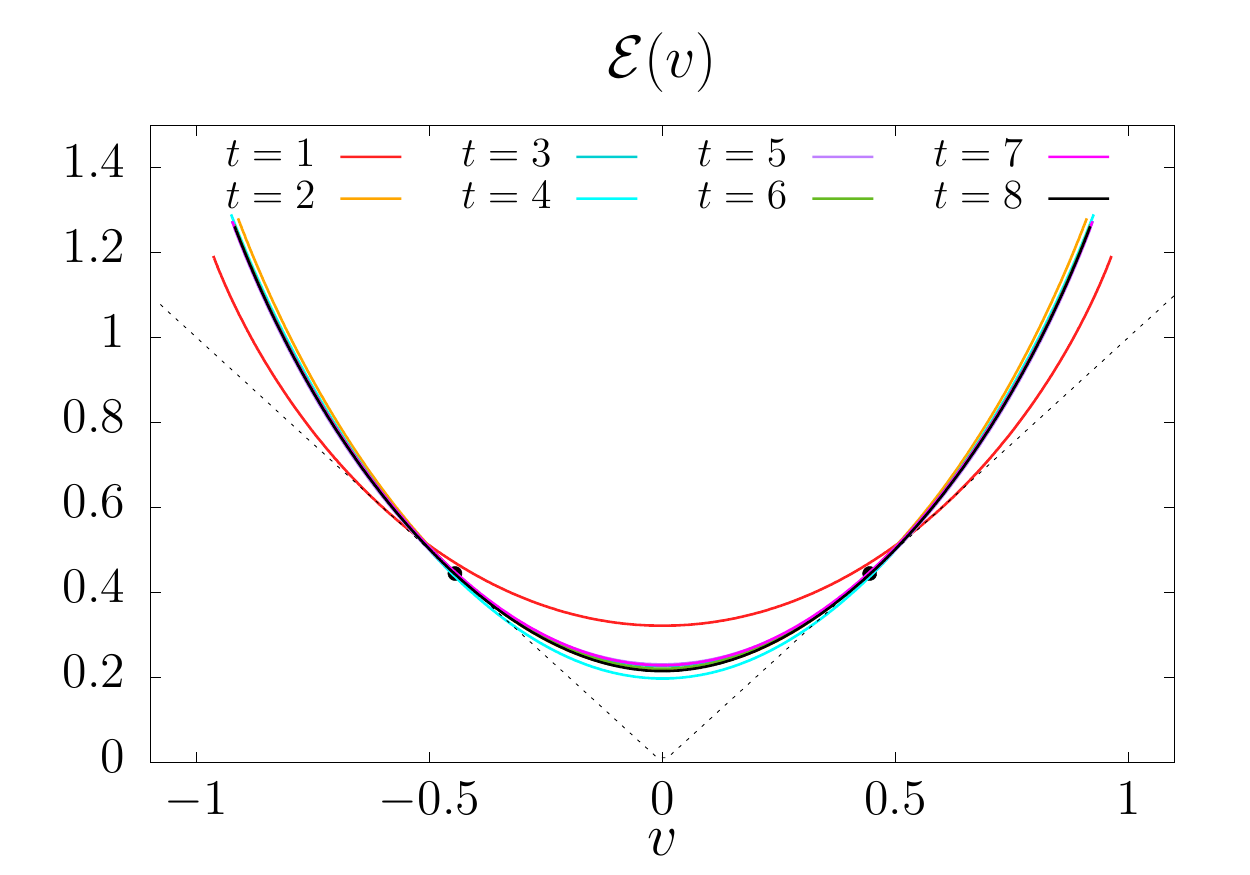}
}\\
\subfigure[]{
  \label{fig:rand_gate}	
  \includegraphics[width=\linewidth]{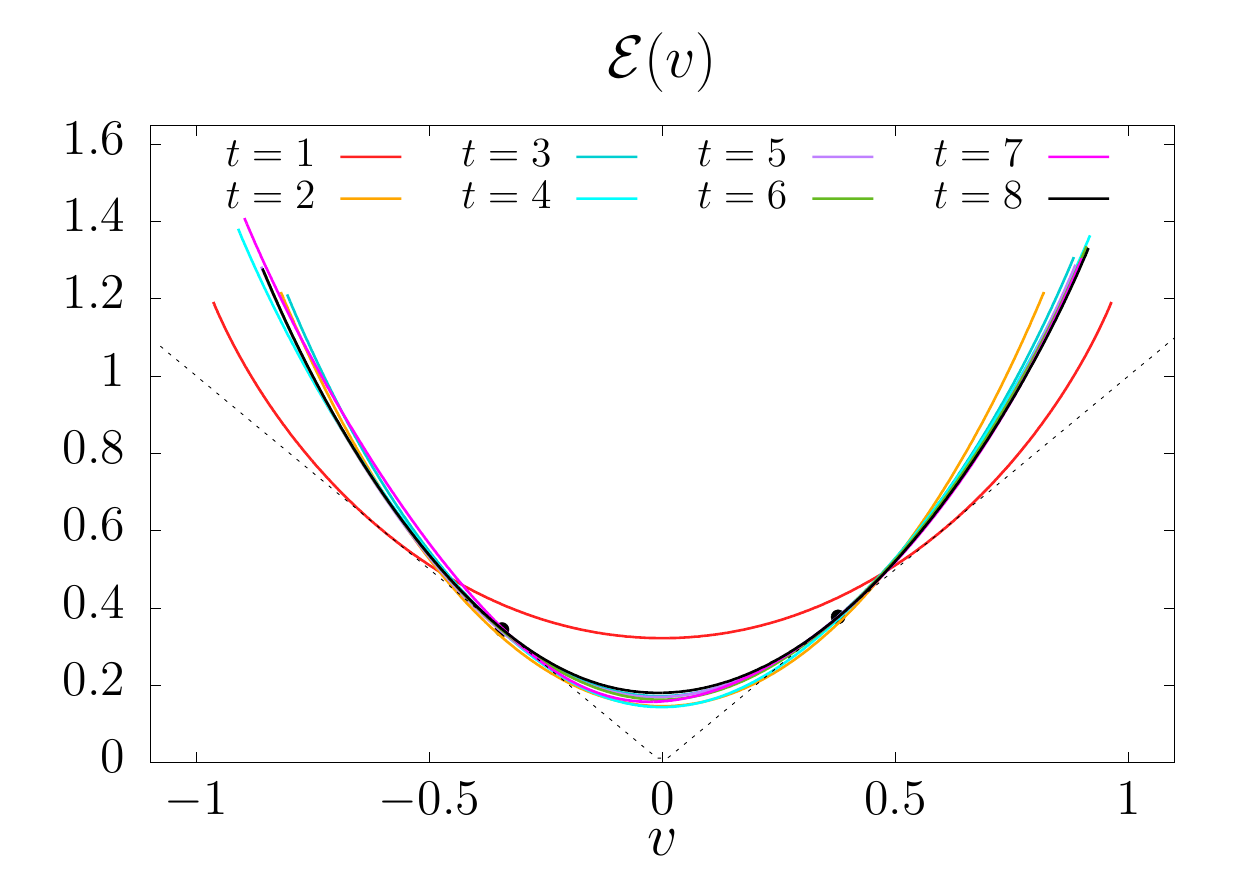}
} 
\caption{Numerical computation of $\mathcal{E}(v)$ with $t_{\rm max}$ up to $8$ for
(a) the reflection-symmetric gate in Eq.~\ref{eq:symmetricgate} with $x=0.8$; (b) a generic non-symmetric gate (specified in Appendix.~\ref{app:num_res}).
Black dots mark estimates of $v_B^{L/R}$ from an independent calculation of the OTOC in App.~\ref{subsec:protocol_vb}.
}
\label{fig:sym_rand_ev}
\end{figure}

\begin{figure}[t]
\subfigure[]{
  \label{fig:huse_ev}
  \includegraphics[width=\columnwidth]{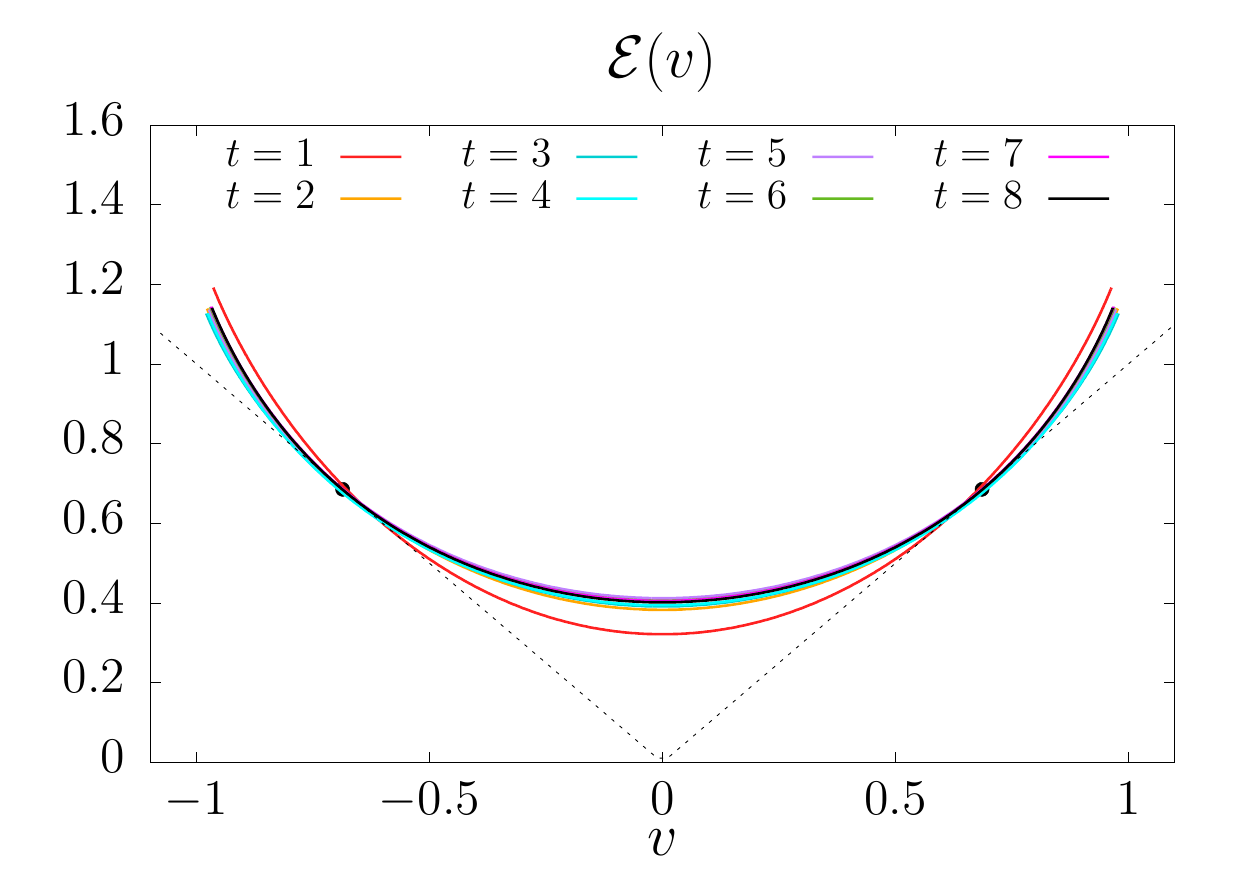}
}
\subfigure[]{
  \label{fig:prosen_gate}	
  \includegraphics[width=\columnwidth]{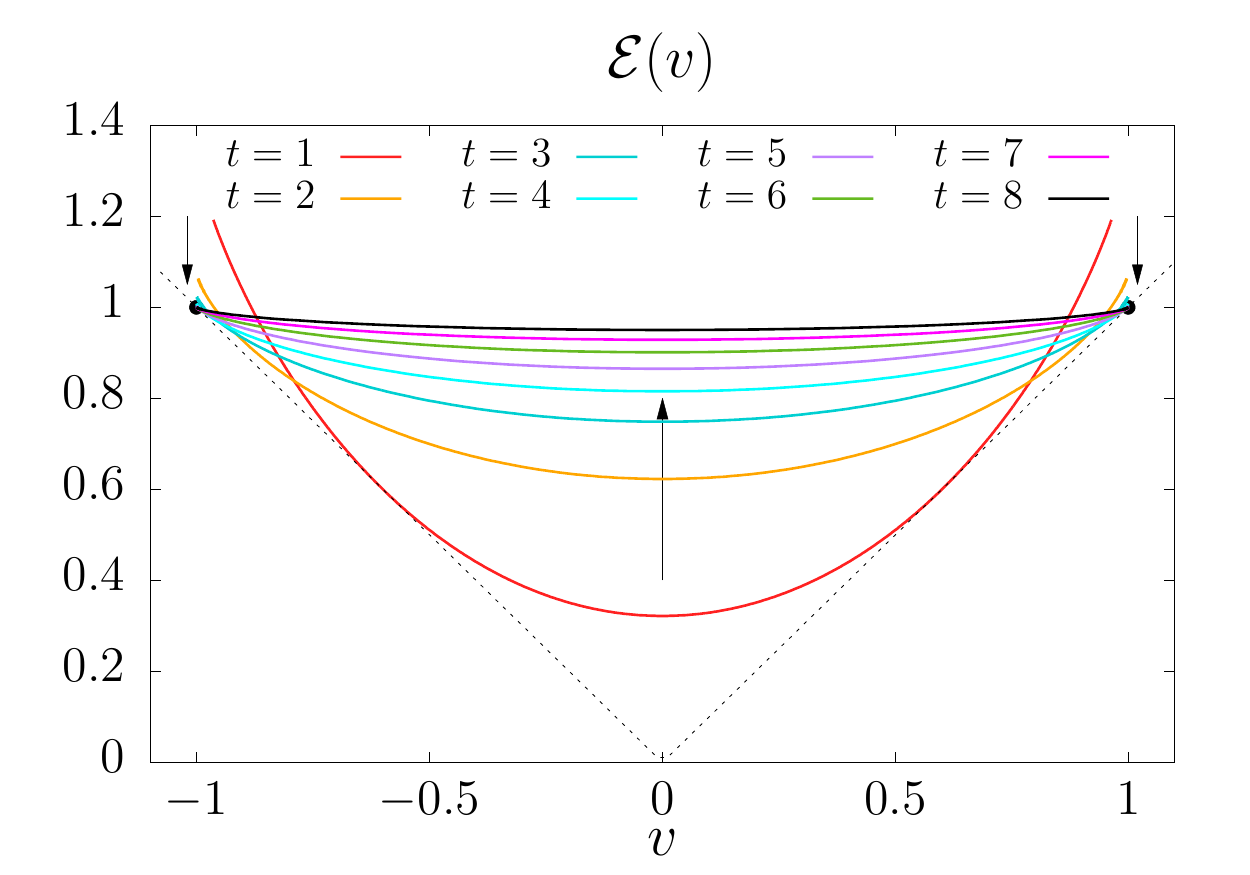}
}
\caption{Numerical calculations of $\mathcal{E}(v)$ for Floquet Ising models. Panel (a) shows the parameters in Eq.~\ref{eq:isinggeneric}. Panel (b) shows the ``dual-unitary'' parameter values in Eq.~\ref{eq:dualunitaryspecific}. 
The bottom arrow in case (b) illustrates the convergence to the expected limiting form $\mathcal{E}(v) = 1$. The intersections with the line $\mathcal{E}(v)=|v|$ are in good agreement with the known value $v_B=1$ (top arrows).}
\label{fig:floqising}
\end{figure}

We are finally ready to apply the numerical scheme in Sec.~\ref{sec:generatingfunctions}
to a variety of Floquet circuit models, described in the following subsections. In Sec.~\ref{sec:beyond_circuit} we will show that the restriction to circuits is not necessary. But circuits are especially convenient for numerics since they have a strict lightcone structure (propagation of information outside the lightcone is not only exponentially suppressed, as in generic spin chains \cite{lieb_robinson}, but exactly zero). The space of circuit models is also already very rich.

For each model we study we will show the sequence of approximations to $\mathcal{E}(v)$, indexed by $t=1, \ldots, t_\text{max}$ with $t_\text{max} =8$: see for example  Fig.~\ref{fig:sym_rand_ev}.
These successive approximations take weights $W$ for longer and longer steps into account.

The curve for $t = 1$ (in red) is independent of the gate defining the circuit: 
this lowest approximation matches the result of the random average computed in Eq.~\ref{eq:ruc_ev}. It can be thought of as a baseline showing how far the R\'enyi  entropy growth in a particular circuit  deviates from the random average.

Our algorithm works if the curve converges  sufficiently rapidly to the actual line tension function. In addition to showing the bare plots, we perform several other checks of convergence.
In Sec.~\ref{sec:Wstructure} we will directly examine the decay of $W(x,t)$ and $W(x,t)/Z(x,t)$ at large $t$.
The assumption on the structure of the domain wall implies that the latter should be negligible at large $t$. 

Additionally, we  check that the line tension function $\mathcal{E}(v)$ satisfies several constraints. It is positive, convex, greater than or equal to $|v|$, and tangent to this line only at $v_B^{\rm L/R}$. We therefore plot the boundary curve $\mathcal{E}=|v|$ with dotted lines. We also mark an  estimate of $v_B$ from an independent numerical computation of the OTOC (the protocol is described in App.~\ref{subsec:protocol_vb}). We show this estimate as a pair of black dots.

Convergence to a form consistent with these constraints is a nontrivial test, because they are not built into the formulas for finite $t_\text{max}$ (indeed we will see that they can be disobeyed by the small--$t_\text{max}$ approximations).

\subsection{Generic symmetric and asymmetric gates}
\label{subsec:sym_rand}

First, we consider the following one-parameter family of reflection-symmetric gates (see Eqs.~\ref{eq:sufour}--\ref{eq:hequal}):
\begin{align}
\label{eq:symmetricgate}
\vec{J} &= x \, \frac{\pi}{4} ( 3, -4, 5) / \sqrt{ 50},&
\vec{h} &= \frac{\pi}{5} ( 1,2,3) / \sqrt{14}.
\end{align}
The parameter $x$ tunes the strength of the interaction between the two qubits.
For small $x$, local scrambling will take a long time: we do not expect our algorithm to converge rapidly in that case, so we will choose $x$ reasonably large.
Otherwise, the numbers above are arbitrary and were chosen to ensure (1) that there is no fine-tuning in the sense of any of the coefficients $J_\alpha$ vanishing, (2) that $u_i$ represents a rotation on the Bloch sphere by a significant angle (here $\pi/5$).

Fig.~\ref{fig:sym_gate} shows results for the gate with $x=0.8$. The results are consistent with a relatively fast convergence of $\mathcal{E}(v)$.
We see that this chosen gate has a smaller $v_B$ and a smaller $v_E^{(2)}$ than the annealed average for the random circuit.

Note that the results are in striking agreement with the general constraints listed above: 
the asymptotic $\mathcal{E}(v)$ is a convex curve that touches the line $\mathcal{E}=v$ at a single point, which is consistent with $(v_B, v_B)$ as obtained from an independent calculation of $v_B$.

Next, Fig.~\ref{fig:rand_gate} shows an example of a generic gate without the reflection-symmetry constraint. We simply picked a random gate from the Haar distribution on ${\rm U}(2)$ and used it to build a translation-invariant circuit. 
The matrix elements of this gate are given explicitly in Appendix~\ref{eq:u_rand}.

Again, the algorithm appears to be working. 
However the convergence is now to an asymmetric curve with {$\mathcal{E}(v) \neq \mathcal{E}(-v)$}. The touching point on the left marks $v_B^R$ and that on the right marks $v_B^L$.

\subsection{Chaotic Ising models}
\label{subsec:chaotic_ising}

Next we consider Floquet ``kicked'' Ising  models \cite{prosen_chaos_2007,kim2013ballistic,kim_testing_2014}  with the Floquet unitary 
\begin{align}
U_{\rm Floq} &= \exp( - i \tau H_z ) \exp( - i \tau H_x)
\end{align}
with a longitudinal field to spoil integrability:
\begin{align}
H_x &= \sum_{i=1}^{L} h_x \sigma_i^x, &
H_z &=  \sum_{i=1}^{L-1} \sigma^z_i \sigma_{i+1}^z + \sum_{i=1}^L h_z \sigma_i^z.
\end{align} 
This can be written in the brickwork circuit form (Fig.~\ref{fig:ruc_struct}) with the gate
\begin{equation}
\label{eq:uz_uz_uz}
u = \tisingu[u_z][u_x][u_x][u_z]
\end{equation}
where:
\begin{align}
u_x &= \exp( - i \tau h_x \sigma^x ), \\
u_z &= \exp( - i \tau [ \sigma^z \otimes  \sigma^z + i  h_z ( \sigma_i^z \otimes \I   + \I \otimes \sigma^z ) / 2 ] ).
\end{align}
We expect that for generic values of the parameters $(\tau, h_x, h_z)$ this gate defines a chaotic model. In special limits, such as $h_z=0$ or $h_x=0$, the model becomes integrable: in those limits we do not expect our algorithm to succeed.

Interestingly, this model also has a ``self-dual'' line in parameter space \cite{akila_particle-time_2016-2,bertini_exact_2018,bertini_entanglement_2018,gopalakrishnan_unitary_2019,bertini2019exact,piroli_exact_2019}:
\begin{align}
\tau & = \f{\pi}{4}, & h_x & = 1, & h_z & \text{ generic},
\end{align}
where some quantities can be computed exactly. In particular, entanglement growth is maximal on this line.
This is related to the fact that for this choice of parameters, the tensor $u$  remains unitary if it is rotated so as to exchange the roles of space and time  \cite{gopalakrishnan_unitary_2019,bertini2019exact,piroli_exact_2019}. This has been referred to as ``dual unitarity''.
Though this property is highly fine-tuned, the model is believed to remain chaotic for generic values of $h_z$ on this line \cite{bertini_exact_2018,piroli_exact_2019}.

The fact that $v_E=1$ for the dual-unitary models implies $v_B=1$ and that the line tension is flat as a function of velocity \cite{jonay_coarse-grained_2018}:
\begin{equation}
\mathcal{E}(v)  = 1.
\end{equation}
This is an interesting test case for our algorithm. 
First, the above form is very far from our perturbative starting point. Second, the flatness of $\mathcal{E}(v)$ implies that for dual-unitary models the large-scale properties of the domain wall are rather different from the generic case, with negative signs in $W$  playing an important role. We describe this in Sec.~\ref{sec:domainwalldualunitary}.

To begin with we check the algorithm for a presumably generic set of parameters within the kicked Ising class in Fig.~\ref{fig:huse_ev}:
\begin{align}\label{eq:isinggeneric}
\tau & = 1, &
h_x & = 0.9045, & 
h_z & = 0.8090.
\end{align}
The algorithm appears to be working, and the curves converging to one that represents a gate more entangling than the random average. 

Kim and Huse studied similar values of the fields,\footnote{More precisely $h_x=(\sqrt{5} + 5 )/8\simeq 0.9045$ and $h_x=(\sqrt{5} + 1)/4\simeq 0.8090$. Here we have used the truncated decimal values.} but with the smaller period $\tau=0.8$ \cite{kim_testing_2014}. However, the Kim--Huse values  $(\tau,h_x)\simeq (0.8,0.90)$ happen to be quite close to the dual-unitary values\footnote{Writing the gate with $(\tau, h_x, h_z) = (0.8, 0.9045, 0.8090)$ in the representation in Eq.~\ref{eq:sufour} gives ${\mathbf J} \simeq (0.8326, 0.8617,0.0001)$. One branch of dual-unitary values is ${\mathbf J} = (\pi/4, \pi/4, J_z)$ with any $J_z$ and any ${\mathbf h^{(i)}}$.}, which are $(\tau,h_x)\simeq (0.785,1)$! This explains the strong entanglement growth in the Kim-Huse model. Our numerical results for this case (not shown) give a line tension function with a $v_B$ close to 1 and a larger $\mathcal{E}(v)$ than for $\tau=1$.

We now check our algorithm for the  dual-unitary case, taking\footnote{In terms of the general parameterization in Eq.~\ref{eq:sufour} \cite{bertini_exact_2018,bertini_entanglement_2018,gopalakrishnan_unitary_2019,bertini2019exact,piroli_exact_2019},
${\mathbf{J}} = (0,\pi/4,\pi/4)$, $u_{1}= $ $ u_{2} =$ $\exp( - i h_z \sigma_z / 2 )$ $ \exp( - i \pi \sigma_x /4 )$, and $u_{3} = u_{4} = \exp( - i h_z \sigma_z / 2 )$.} 
\begin{align}\label{eq:dualunitaryspecific}
\tau & = \f{\pi}{4}, & h_x & = 1, & h_z & = 0.6.
\end{align}
The value of $h_z$ is not important for the dual-unitarity condition, but we should have $\tau h_z \ne 0 \text{ mod } 2\pi$ to avoid the model being free. Empirically we only see very small differences for the line tension function among different choices of $h_z$, including the special point $h_z=1$ where the dynamics is Clifford as well as being dual-unitary.

Results are shown in  Fig.~\ref{fig:prosen_gate}.
The curves seem to converge to the expected flat line $\mathcal{E}(v) = 1$.
The touching points where $\mathcal{E}(v_B) = |v_B|$ seem to converge to the expected value $v_B = 1$ very fast.

Therefore the domain wall structure appears to make sense even for dual-unitary gates, which represent a limiting case. 
However in this case negative signs in $W$ are important: without these it is impossible to have a flat $\mathcal{E}(v)$, see Sec.~\ref{sec:domainwalldualunitary}.

\subsection{Structure of $W$ matrix}
\label{sec:Wstructure}

\begin{figure}[t]
\centering
\subfigure[]{
  \label{fig:sym_gate_WZ}	
  \includegraphics[width=0.9\linewidth]{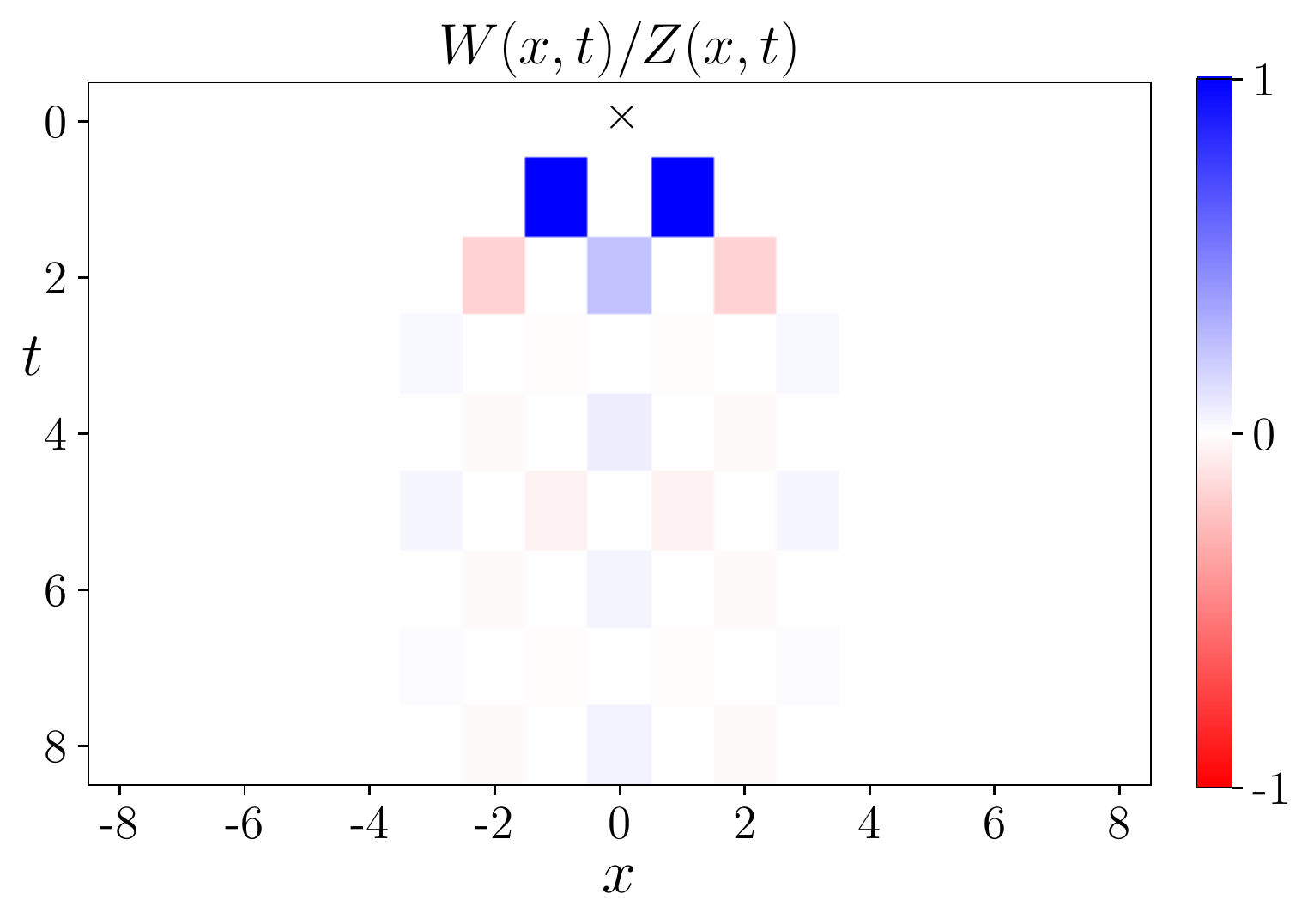}
}
\subfigure[]{
  \label{fig:prosen_WZ}	
  \includegraphics[width=0.9\linewidth]{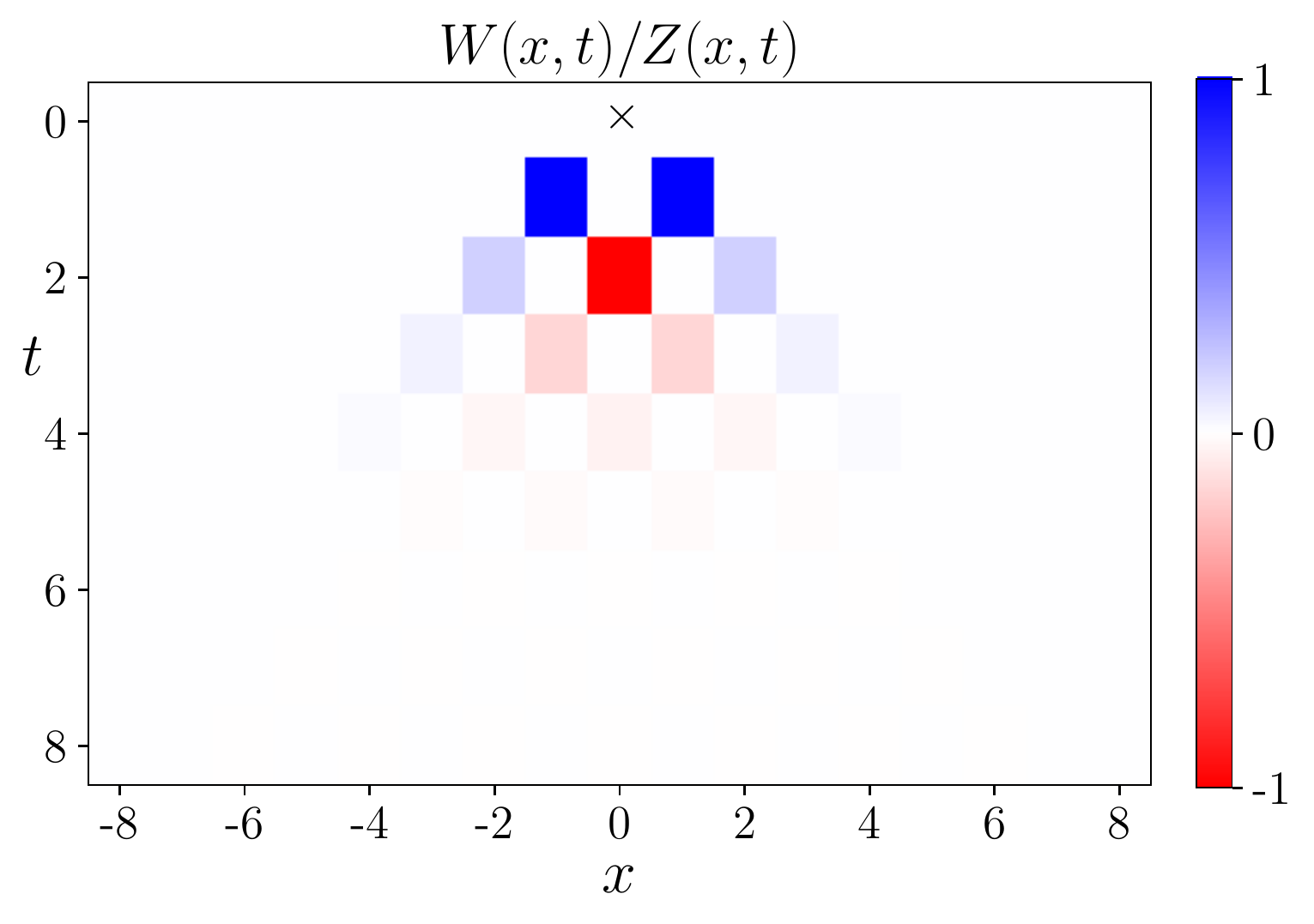}
}
\caption{The space-time structure of $W(x,t)/Z(x,t)$ for (a) the generic reflection-symmetric gate in Eq.~\ref{eq:hequal} [Fig.~\ref{fig:sym_gate}]
and (b) the maximally-entangling dual-unitary gate specified in 
Eq.~\ref{eq:dualunitaryspecific} [Fig.~\ref{fig:prosen_gate}]. The cross marks the coordinate $(0,0)$. In our construction, only steps with even $x + t $ are possible.
}

\label{fig:WZ}
\end{figure}

We now examine the temporal decay of the step weights $W$ defined in  Sec.~\ref{sec:domainwallstructure}. 
In order to examine the importance of long steps, we consider the normalized quantity $W(x,t)/Z(x,t)$: this compares the weight of a single long step, with displacement $x$ and duration $t$, with the total weight of all paths with the same two endpoints. This ratio should decay with $t$ if asymptotically long steps can be neglected.
(Individually, both $W(x,t)$ and $Z(x,t)$ decay exponentially with $t$.)
A slight extension of the reasoning in Sec.~\ref{sec:expsup} shows that in a typical realization of a \textit{random} circuit at large $q$, the typical value of this ratio (for a given starting point of the domain wall in spacetime) decays like $q^{-t}$.

Fig.~\ref{fig:WZ} shows heat maps of $W(x,t)/Z(x,t)$ for examples of translation-invariant Floquet models for spin-1/2s.
Panel~(a) is for the generic reflection-symmetric gate in Eq.~\ref{eq:hequal}, whose line tension function was shown in Fig.~\ref{fig:sym_gate}, and Panel~(b) shows the dual-unitary  gate of 
Eq.~\ref{eq:dualunitaryspecific}  and Fig.~\ref{fig:prosen_gate}.

In both cases the weight is small at large $t$, though finite-time effects in this quantity seem to be large for the former gate at displacement $x=0$.
We also find decay of $W/Z$ for the generic asymmetric gate discussed in Sec.~\ref{subsec:sym_rand} and for the kicked Ising model with the generic parameters in Eq.~\ref{eq:isinggeneric} (data not shown). 

The domain wall picture appears to be well-defined both for the generic models and for the dual-unitary model. However there are key differences between the two cases which we discuss in Sec.~\ref{sec:domainwalldualunitary}. 
Note that the weight at $x=0$, $t=2$ is positive for the gate in Panel~(a) and negative for the dual-unitary gate in Panel~(b). In the former case this extra positive step weight increases $Z(0,t)$ compared to $Z_{0\perp}(0,t)$, leading to decreased $v_E^{(2)}$. In the latter case the negative step weight decreases $Z(0,t)$ compared to $Z_{0\perp}(0,t)$, helping the dual-unitary gate to attain the fastest possible decay of $Z(0,t)$, i.e. the most rapid possible entropy growth.

Let us also show an example where local gate is very weakly entangling, resulting in slower convergence of the algorithm. This is the reflection-symmetric gate
defined in Eq.~\ref{eq:symmetricgate}, but with the smaller interaction constant ${x = 0.5}$  (in Sec.~\ref{subsec:sym_rand} we showed results for $x=0.8$).

\begin{figure}[t]
\centering
  \includegraphics[width=0.9\linewidth]{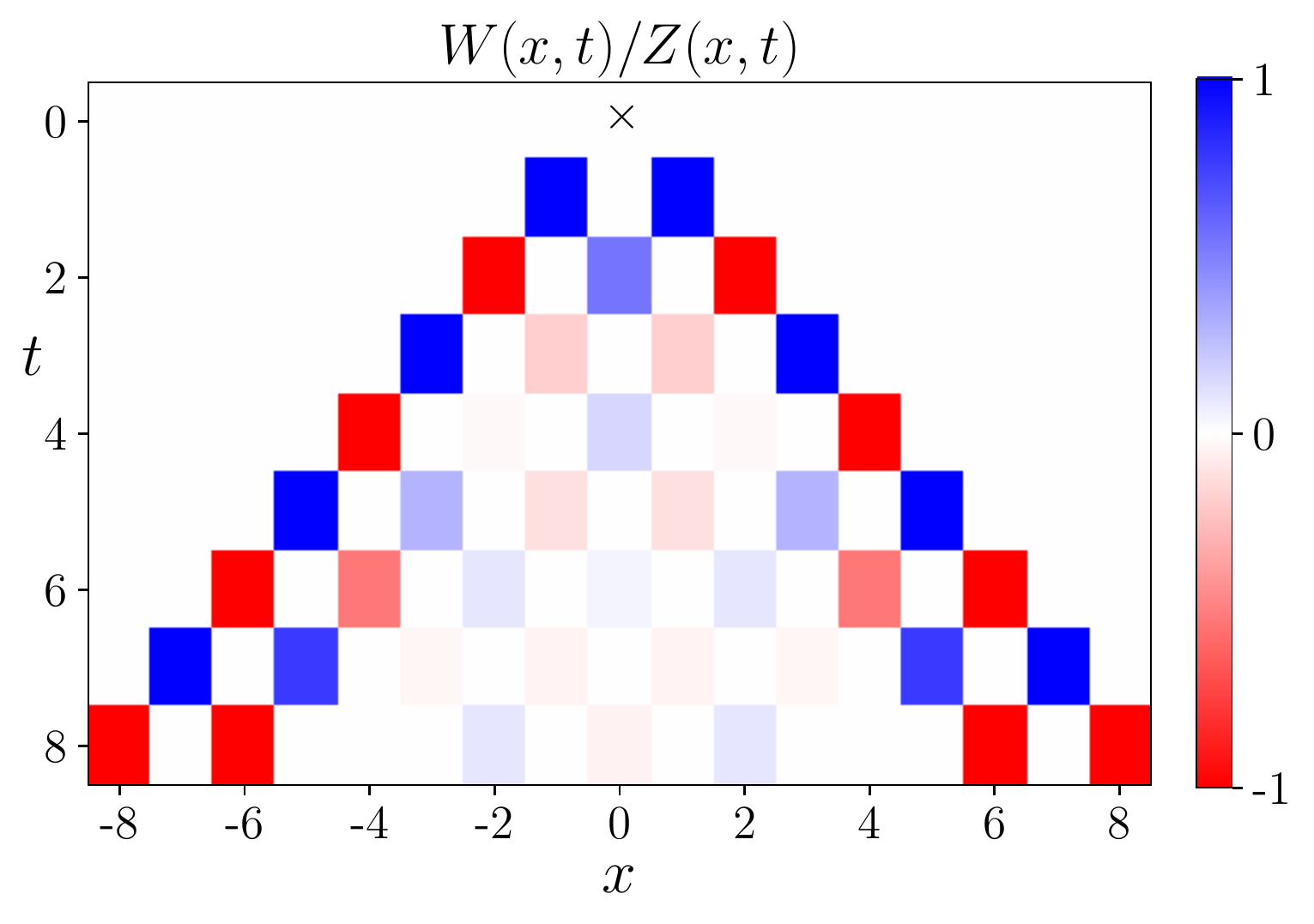}
\caption{Weakly entangling gate with ${x=0.5}$ (Eq.~\ref{eq:symmetricgate}): Heatmap showing relative importance of long steps. }
\label{fig:sym_fail_1}
\end{figure}

\begin{figure}[t]
\centering
\subfigure[]{
  \label{fig:sym_gamma}	
    \includegraphics[width=0.9\columnwidth]{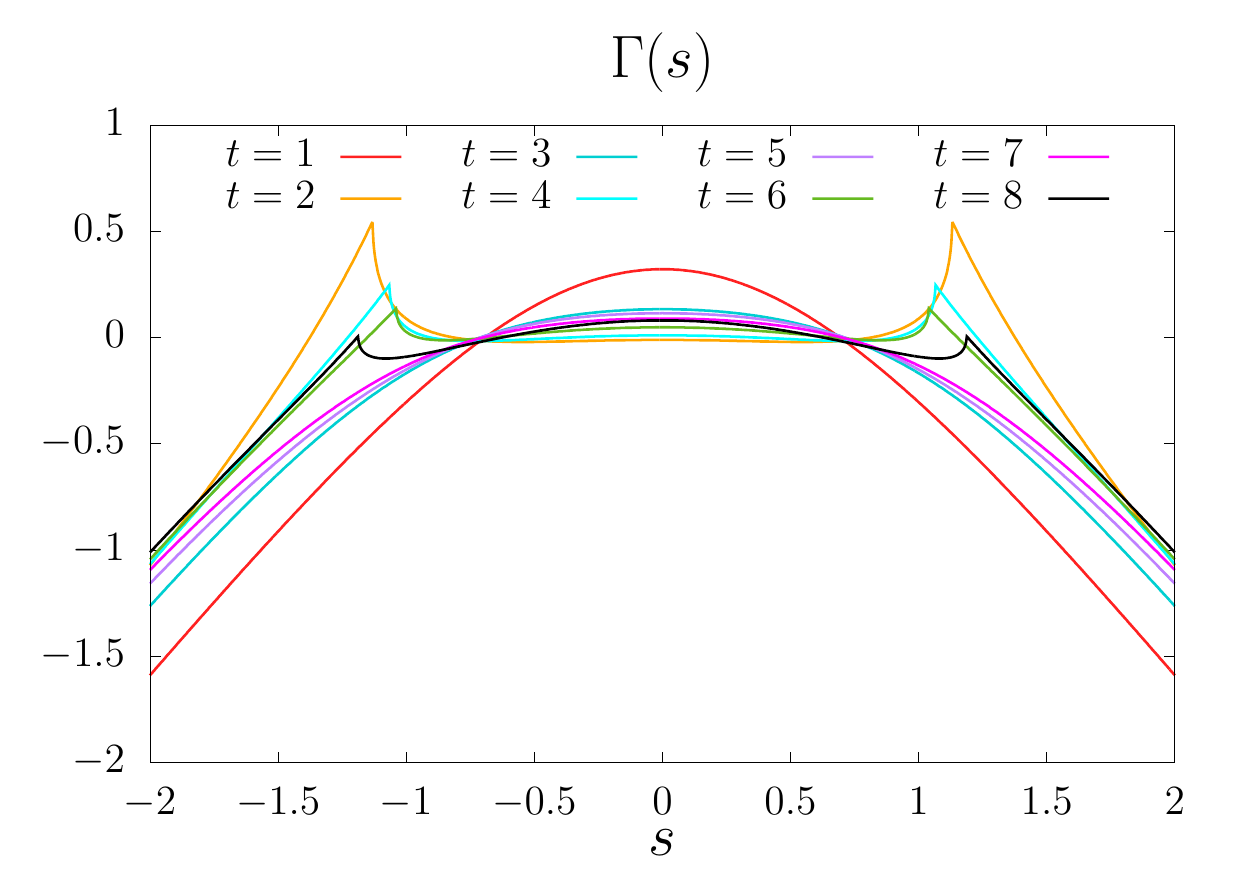}
}
\subfigure[]{
  \label{fig:sym_gate_fail}	
  \includegraphics[width=0.9\linewidth]{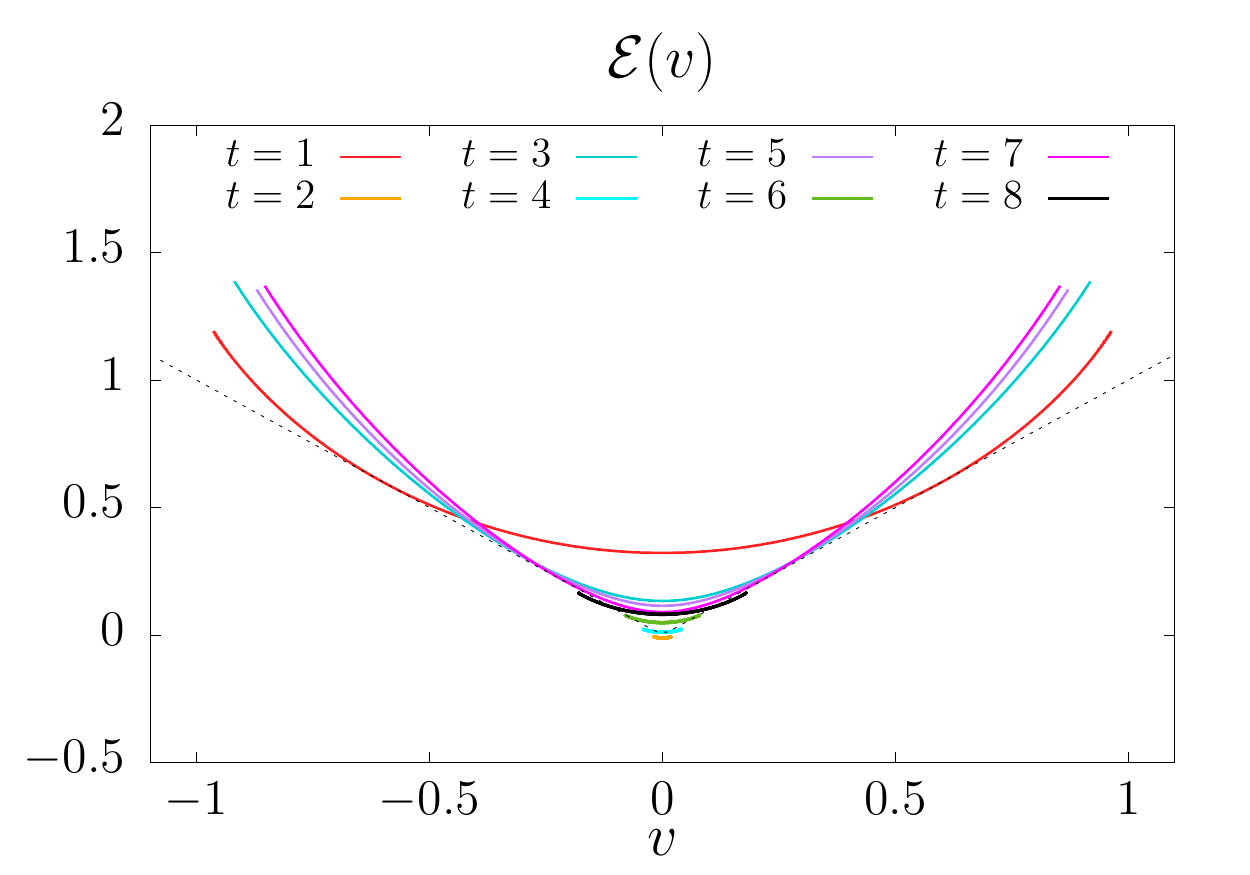}
}
\caption{Weakly entangling gate with ${x=0.5}$ (Eq.~\ref{eq:symmetricgate}): (a) Successive approximations to 
$\Gamma(s)$ (Sec.~\ref{sec:generatingfunctions}). For some small  $t$ values $\Gamma(s)$ is negative around $s = 0$. (b) Approximations to $\mathcal{E}(v)$ obtained by Legendre transform of (a). Note the small $\mathcal{E}(0)=\Gamma(0)$.}
\label{fig:sym_fail_2}
\end{figure}

First we show the ratio $W/Z$ in Fig.~\ref{fig:sym_fail_1}. We see that there are large values close to the edges of the lightcone. However, the apparent decay inside the lightcone suggests that the results may still converge, just more slowly than in the cases examined above.

The slow convergence in this case is related to the fact that the gate is weakly entangling. First we examine the function $\Gamma(s)$ defined in Sec.~\ref{sec:generatingfunctions} (whose physical meaning, for $|s|<\ln 2$, is as an entanglement entropy growth rate in a state with a nonzero gradient in the entanglement $S(x)$ across a cut at position $x$, with $\partial S/\partial x = s$).
The approximations to $\Gamma[s]$ for $t = 2, 4, 6, 8$ are shown in Fig.~\ref{fig:sym_gamma}. Note that they are  no longer concave functions over the whole range $s$. However, this effect gets less severe with increasing $t$. If we restrict to the concave region in performing the Legendre transform, we get a reasonable result for $\mathcal{E}(v)$: see  Fig.~\ref{fig:sym_gate_fail}. This is consistent with slow convergence to a function that satisfies the general constraints, with a very small entanglement growth rate.

For a quantitative analysis of convergence rates in the various models, See App.~\ref{subsec:W_decay}.


\section{Coarse-graining: generic models versus dual-unitary circuits}
\label{sec:domainwalldualunitary}

The thin domain wall conjecture implies that the entanglement membrane is well-defined once we coarse-grain sufficiently in space and time. 
The relevant lengthscale, the width of the domain wall, is microscopic, in the sense that it remains finite as the system size and the total time $t$ of the evolution go to infinity.
In the models investigated here, which do not contain any small parameter, the width of the domain wall is an order 1 multiple of the lattice spacing. In cases where the dynamics is tuned close to a special point, a large lengthscale may emerge.

In this section we discuss this coarse-graining in slightly more detail, in order to make a distinction between two universality classes.
 
Microscopically, the step weights $W(x,t)$ can be of either sign. 
But we conjecture that, for generic models, the minus signs do not alter the universality class from that of a classical directed path.
Heuristically, we can think of the step weights becoming positive after coarse-graining, so that the membrane is effectively a classical directed path with diffusive wandering (at least for $v<v_B$). 
We will restate this in terms of the recursion relation for $Z(x,t)$ below.

\begin{figure}[t]
\centering
\subfigure[]{
  \label{fig:huse_Z_tilde}	
  \includegraphics[width=0.9\columnwidth]{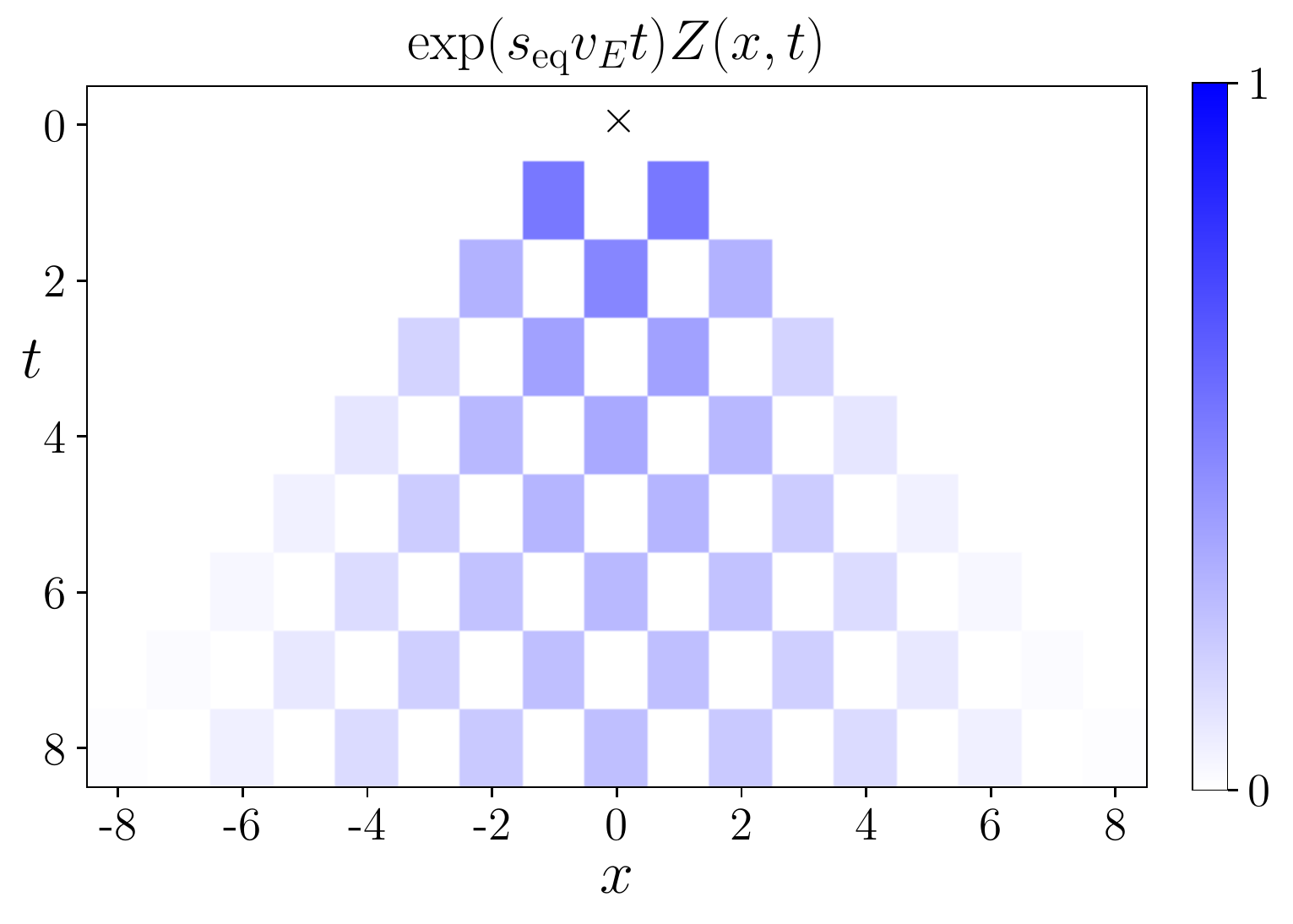}
}
\subfigure[]{
  \label{fig:prosen_Z_tilde}	
  \includegraphics[width=0.9\columnwidth]{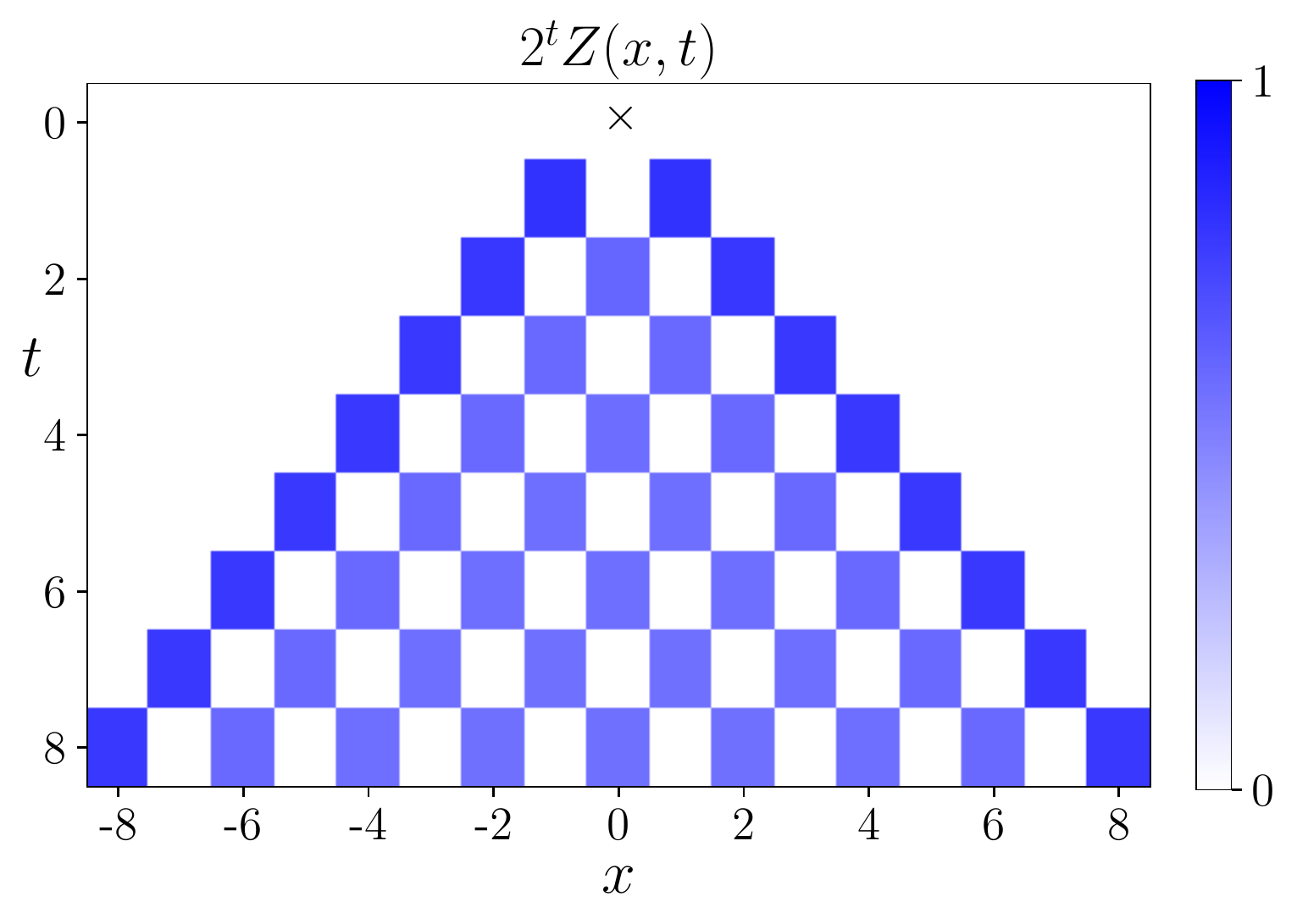}
}
\caption{Heat map of $\widetilde{Z}(x, t) = \exp( s_{\rm eq} v_E t) Z(x, t)$. (a) Floquet Ising model as in Fig.~\ref{fig:huse_ev}. $\widetilde{Z}(x, t)$ is consistent with the solution of a parabolic equation. For example, the decay at ${x=0}$ is compatible with the  $t^{-1/2}$ scaling in Eq.~\ref{eq:Zdiffusion}: this is demonstrated in App.~\ref{subsec:diffusivescaling}. (b) Dual-unitary model as in Fig.~\ref{fig:prosen_gate}. Unlike the previous model, $\widetilde{Z}(x, t)$ is approximately constant in the interior of the light cone and zero outside.}
\label{fig:Z_tilde}
\end{figure}

This coarse-grained picture implies a strictly convex $\mathcal{E}(v)$, with a positive second derivative that is related to the diffusivity of the path. 
[This diffusivity can depend on the coarse-grained velocity $v$ that we condition on.] 
This implies for example that  the scaling of $Z(x,t)$ is:
\begin{align}\label{eq:tonehalf}
Z(x,t) & \sim 
t^{-1/2} \, e^{- s_\text{eq} \mathcal{E}(v) t}.
\end{align}
The power-law prefactor is universal and comes from the fact that  both endpoints of the random path are fixed. For example, consider the case where $x\ll t$, so that the coarse-grained velocity is close to zero, and let the model be parity-symmetric.
Expanding in ${v=x/t}$, the above is then
\begin{align}\label{eq:Zdiffusion}
\widetilde Z(x,t) & \sim 
t^{-1/2} \, \exp\lf -\f{s_\text{eq} \mathcal{E}''(0)  x^2}{2t} \ri 
&
(x&\ll t),
\end{align}
where we have factored out the extensive free energy of the path by defining (recall ${v_E = \mathcal{E}(0)}$)
\begin{equation}\label{eq:Ztildedef}
Z(x,t)= e^{-s_\text{eq} v_E t}  \widetilde Z(x,t).
\end{equation}
Up to a normalization constant, Eq.~\ref{eq:Zdiffusion} is the probability for a random walker with diffusion coefficient
\begin{equation}\label{eq:diffusivity}
D=\f{1}{2\seq \mathcal{E}''(0)}
\end{equation}
to be at position $x$ at time $t$.
In Fig.~\ref{fig:huse_Z_tilde} we show $\widetilde Z(x,t)$ for a generic kicked Ising gate: these numerical results are in good agreement with the $t^{-1/2}$ scaling in Eq.~\ref{eq:Zdiffusion} (App.~\ref{subsec:diffusivescaling}).

The dual-unitary model behaves differently. The membrane and its line tension function still appear to make sense, but its wandering is not diffusive. For example, we can contrast the scaling of $Z(x,t)$ with the formula above. The numerical results in Fig.~\ref{sec:Zdualunitary} are consistent with the large $t$ scaling 
\begin{align}
Z(x,t) & \sim e^{- s_\text{eq} t} = 2^{-t},
\end{align}
with a trivial $x$-dependence inside the lightcone (we neglect the even-odd effect at the lattice scale). This is also consistent with the analytical calculation in Sec.~\ref{sec:Zdualunitary} below.

Evidently, in the dual-unitary case, the membrane is not a classical random walk. Indeed, since for the dual-unitary model $\mathcal{E}(v)$ is constant (inside the lightcone) and $\mathcal{E}''(v)$ vanishes, the diffusivity in Eq.~\ref{eq:diffusivity} must diverge in the limit that a model becomes dual unitary.

The distinction between the two cases
may be clearer if we think of the recursion relation as a discrete analogue of a linear partial differential equation.
In the limit of large $t$, we may write Eq.~\ref{eq:Wbyrecursion}  as
\begin{equation}
\label{eq:discretedifference}
Z(x,t) = \sum_{\Delta t=1}^{\infty}\sum_{\Delta x} W(\Delta x, \Delta t) Z(x-\Delta x, t-\Delta t).
\end{equation}
We would like to relate this to a continuum equation in the limit of large times.
We cannot immediately perform a derivative expansion of $Z$: since it is exponentially decaying, higher derivatives are not parametrically smaller than lower ones. 
For simplicity, consider first the regime $x\ll t$ discussed above (close to the $v=0$ ray) and assume reflection symmetry (so $\mathcal{E}'(0)=0$). 
Then $\widetilde Z$ in Eq.~\ref{eq:Ztildedef} has a sensible derivative expansion. 
Let us define the following ``average'' for an arbitrary function $F$,
\be \label{eq:centralrayaverage}
\<F(y,\tau)  \>_0 = 
\sum_{y,\tau} W(y,\tau) e^{ 
 s_\text{eq} v_E \tau
}  F(y,\tau).
\end{equation}
(We have replaced $\Delta x$, $\Delta t$ with $y$, $\tau$ to avoid clutter. The subscript on the average specifies the spacetime ray whose vicinity we are considering.) This satisfies ${\<1\>_0=1}$, by the relations in Sec.~\ref{sec:generatingfunctions}.
The derivative expansion of Eq.~\ref{eq:discretedifference} yields:
\begin{equation}\label{eq:centralrayexpansion1}
\langle \tau \rangle_0 \, \partial_t \widetilde Z = 
\f{1}{2} \langle  y^2\rangle_0 \, \partial_x^2 \widetilde Z,
\end{equation}
plus less-relevant terms. The thin domain wall assumption implies that the two coefficients shown are finite. 
While the weights $W$ defining the averages are not necessarily positive, the averages shown will be nonzero in the absence of fine-tuning.
The higher terms can then be dropped, leading to the diffusive (random walk) scaling discussed above. We can also check using the relations between generating functions that
\begin{equation}\label{eq:tauysquared}
 \<\tau\>_0 = \seq \mathcal{E}''(0) \< y^2\>_0,
\end{equation}
so that the diffusion constant in Eq.~\ref{eq:centralrayexpansion1} agrees with that in Eq.~\ref{eq:tonehalf}.
We emphasize that the above equation is valid only close to the $v=0$ ray: see below for a more general equation. 
The information contained in  Eq.~\ref{eq:centralrayexpansion1} is already contained in $\mathcal{E}(v)$, and the spacetime picture is usually more convenient.

The universality class changes if $\<\tau\>_0$ vanishes due to fine tuning.
It is clear from the definition in Eq.~\ref{eq:centralrayaverage} that this requires negative step weights (which are absent in the large $q$ limit described in Sec.~\ref{sec:tractable_limit}).
When this vanishing occurs, we must take into account the next time derivative term, which was previously irrelevant:
\begin{equation}\label{eq:centralrayexpansion2}
\langle \tau^2 \rangle_0 \, \partial_t^2 \widetilde Z 
=  \langle  y^2\rangle_0 \, \partial_x^2 \widetilde Z.
\end{equation}
This is what occurs in the dual unitary model above. We see that the equation becomes a wave equation, consistent with the constancy of $\widetilde Z$ inside the lightcone (Sec.~\ref{sec:Zdualunitary}). We checked that $\< \tau \>_0$ is consistent with zero in the dual unitary circuit: the finite $t$ approximation to $\< \tau \>_0$ from the truncated $W(x,t)$ matrix decays exponentially in $t$, with a characteristic time $\sim 1.8$ steps.

To simplify further, let us consider a toy model, which only takes a single additional nonzero $W$ element (beyond the ones present in $Z_{0\perp}$). We parameterize its elements in terms of a constant $\mu$
\begin{align}
W(\pm 1, 1) & = K,
& 
W(0, 2) & = - \mu K^2.
\end{align}

When $\mu<1$ we are in the diffusive class. However when $\mu=1$ (in fact the dual unitary model is numerically quite close to this simplified model), removing the factors of $K$ by defining  ${Z(x,t) = K^t \widetilde Z(x,t)}$
gives
\begin{equation}
\widetilde Z(x,t) = \widetilde Z(x+1,t-1)+ \widetilde Z(x-1,t-1)- \widetilde Z(x,t-2),
\end{equation}
which is a discrete version of the wave equation, ${(\partial_t^2 - \partial_x^2) \,\widetilde Z = 0}$. This gives a $\widetilde Z(x,t)$ which is constant inside the lightcone and zero outside.

Numerically, we find that the structure in the specific dual unitary kicked Ising model that we studied above is similar:  $2^t Z(x,t)$ is approximately constant within the lightcone, and zero outside. Fig.~\ref{fig:Z_tilde} contrasts the spacetime pattern of $e^{s_\text{eq} v_E t} Z(x,t)$ for the generic kicked-Ising model and the dual-unitary kicked Ising model.

In the derivations of the continuum equations (Eqs.~\ref{eq:centralrayexpansion1},~\ref{eq:centralrayexpansion2}), we restricted to the vicinity of a particular ray.
For a more general equation, we may write ${Z(x,t) = \exp( -S(x,t))}$, and expand in derivatives of $S(x,t)$.
The resulting dynamical equation \cite{jonay_coarse-grained_2018} also applies to the second Renyi entropy $S(x,t)$ of a time-evolving quantum state in a 1D system, for an entanglement cut at $x$.\footnote{This corresponds to a different lower boundary condition, where the domain wall endpoint at $t=0$ is free and weighted with $Z(x,0)=e^{-S(x,0)}$. The recursion relation is the same.} As described in Eq.~\eqref{app:eq_Z_tilde}, this gives the equation 
\begin{align}
\f{\dd S}{\dd t} = 
\seq \lf
\Gamma\lf \f{\dd S}{\dd x} \ri  - \f{1}{2} \Gamma'' \lf \f{\dd S}{\dd x} \ri  \f{\dd^2 S}{\dd x^2}
+ \ldots
\ri
\end{align}
where $\Gamma$ is the Legendre transform of $\mathcal{E}$ (Sec.~\ref{sec:generatingfunctions}).
The first term is the leading-order dynamics that captures the $\mathcal{O}(t)$ term in the entanglement. 
Here $\dd S/\dd x$ and $\dd S/\dd t$ are generally of order 1 at late times, so we keep all powers of $\dd S/\dd x$, but higher derivatives are small and we can expand in them. The second term gives subleading corrections \cite{jonay_coarse-grained_2018}. The random walk picture implies that the two coefficients are related.
The above equation can also be written in terms of $\mathcal{E}(v)$, giving a form consistent with Eqs.~\ref{eq:tauysquared},~\ref{eq:centralrayexpansion1} (App.~\ref{app:eq_Z_tilde}).

\subsection{Partition function for dual unitary models}
\label{sec:Zdualunitary}

Let us confirm by an explicit calculation that dual-unitarity implies
the abovementioned approximate  constancy of $Z(x,t)$ within the lightcone [which is related to the wavelike structure of the recursion equation for $\widetilde Z$, and which implies constancy of $\mathcal{E}(v)$ as a function of $v$].

\begin{figure}[b]
\centering
\subfigure[]{
  \label{fig:dual_unitary_a}	
  \includegraphics[width=0.46\columnwidth]{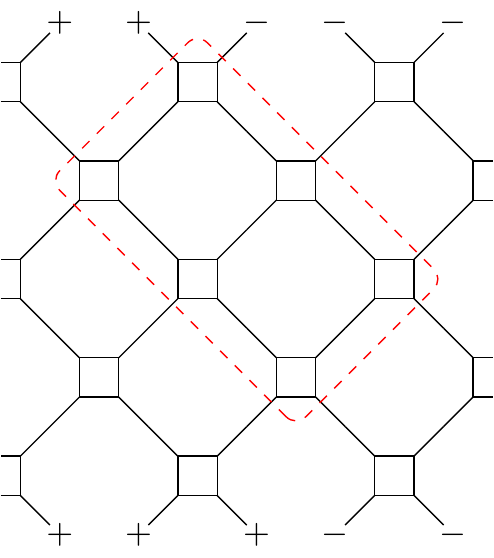}
}
\subfigure[]{
  \label{fig:dual_unitary_b}	
  \includegraphics[width=0.46\columnwidth]{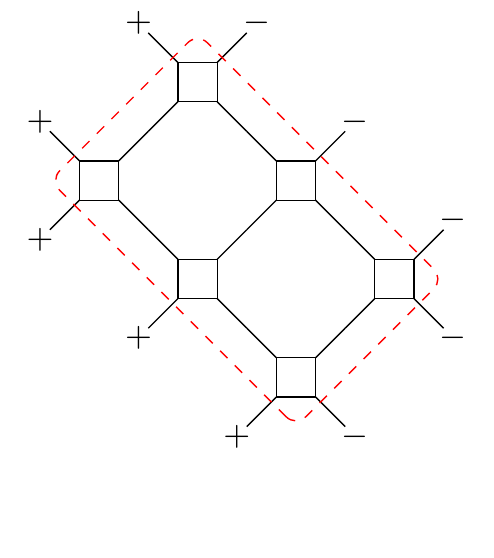}
}
\caption{ The partition function $Z_{\rm op}(x,t)$ for a dual-unitary circuit. (a) $ q^{2L} Z_{\rm op}(x = 1, t = 5)$. By unitarity (the first identity in Eq.~\ref{eq:uni_duluni}) we can remove the unitaries outside the lightcone, giving $q^{2t} Z_{\rm op }(x = 1, t = 5)$ as expression (b). Then Eq.~\ref{eq:uni_duluni} allows the remaining gates to be removed.
}
\label{fig:dual_unitary}
\end{figure}

This fact can be seen by considering a slightly different partition function,
in which we place states $\ket{\pm}$ at the lower boundary, rather than the dual states $\ket{(\pm\pm)^*}$ used in Eq.~\eqref{eq:Z_x_y_t}:
\begin{equation}
Z_{\rm op}( x, y, t ) = \langle \cdots +  - \cdots | U^{(2)}(t)   | \cdots  + - \cdots \rangle  / q^{2L}.
\end{equation}
Here $x$ and $y$ label the positions of the domain walls  between $+$ and $-$ on the top and bottom boundaries respectively.  Fig.~\ref{fig:dual_unitary_a} 
shows an example of $q^{2L} Z_{\rm op}(x,t)$.
In contrast to the domain wall partition function $Z(x,t)$ defined in Eq.~\ref{eq:Z_x_y_t}, $Z_{\rm op}(x,y,t)$ does not force the end point of the domain wall to lie precisely at $y$, but instead penalizes configurations whose end point deviates from $y$ (with a negative power of $q$ in the weight). 
However, in analogy to the random circuit case, we expect that this difference only leads to an $\mathcal{O}(1)$ boundary term in the free energy (so long as $|x-y|\leq v_B t$, which in the dual-unitary circuit is a trivial restriction since $v_B=1$), so that
\begin{equation}
\begin{aligned}
\ln Z_{\rm op} (x, y, t ) &
\sim  \ln Z(x, y, t )  
\sim  - \mathcal{E}(v) \, t \, \ln q.
\end{aligned}
\end{equation}

Physically, the above partition function $Z_\text{op}$ simply computes the operator entanglement of the time evolution operator itself, for a certain partition of the ``legs'' of the time evolution operator.
This operator entanglement is maximal in the dual unitary circuits, like various other measures of entanglement in these systems \cite{bertini_exact_2018, gopalakrishnan_unitary_2019, bertini2019operator,piroli_exact_2019}. This is easily seen by manipulations similar to those in \cite{gopalakrishnan_unitary_2019, bertini2019exact,piroli_exact_2019}.
Specifically (setting $y=0$) we show that  ${Z_{\rm op}(x,t) = q^{-t}}$ 
for a dual-unitary circuit, so long as $x$ is within the light cone of the origin.

We represent the multi-layer two-site unitary evolution matrix as a four-leg tensor:
\begin{align}
\label{eq:u_u_dual}
u^{(N)}& = 
\fineq[-0.8ex][0.8][0.8]{
\dualgate[0][0][][A][B][C][D]
},
&
u_{\rm dual}^{(N)}
&= 
\fineq[-0.8ex][0.8][0.8]{
\dualgate[0][0][][B][C][A][D]
}.
\end{align}
We have labelled the external legs of this tensor by $A, B, C, D$.
As a unitary, the legs $A,B$ correspond to the row index and $C,D$ to the column index.
We rotate the tensor by $90^{\circ}$ to obtain the dual tensor. Dual-unitarity requires the dual tensor to be unitary (with $B,C$ as row legs and $A,D$ as column legs) as well. 
To contrast with the generic gate in Eq.~\ref{eq:u_tensor}, we have drawn the dual-unitary tensor as a (fourfold symmetric) square in Eq.~\ref{eq:u_u_dual}. For $N=2$,  unitarity and dual-unitarity can be used to propagate $++$ (or $--$) states from the top to bottom, right to left, and vice versa. Graphically, we have (the same holds for arbitrary $N$ with $+$ replaced by any $\sigma$)
\begin{align}
\label{eq:uni_duluni}
\fineq[-0.8ex][0.8][0.8]{\dualgate[0][0][][+][+][][]}& = 
\fineq[0ex][0.8][0.8]{
  \node () at (-0.6,1.25) {$+$};
  \node () at (0.6,1.25) {$+$};
  \draw (-0.6,1)--++(0,-0.4);
  \draw (0.6,1)--++(0,-0.4);
},
&
\fineq[-0.8ex][0.8][0.8]{\dualgate[0][0][][][+][][+]} &= 
\fineq[-0.8ex][0.8][0.8]{
  \node () at (-0.6,0.6) {$+$};
  \node () at (-0.6,-0.6) {$+$};
  \draw (-0.85,0.6)--++(-0.4, 0);
  \draw (-0.85,-0.6)--++(-0.4, 0);
  \node () at (-0.85,0) {};
}
\end{align}
The first of these relations (unitarity) 
allows us to remove the unitary gates outside the lightcones in Fig.~\ref{fig:dual_unitary_a}.
Keeping track of the $q$ factor,   $q^{2t} Z_{\rm op}(x, t)$ equals the expression in  Fig.~\ref{fig:dual_unitary_b}.
Next,  dual-unitarity (the second relation) can be used to  propagate the  $+$ states to the right, so that all the unitary gates are removed and  Fig.~\ref{fig:dual_unitary_b} is reduced to ${\braket{+}{-}^t=q^t}$.
Hence $q^{2t} Z_{\rm op}(x, t) = q^t$, so that $Z_{\rm op}(x, t) = e^{- s_{\rm eq} t}$ within the lightcone.


\section{Operator spreading}
\label{sec:opspread}

We have focused on the second R\'enyi entropy as a test case, but the present formalism can be applied to almost any observable expressible in terms of $U^{(N)}$. 
One of the simplest is the out-of-time order correlation function, which requires $N=2$, and which on large scales we expect to reduce to a   stochastic growth process
\cite{nahum_operator_2017, von_keyserlingk_operator_2017}.

\begin{figure}[t]
\centering
\includegraphics[width=\columnwidth]{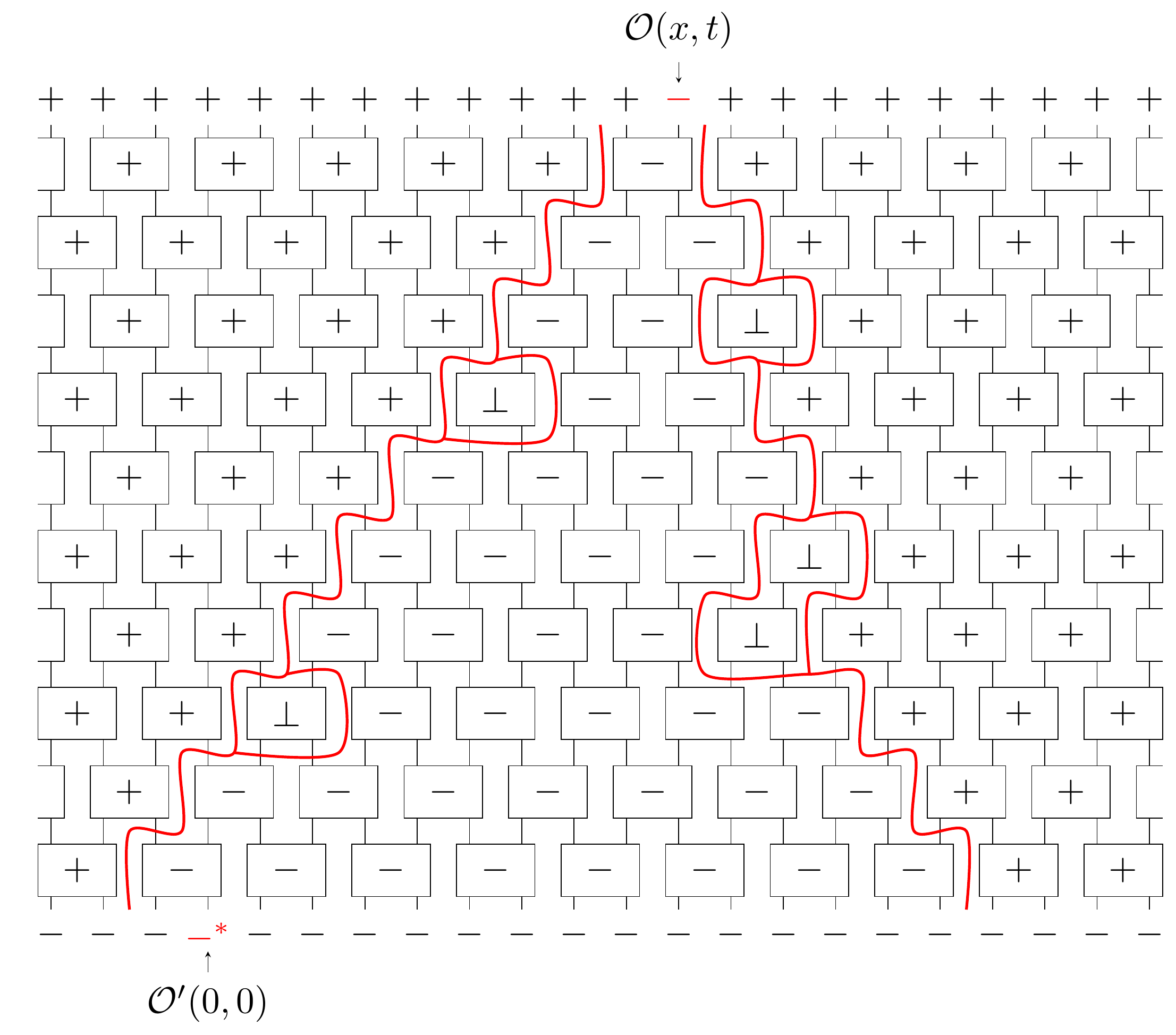}
\caption{The membrane picture for the out-of-time ordered commutator $\tilde{C}(x,t)$. The top boundary condition creates two domain walls. The bottom boundary condition favors $-$,  providing an outward force on the lower endpoints of the two domain walls. The right domain wall thus moves with the optimal speed $v_B$. Here, the left wall is forced by the dual state $|-^* \rangle$ at the operator insertion point to travel with $v > v_B$.}
\label{fig:otoc}
\end{figure}

The OTOC can be written in terms of the same partition sum as for $e^{-S_2}$, but with different boundary conditions. 
In the Haar-averaged circuit this reduces to a simple partition function for a pair of domain walls. Since we have now made sense of the domain wall in a generic, fixed circuit, this picture from the random circuit carries over. The modification is that the domain walls now have a nontrivial internal structure, including finite regions of $\perp$, with an associated microscopic timescale $t_0$. A typical configuration from the partition sum is illustrated in Fig.~\ref{fig:otoc}.

For a brief heuristic summary, we consider the OTOC in the form of a commutator squared of two traceless operators,
\begin{equation}
\tilde C(x, t ) = - \frac{1}{2} \langle  [\mathcal{O}(x, t), \mathcal{O}'( 0, 0 )]^2 \rangle .
\end{equation}
We take the expectation value at infinite temperature, and to simplify the boundary conditions we average over local unitary rotations of the local operators ${\mathcal{O}\rightarrow V\mathcal{O}V^\dag}$, ${\mathcal{O}'\rightarrow V'\mathcal{O}'V'^\dag}$.
Carrying out manipulations similar to Sec.~VI B of \cite{nahum_operator_2017}, we rewrite $\tilde{C}(x,t)$ using the present notation (here $\tr$ is a single-site trace)
\begin{equation}
\begin{aligned}
  &\tilde{C}(x,t) =  \frac{\tr( \mathcal{O}^2) \tr(\mathcal{O}'^2)}{q^2 - 1} \times \\
  &\,\,\langle \dots ++ - ++ \dots  | U^{(2)} | \dots -- -^* -- \dots  \rangle  / q^{L-1}.
\end{aligned}
\end{equation}
The operator insertion at the top initiates a domain of $-$ in the bulk. Because of the $-$ boundary condition at the bottom, this domain prefers to expand at rate $\pm v_B$ (assuming reflection symmetry).\footnote{The lower boundary condition contributes a weight proportional to ${q^{d}=e^{\seq d}}$, which tries to increase the width $d$ of the $-$ domain at the lower boundary. The boundary weight means that the optimal velocity $v_L$ for the left domain wall, minimizing the free energy, is $v_B$, since by the properties of $\mathcal{E}$ this minimizes ${\seq \left( - v_L + \mathcal{E}(v_L) \right)t}$.}

However, the operator insertion at the initial time cannot lie in the $+$ region exterior to the $-$ domain, because this would lead to a factor of $\langle +| -^* \rangle =0$.
If $|x|< v_B t$ then $\tilde C(x, t )$ is constant at leading order,  ${\tilde C(x, t ) = q^{-2}\tr\mathcal{O}^2\tr\mathcal{O}'^2}$.
But if $|x|>v_B$, the lower operator insertion ``stretches'' one of the domain walls, forcing it to move at a speed greater than $v_B$. 
This is illustrated for the  left domain wall in Fig.~\ref{fig:otoc}. The additional free energy cost means that, outside the lightcone, for ${x/t>v_B}$, the OTOC scales as  \cite{khemani_velocity-dependent_2018}
\begin{align}
\tilde C(x,t) & \sim e^{-\mu(v) t}, &
\mu(v) & \equiv \seq \lf \mathcal{E}(v) - v \ri.
\end{align}

In random models these results can also be obtained from a stochastic growth process for the operator cluster \cite{von_keyserlingk_operator_2017, nahum_operator_2017,rowlands2018noisy}.
In this picture, $\mu(v)$ is the large deviation function describing the probability for the boundary of the operator cluster to propagate at speed $v$ \cite{correlatorpaper}.
From the above discussion, we expect that in generic models this stochastic picture applies after coarse-graining to timescales greater than $t_0$.

Forthcoming work using the memory matrix formalism will give a complementary view on the operator spreading problem \cite{mccullochforthcoming}.


\section{Beyond circuit models}
\label{sec:beyond_circuit}

In Sec.~\ref{sec:def_spin_model} we described how to introduce the pairing field (the spin ${s \in S_N \cup \{\perp\}}$ in the effective partition function) in the context of models with a circuit structure. This was a natural starting point, since the idea of inserting permutation states was inspired by the random unitary circuit. However there is no need to restrict to circuits. 

In this section we study generic Floquet systems, i.e. models where the Hamiltonian is periodic in time, but which do not have the form of a circuit.
We are still free to insert resolutions of identity as in Eq.~\eqref{eq:def_Z}:
\begin{equation}
\label{eq:Z_c_floq_proj}
Z = 
\fineq[0][0.4][1]{
  \partitionZc;
} = 
\sum  
\fineq[0][0.4][1]{
  \partitionZtwosite;
}
\end{equation}
In this graphical equation, the shaded region denotes the Floquet operator (evolution operator for one time period), which in general will not have an exact circuit decomposition. 
As in Eq.~\ref{eq:def_Z}, the horizontal bars  represent projectors onto either a permutation state or the $\perp$ subspace, these possibilities being summed over.

For models that are translationally invariant by 1 lattice spacing, an alternative that may appear at first sight more natural is to insert single-site, rather than two-site projectors: 
\begin{equation}
\label{eq:Z_c_floq_one_site_proj}
Z = 
\fineq[0][0.4][1]{
  \partitionZc;
} = 
\sum  
\fineq[0][0.4][1]{
  \partitionZonesite;
}
\end{equation}
Both rewritings of $Z$ are exact. In the following we consider both schemes.

\begin{figure}[h]
\centering
\subfigure[]{
  \label{fig:c_floq_dw_two_site}	
  \includegraphics[width=0.45\columnwidth]{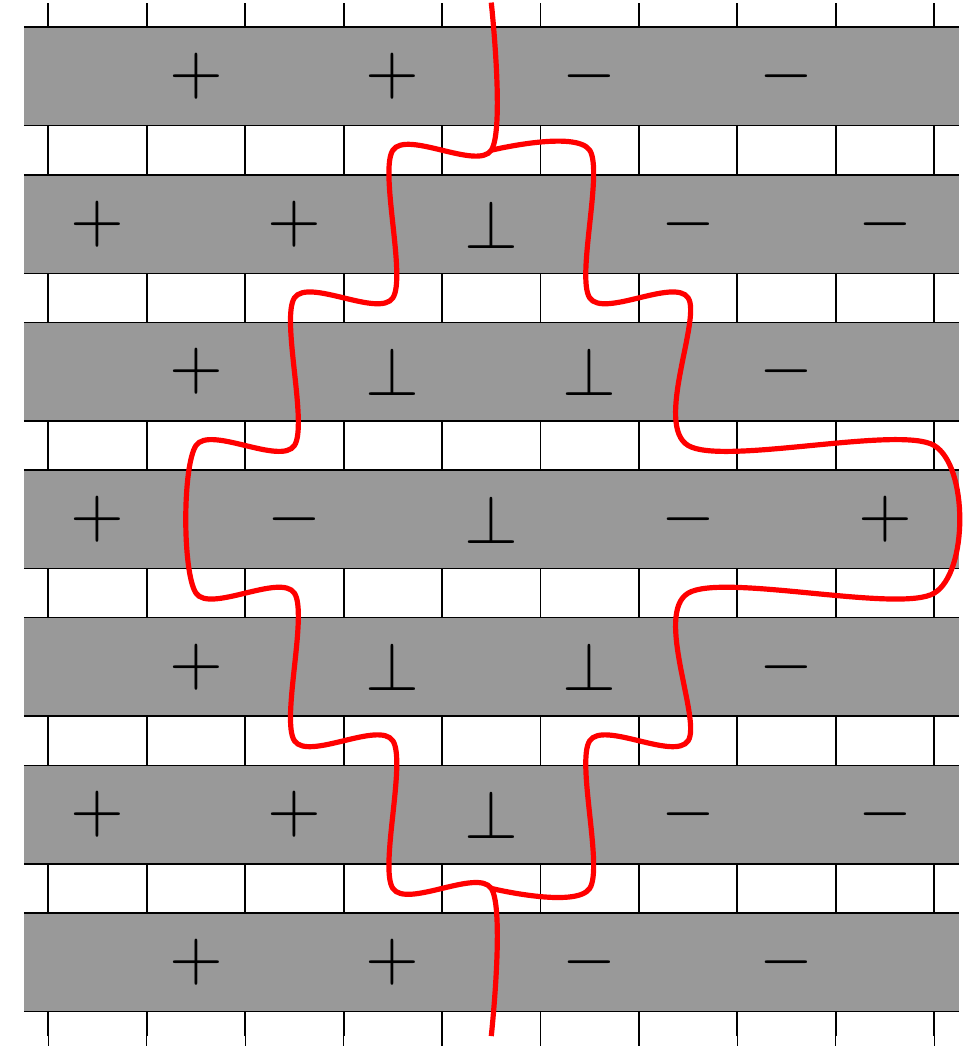}
}
\subfigure[]{
  \label{fig:c_floq_dw_two_site}	
  \includegraphics[width=0.45\columnwidth]{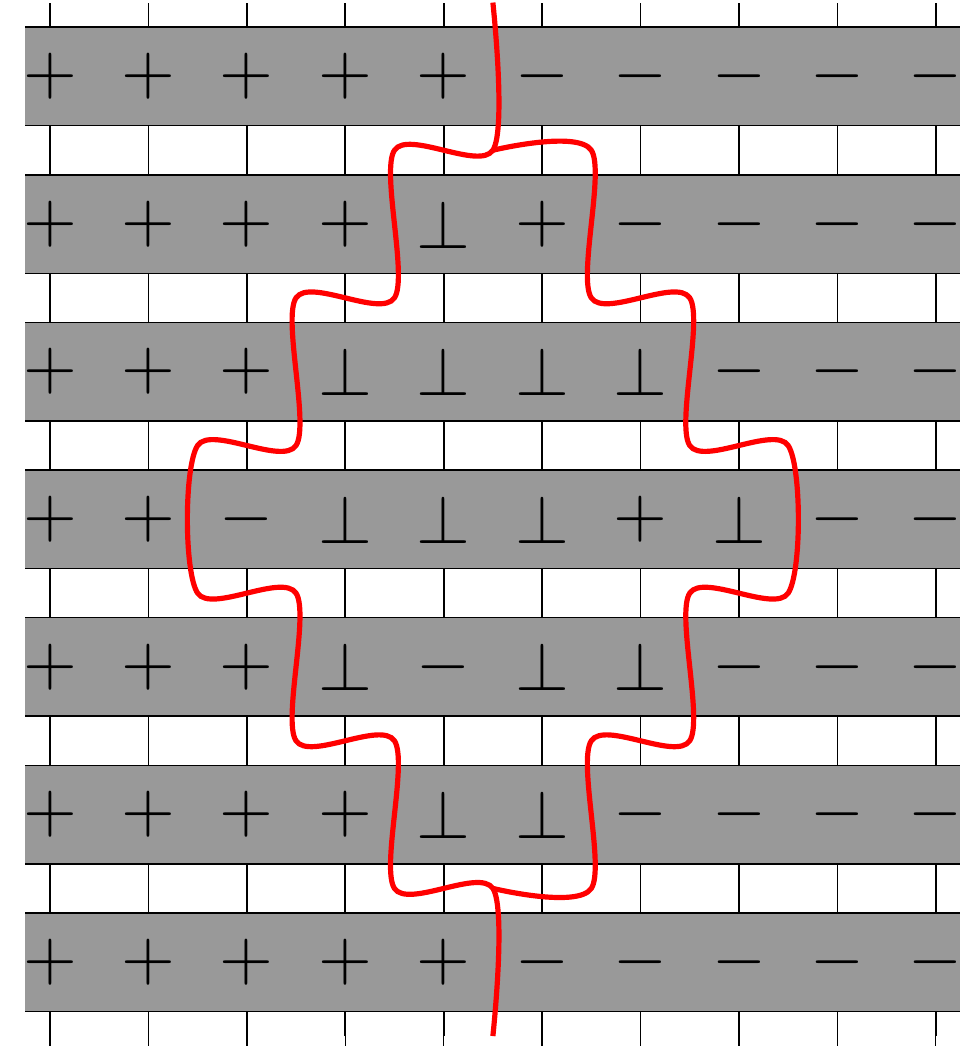}
}
\caption{Domain wall (membrane) in Floquet evolution without a circuit structure. The spin variables $+$, $-$ and $\perp$ denote the three choices of projection operators inserted on the horizontal bars beneath them, see Eq.~\eqref{eq:Z_c_floq_proj} and Eq.~\eqref{eq:Z_c_floq_one_site_proj}. Figures show samples of thick domain walls for (a) two-site and (b) one-site insertion. }
\label{fig:c_floq_dw}
\end{figure}

The partition functions are sums over spins $s_{x,t}$, taking $N! + 1$ values, associated with each horizontal bar.
We focus on the case of the purity ($N = 2$), where
$s_{x,t}$ can be $\pm$ or $\perp$.
We choose a graphical notation that resembles that for the circuit, labelling the shaded region immediately above the horizontal bar with the bar's spin value: see  Fig.~\ref{fig:c_floq_dw}.
This figure shows possible spin configurations for a section of domain wall of finite width in the two schemes.

The numerical algorithm in Sec.~\ref{sec:generatingfunctions} can be carried over with little modification. 
We first define the partition function $Z(x,t)$ the same way as in Eq.~\eqref{eq:Z_x_y_t} --- the difference here is that $Z(x,t)$ may be nonzero even when $x$ is arbitrarily large. We then solve for the irreducible step weights $W(x,t)$ recursively, using Eq.~\eqref{eq:Wbyrecursion}. 

The approach is feasible because the Floquet evolution has
an emergent Lieb-Robinson light cone \cite{lieb_finite_1972} which  suppresses $Z(x,t)$ for large $x$.
For practical computation of $Z(x,t)$, we must truncate the Floquet operator to a chain of finite length $L$, so at the end we must check that our results are converged in $L$. For a fixed $t$, the error introduced due to the finiteness of $L$ should be exponentially small at large $L$, but of course the required $L$ grows linearly with $t$.

\begin{figure}[t]
\centering
\subfigure[]{
  \label{fig:two_site}	
  \includegraphics[width=0.9\columnwidth]{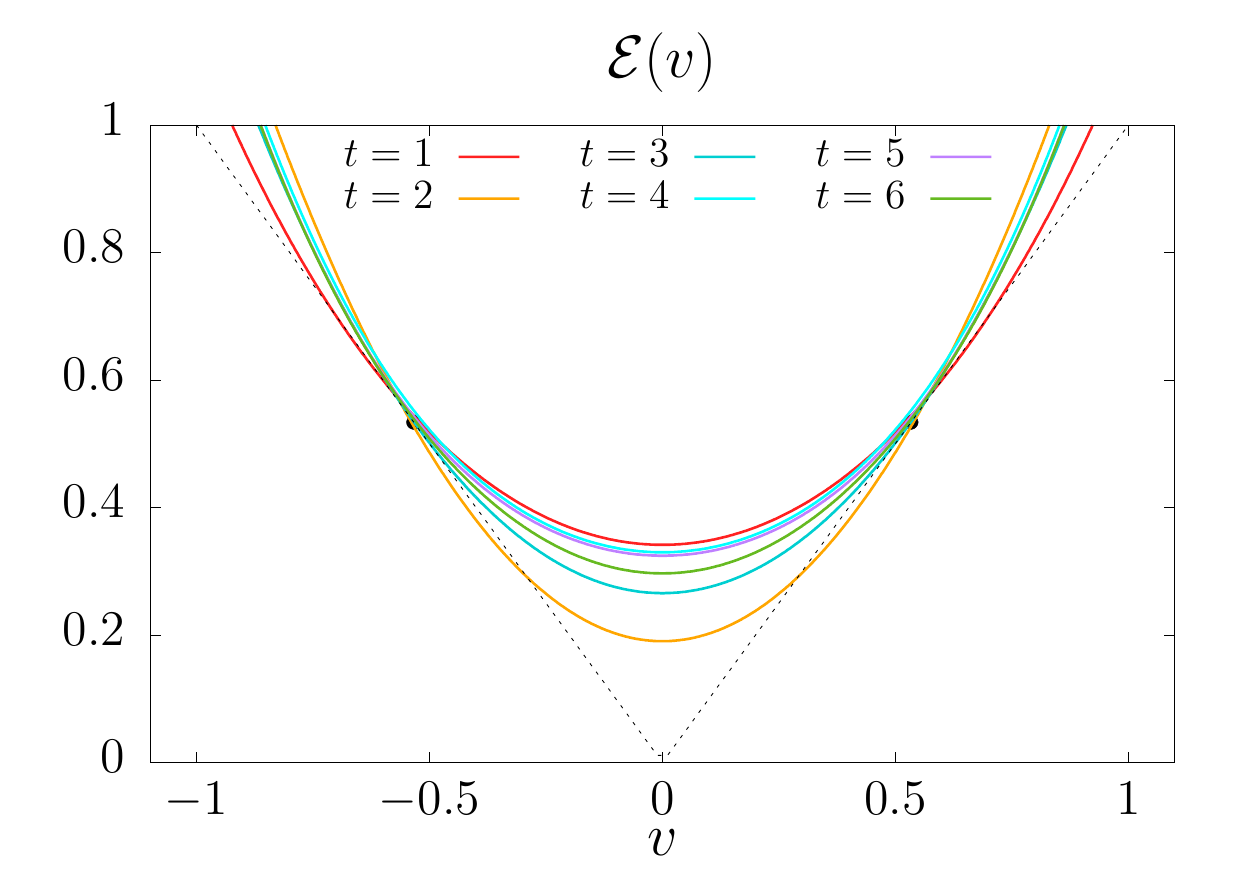}
}
\subfigure[]{
  \label{fig:one_site}	
  \includegraphics[width=0.9\linewidth]{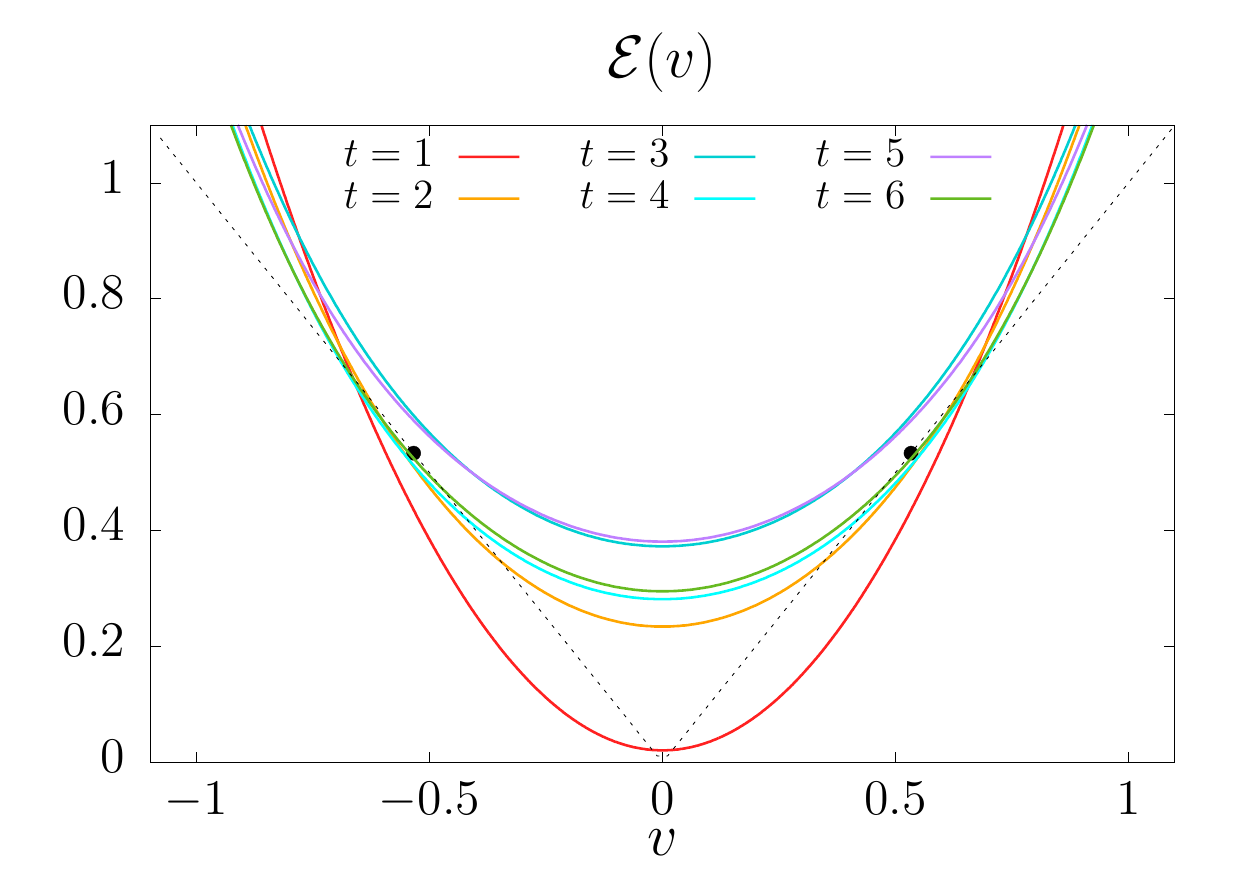}
}
\caption{Numerical calculations of $\mathcal{E}(v)$ for Floquet evolution with the square-wave $h_x(t)$ (Eqs.~\ref{eq:floq_cos_w},~\ref{eq:hJw}) for (a) two-site and (b) one-site schemes.}
\label{fig:c_floq}
\end{figure}

\begin{figure}[t]
\centering
\includegraphics[width=0.9\columnwidth]{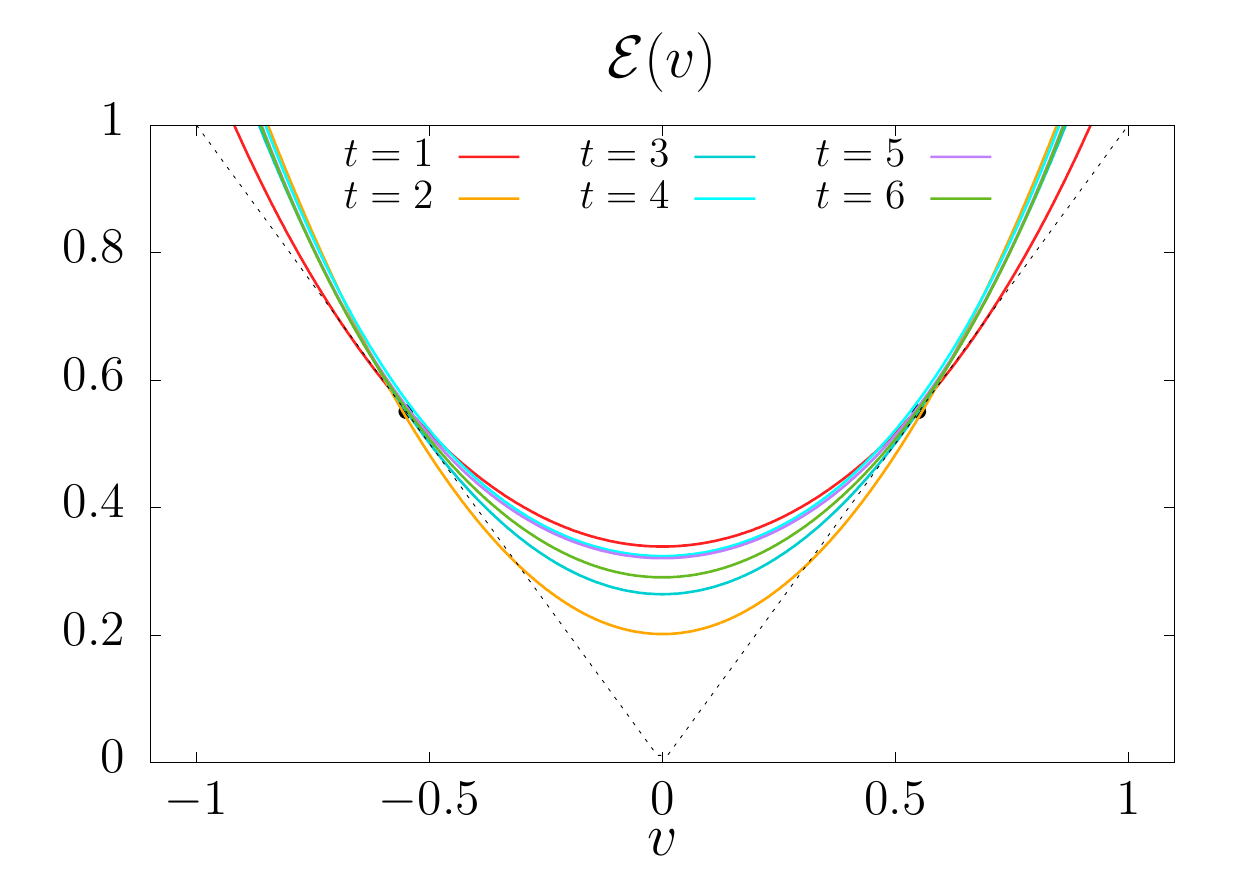}
\caption{Numerical calculations of $\mathcal{E}(v)$ for Floquet evolution with the sinusoidally varying $h_x(t)$ in  Eq.~\eqref{eq:cosinehx} for the two-site scheme.}
\label{fig:cos_floq}
\end{figure}

We consider two examples of an Ising model with an oscillating transverse field $h_x(t)$,
\begin{equation}
\label{eq:floq_cos_w}
H(t)  = \sum_{i=1}^{L-1} J \sigma_i^z \sigma_i^{z+1} +\sum_i h_x(t) \sigma_i^x
 + \sum_i h_z \sigma_i^z.
\end{equation}
In the first example $h_x(t)$ is a square wave with period  $\tau = 1$, taking the values $h^{(1)}$ and $h^{(2)}$ for half the period each:
\begin{align}
\label{eq:hJw}
  h^{(1,2)}_x & = 0.9045 \pm w,
  &  w &= 0.3,
  \\
J & = 0.5, & h_z & = 0.8090.
\notag
\end{align}
The Floquet operator is then 
\begin{equation}
\label{eq:floq_w}
U_{\text{Floq}} = \exp( -i H_1 \tau/2 ) \exp( - iH_2 \tau/2 )
\end{equation}
with $H_{1,2}$ equal to (\ref{eq:floq_cos_w}) with the appropriate value of $h_x$.

In the second example $h_x(t)$ oscillates sinusoidally, with period $\tau=1$,
\begin{align}\label{eq:cosinehx}
    h_x = 0.9045 \left[ 1  + w \cos( 2 \pi t / \tau ) \right],
\end{align}
the parameters $J$, $h_z$ and $w$ being the same  as in Eq.~\eqref{eq:hJw}.
Neither of these two examples allows a circuit representation.

For the first system (square wave) we compare the 1 and 2 site insertion schemes.
Fig.~\ref{fig:c_floq} shows approximations to the line tension function, for increasing values of $t_{\rm max}$, in a system of $L = 12$ spins, for the two-site and one-site schemes respectively.
We position the domain wall at the top boundary so as to minimize finite--$L$ effects, 
see the numerical details in App.~\ref{subsec:protocol_WZ}. (Performing calculations for a given $L$ with multiple choices of the top boundary condition incurs an additional cost, but this is a subleading factor.)
We compute up to 
$t_{\rm max} = 6$. 
We estimate the error due to the finiteness of $L$ by computing the difference in the  ${t=5}$ result for $\mathcal{E}(0)$ for ${L = 10}$ and for ${L = 12}$. We find a relative error ${0.2\%}$ for the two-site scheme and $1\%$ for the one-site scheme.

For both schemes, the $t=6$ curves in Fig.~\ref{fig:c_floq} are consistent with reasonable convergence to a legitimate line tension function, with a value of $v_B$ consistent with that computed directly from the OTOC.

One counter-intuitive feature of Fig.~\ref{fig:two_site} is that the two-site scheme performs better than the one-site scheme for small $t$ (the one-site scheme is also more expensive computationally).
The contrast is most obvious at ${t = 1}$, where the curve for the one-site scheme (Fig.~\ref{fig:two_site}) strongly violates the constraints of the line tension.
Despite this, the two curves for $t=6$ are in remarkably good agreement, with $0.5\%$ relative difference in the entanglement velocity estimates.

Finally, in Fig.~\ref{fig:cos_floq} we show results for the sinusoidally oscillating field, Eq.~\ref{eq:cosinehx}. These are qualitatively similar to Fig.~\ref{fig:two_site}.


\section{Outlook}
\label{sec:outlook}

In random circuits, the ``pairing field'' and the membrane picture emerge in a simple way from the average over unitaries whenever we deal with dynamical quantities that require multiple forward and backward evolution operators (such as R\'enyi entropies or OTOCs).
Here we have shown how to make sense of the
pairing field,
which describes a pairing of trajectories in the multi-sheeted path integral,
in realistic systems that do not involve randomness.  
This opens the way to extending the various results that have been obtained in random circuits to a much broader class of more realistic models. 
More generally, by identifying appropriate effective fields, it provides a framework for applying the renormalization group to chaotic real-time dynamics.

We separated the local multi-copy Hilbert space into ``paired'' states, and states in their orthogonal complement, $\perp$.  Dynamical quantities then became partition functions  for an effective field ${s(x,t)\in \{ \perp\} \cup S_N}$. 
The state $\perp$ of the effective spin here is reminiscent of the  state $S=0$, representing a vacancy, in the Blume-Capel model (the classical Ising model with vacancies, where the spin takes the values  ${S = 0, \pm 1}$).  
We have shown that in strongly chaotic Floquet systems the ``vacancies'' $\perp$ dress the structure of domain walls between different pairing states, but leave the basic structure of the domain wall intact on large scales. 

After using a typical realization of a random circuit as a test case, 
we applied this picture to Floquet systems, for spin-1/2, that have no small parameter at all and no randomness. 

The method gave convincing results for the line tension function for $N=2$ (which determines both the second R\'enyi entropy  growth rate and the butterfly velocity) from calculations on relatively small lengthscales. A simple but important ingredient was a resummation of the weights in the spin model into a single function $W(x,t)$, describing the amplitude for the membrane --- which in 1+1D can be viewed as a directed path --- to take a step of temporal extent $t$ and spatial displacement $x$. Interestingly, there was a qualitative difference in the structure of $W$ for generic Floquet models (whether circuits or not) and for dual-unitary, maximally entangling circuits.

Many questions remain for the future.  First, there are many kinds of extra structure that could be incorporated and which might be expected to have interesting effects.

We have focused on Floquet models without any conserved quantities. 
Any slow mode, for example a conserved $\mathrm{U}(1)$ charge, will interact with the dynamics of the membrane. 
(A single physical conserved charge gives rise to multiple conserved charges in the replicated Hilbert space.) Results on random circuits with conservation laws provide a starting point \cite{rakovszky_diffusive_2017,khemani_operator_2017}. 
Our approach can be extended to models with conservation laws by refining the resolution of the identity to specify information about the charge as well as the permutation: we will discuss this elsewhere.
One cautionary note is that it has recently been shown that  the dynamics of the higher R\'enyi entropies, for which the present method is suitable, can be qualitatively different from the that of the von Neumann entropy in models with conservation~laws~\cite{rakovszky_sub-ballistic_2019,huang_dynamics_2019,zhou2019diffusive,znidaric_entanglement_2019}. 

In the present approach, the von Neumann entropy corresponds to a nontrivial replica-like limit, which it would be interesting to understand better even in the absence of conserved quantities.

Also interesting are continuous spatiotemporal, as opposed to internal, symmetries, for example Lorentz invariance. Such symmetries will constrain the effective field theory of the membrane, as well as giving new hydrodynamic modes \cite{liu_lectures_2018,rangamani_gravity_2009}.

There is considerable freedom in the precise way in which the effective spins are introduced. We focused mainly on inserting a resolution of the identity for pairs of sites, 
since this was convenient for making contact with the Haar-random circuits studied previously. 
We also explored the use of single-site resolutions of the identity: this appeared to be less efficient for the model we studied. It would be useful to understand this better in order to optimize the efficiency of the scheme. 

The basic idea of this paper is not restricted to 1+1D:  the effective partition function can be introduced in the same way in any number of dimensions, and we may again argue for the  coarse-grained picture in terms of a codimension-1 membrane dressed with $\perp$ states.
However, the additional computational cost means that in higher dimensions we are likely to need either special structure, or a small parameter, in order to obtain quantitative results.
{The method may also be applied to dynamics without spatial structure, in which any qubit can interact with any other.}

For the observables studied here, the pairing between layers of the circuit is local in time. For some dynamical quantities, for example the spectral form factor, pairings that are non-local in time play a role \cite{chan_solution_2018,chan2018spectral,kos2017many}.  It would be interesting to  extend the present approach to this setting.

Our method for defining the effective spin model might illuminate other interesting universal dynamical phenomena. 
First, we may consider how the description in terms of an entanglement membrane breaks down on approaching a many-body-localized phase. We hope to return to this elsewhere. 
There are other failure modes that could be investigated, for example the approach to a noninteracting or integrable point.

Within the more limited domain of random circuits, the effective spin model here may also give a way of thinking about entanglement in random Clifford circuits which is more generalizable (e.g. to higher dimensions) than the  approach of mapping the evolution of stabilizers to an effective classical stochastic dynamics \cite{nahum_quantum_2017,li2019measurement, gullans2019dynamical}.

The approach could also be used to study questions motivated by quantum information such as the emergence of the ``design'' property in random circuits with non-Haar--distributed gates \cite{harrow2009random, brandao2016local,roberts2017chaos,
cotler2017chaos,
hunter-jones_unitary_2019}.

The effective spin model may also  be extended to dynamics with measurement, or other types of interaction with an environment, for example to clarify the relationship between different universality classes of measurement-induced criticality \cite{skinner2018measurement,li2018quantum, chan2018weak,choi2019quantum,szyniszewski2019entanglement,li2019measurement,cao2018collective,gullans2019dynamical,gullans2019scalable,tang2019measurement,bao2019theory,jian2019measurement,zabalo2019critical}.

The approach could also be adapted to study  tensor networks that do not involve a time direction, for example in the context of the AdS-CFT correspondence where random tensor networks have been used \cite{hayden2016holographic,vasseur2018entanglement}. 

For chaotic dynamics, further numerical tests and further development of the numerical scheme, extending the range of models to which it could be applied usefully, would also be worthwhile and might give new insights into the physics on the scale of the membrane thickness (which, in many natural situations where there  is a small parameter at hand, may be  much larger than the lattice spacing).


\acknowledgements
We thank John Chalker, Tibor Rakovszky, and in particular Curt von Keyserlingk for very useful discussions.
TZ was supported by a postdoctoral fellowship from the Gordon and Betty Moore Foundation, under the EPiQS initiative, Grant GBMF4304, at the Kavli Institute for Theoretical Physics. 
This research is supported in part by the National Science Foundation under Grant No. NSF PHY-1748958.
We acknowledge support from the Center for Scientific Computing from the CNSI, MRL: an NSF MRSEC (DMR-1720256) and NSF CNS-1725797.
AN was supported by a Royal Society University Research Fellowship.

\appendix

\section{Properties of the spin model}
\subsection{Projectors onto permutation states}
\label{app:P_sigma}

The operator $P_\sigma$ defined in Eq.~\ref{eq:p_sigma} is a projection operator onto the $\sigma$ permutation state (it is a nonorthogonal projector, i.e. $P_\sigma^\dag\neq P_\sigma$). Here we prove the projector property in Eq.~\ref{eq:proj_prop}. According to the definition of $P_{\sigma}$ in terms of the Weingarten function and permutation states in Eq.~\ref{eq:p_sigma}, we have 
\begin{equation}
P_{\sigma_1}P_{\sigma_2} = \sum_{\tau_1,\tau_2} \text{Wg}(\tau_1^{-1} \sigma_1) |\tau_1  \rangle \langle \sigma_1 | \tau_2 \rangle  \text{Wg}(\tau_2^{-1} \sigma_2) \langle \sigma_2 | 
\end{equation}
We have simplified the notation by writing ${\ket{\tau}}$ as a shorthand for the two-site state ${\ket{\tau\,\tau}}$.  This emphasizes that the formulas here apply to an arbitrary Hilbert space, using the Weingarten functions for the appropriate value of the Hilbert space dimension (equal to $q^2$ here). We will return to the previous notation after Eq.~\ref{eq:appPP}. 
Note that states associated with distinct permutations are not orthogonal.

Unless $q^2 = 0, \pm 1, \pm 2, \cdots, \pm (N-1)$, the function $\text{Wg}( \tau^{-1} \sigma )$, viewed as a matrix with $\tau$ and $\sigma$ as its indices, is the inverse of the Gramian matrix $\langle \sigma | \tau \rangle $ \cite{gu_moments_2013,collins_integration_2006}:
\begin{equation}
\label{eq:weingarten_inv}
  \sum_{\tau } \langle \sigma_1 | \tau  \rangle \text{Wg}(\tau^{-1} \sigma_2) = \delta_{\sigma_1 , \sigma_2  }. 
\end{equation}
This means the states $| \sigma^* \rangle $ defined in Eq.~\ref{eq:dual_basis} indeed form the dual basis, 
\begin{equation}
\langle \sigma_1^* | \sigma_2  \rangle  = \delta_{\sigma_1, \sigma_2} . 
\end{equation}
So for $q^2 \ge N$, we have 
\begin{equation}\label{eq:appPP}
P_{\sigma_1}P_{\sigma_2} = \sum_{\tau_1,\tau_2} \text{Wg}(\tau_1^{-1} \sigma_1) |\tau_1  \rangle \delta_{\sigma_1, \sigma_2} \langle \sigma_2 |  = P_{\sigma_1}\delta_{\sigma_1 \sigma_2 }.
\end{equation}

Any four-layer unitary gate with the projector  $P_{\sigma}$ attached at the bottom is equal to $P_{\sigma}$ (Eq.~\ref{eq:un_sigma}). In Eq.~\ref{eq:p_box}, we represent the projector as a tensor  (box) labelled  with a $\sigma$ inside. Given the expression in Eq.~\ref{eq:p_sigma}, it can be decomposed as
\begin{equation}
\label{eq:ptensor}
\ptensor[\sigma] = \pdecompeq[0][0][\sigma^*][][\sigma][][\sigma][]
\end{equation}
where the ket ${| (\sigma\,\sigma)^* \rangle = \sum_{\tau} \text{Wg}( \tau^{-1} \sigma ) | \tau \, \tau\rangle}$ is the two-site dual state of ${|\sigma \sigma \rangle}$, and the bras at the bottom  just give ${\bra{\sigma\,\sigma}=\langle \sigma | \otimes \langle  \sigma |}$ on the two sites. 

This structure helps us to clarify the definition of the triangle weights $J( \sigma_b, \sigma_c; \sigma_a )$ as the result of contracting the red part of the following tensor: 

\begin{align}
  \tribox[\sigma_b][\sigma_c][\sigma_a]  &\equiv
\text{shaded part of}\lf\,
  \triboxdecomp[\sigma_b][\sigma_c][\sigma^*_a] \, \ri\\
  &= \langle \sigma_b \, \sigma_c | (\sigma_a \sigma_a)^*\rangle 
\end{align}
which yields the expression of $J( \sigma_b, \sigma_c; \sigma_a) $ in Eq.~\ref{eq:J_in_weingarten}.


\subsection{$N$-independence of the weights} 
\label{app:N_indep}

The triangle weights
\begin{equation}
J( \sigma_b, \sigma_c; \sigma_a ) = \tribox[\sigma_b][\sigma_c][\sigma_a]  = \jabc[\sigma_a][\sigma_b][\sigma_c] 
\end{equation}
define the interactions of the effective spin models arising from the random untiaries. 
The spin $\sigma$ is an element of $S_N$, where $N$ is the number of copies of $U$ and of $U^*$ in the multi-layer circuit.

In some cases of interest to us the spins $\sigma_a$, $\sigma_b$,  $\sigma_c$ are all in the smaller permutation group $S_M$, with ${M\leq N}$, corresponding to permutations involving the indices ${1,\ldots, M}$ only. 
For the corresponding bras, the first $2M$ layers ($M$ ``$U$'' layers and $M$ ``$U^*$'' layers) can be paired arbitrarily, with the pairing labelled by a permutation in $S_M$, and the remaining $2(N-M)$ layers are paired in the manner prescribed by the identity permutation.
For example, consider the case ${\sigma_a = \sigma_b = \mathbb{I}}$ and $\sigma_c=(12)$:
\begin{equation}
\label{eq:J_K}
\jabc[\I][\I][(12)] = K.
\end{equation}
This weight is independent of $N$ (proved algebraically in \cite{zhou2019emergent}).
This $N$-independence holds whenever all the spins are in the permutation group $S_M$, 
as conjectured in \cite{zhou2019emergent} and proved by Hunter-Jones in Ref.~\cite{hunter-jones_unitary_2019}. Here we review this result.

Algebraically, the triangle is the coefficient in the expansion (\ref{eq:triangleasexpansioncoeff}):
\begin{equation}\label{eq:triangleasexpansioncoeffappendix}
\bra{\sigma_b\, \sigma_c} P_\parallel^{(N)}
=\sum_{\sigma_a\in S_N} 
\jabc[\sigma_a][\sigma_b][\sigma_c] \bra{\sigma_a\,\sigma_a},
\end{equation}
(the two spins refer to the two spatial states), which is equivalent to the definition in Eq.~\ref{eq:J_in_weingarten},
\begin{equation}\label{eq:Jexplicit}
\jabc[\sigma_a][\sigma_b][\sigma_c]  = \sum_{\tau\in S_N} \langle \sigma_b\,  \sigma_c  | \tau \,\tau \rangle  \text{Wg}( \tau^{-1}\sigma_a; q^2 ),
\end{equation}
as we see by using Eq.~\ref{eq:weingarten_inv} to extract a particular coefficient on the right hand side of Eq.~\ref{eq:triangleasexpansioncoeffappendix}.

Now consider the case where ${\sigma_b,\sigma_c\in S_M}$, where $S_M$ is the subgroup mentioned above.\footnote{By the symmetry in Eq.~\ref{eq:symmgp}, the following also applies to other subsets of states related to this one by left and right multiplication with fixed elements of $S_N$.}
The fact that the resulting weights are independent of $N$ is not immediately apparent from Eq.~\ref{eq:Jexplicit}, but can be seen from Eq.~\ref{eq:triangleasexpansioncoeff}. When ${\sigma_b,\sigma_c\in S_M}$, we have
\begin{equation}\label{eq:reducetoPM}
\bra{\sigma_b\, \sigma_c} P_\parallel^{(N)}= 
\bra{\sigma_b\, \sigma_c} \lf P_\parallel^{(M)}\otimes \mathbb{1}^{(N-M)}\ri.
\end{equation}
For example, we can see this by writing the projection operator $P_\parallel^{(N)}$ as the average of a stacked random unitary. The unitaries in the layers with index greater than $M$ cancel, since the bra contracts them together in the same way on each site. This leaves the average of a stack acting only on the first $2M$ layers, which is equal to $P_\parallel^{(M)}$ acting on those layers. The identity acts on the remaining $2(N-M)$ layers.

This shows that in Eq.~\ref{eq:reducetoPM} the only nonvanishing terms on the right-hand-side have $\sigma_a\in S_M$ as well. As a result Eq.~\ref{eq:reducetoPM} becomes equivalent to the same equation for the $2M$-layer case (except that both sides are tensored with the ``identity''   bra for the remaining $2(N-M)$ layers). This shows that the triangle weights are the same as in the $2M$-layer case, i.e. independent of $N$.

Note however that if all of the spins are in $S_{M} \otimes S_{N-M} \subset S_N$, where the subgroups refer to the first $2M$ and the last $2(N-M)$ layers respectively, the weight does \textit{not} factorize in general. That is, if we have 
\begin{align}
\sigma_a &= \sigma_{a1} \sigma_{a2}, &
\sigma_b &= \sigma_{b1} \sigma_{b2}, &
\sigma_c &= \sigma_{c1} \sigma_{c2},
\end{align}
where $\sigma_{b1}\in S_M$ and $\sigma_{b2}\in S_{N-M}$ {\it etc.}, the triangle 
\begin{equation}
\jabc[\sigma_a][\sigma_b][\sigma_c] 
\end{equation}
is in general not equal to 
\begin{equation}
\jabc[\sigma_{a1}][\sigma_{b1}][\sigma_{c1}] \times \jabc[\sigma_{a2}][\sigma_{b2}][\sigma_{c2}]. 
\end{equation}
The two are equal if, for example, $\sigma_{b2}=\sigma_{c2}$.


\subsection{Suppression of large $\perp$ clusters}
\label{app:supp_perp}

We analyze the spin model that arises in computing the mean square weight $\overline{\Omega_C^2}$ of a $\perp$ cluster like that in Fig.~\ref{fig:supp_perp_cluster}. (The average is over Haar-random unitaries.)
The spins take values in $S_4$, because in total we need four $U$ layers and four $U^*$ layers to write $\Omega_C^2$.
We will see that there is a distinction between spin values in the $S_2\times S_2$ subgroup generated by $(12)$ and $(34)$, and spin values outside this subgroup. At leading order, the former are forbidden inside the cluster.

We first look at the region outside the $\perp$ region.
Each block there is labelled either as $+$ or $-$, indicating a local insertion of $P_\pm$ before squaring. Recall that this projection operator ``absorbs'' the local unitary gate: equivalently, in the squared system a given $+$ block is (recall that each $P_+^{(2)}$ acts on two physical sites):
\begin{equation}
  \label{eq:u_outside_perp}
\overline{
\uc[u][4]}
\lf \uc[P][2][+] \otimes \uc[P][2][+] \ri   =
\uc[P][2][+] \otimes \uc[P][2][+] 
\end{equation}
and similarly for $-$. This defines an interaction triangle for the case where the lower spin in the triangle is from a plus block, in analogy to the triangle interaction defined in Sec.~\ref{sec:spinmodelgeneralities} for the case without the $P^{(2)}_+\otimes P^{(2)}_+$ factor. 
However, because of the projection operator, the spin associated with the block (the lower spin of the triangle) is \textit{fixed} to be $\mathbb{I}$.
Similarly, the spin associated with a $-$ block is fixed to be $\sigma=(12)(34)$.
Therefore in computing $\Omega_C$, the spins outside the cluster are not free to fluctuate. Their role is to provide a boundary condition for the spins inside the cluster.

Explicitly, the $+$ weights are (from Eq.~\ref{eq:triangleasexpansioncoeff} with $P_\perp^{(4)}$ replaced by $P_+^{(2)}\otimes P_+^{(2)}$; the $-$ weights are related to the following by symmetry)
\begin{align}
& \jabc[\mathbb{I}][\sigma_b][\sigma_c][+] 
  = q^{-8}
\bra{\sigma_b \,\sigma_c}
\lf \uc[P][2][+] \otimes \uc[P][2][+] \ri
\ket{\mathbb{I}\,\mathbb{I}}.
\end{align}
Using the expression for the projector in Eq.~\ref{eq:p_pm}, this $+$ triangle is equal to
\begin{equation}
\frac{q^8}{(q^4 - 1)^2}  \sum_{\tau \in S_2} q^{ - |\sigma_b \tau| - |\sigma_c \tau | - 2|\tau|  } (-1)^{|\tau|} 
\end{equation}
Since $|\sigma_b \tau| + |\tau| \ge |\sigma_b|$, in the large $q$ expansion, the $+$ triangle is at most of order 
\begin{equation}
\label{eq:plus_tri_expansion}
\jabc[\mathbb{I}][\sigma_b][\sigma_c][+]  \lesssim q^{- |\sigma_b| - |\sigma_c| } 
\end{equation}
and similarly the $-$ triangle can be bounded by
\begin{equation}
\label{eq:minus_tri_expansion}
\jabc[(12)(34)][\sigma_b][\sigma_c][-]  \lesssim q^{- |\sigma_b(12)(34)| - |\sigma_c(12)(34)| } 
\end{equation}
Here $|\sigma_b|$ and $|\sigma_b (12)(34)|$ can be understood as the numbers of elementary domain walls (i.e. number of transpositions) on the left edge of the $+$ and $-$ triangle respectively. And similarly the corresponding $\sigma_c$ terms are for the right edge. We thus interpret Eq.~\ref{eq:plus_tri_expansion} and Eq.~\ref{eq:minus_tri_expansion} as costing $q^{-1}$ per out-going domain wall on the two edges. 

Inside the cluster we have $\perp$ blocks. These lead to a ``$\perp$'' interaction triangle defined by:
\begin{align}\label{eq:perptridefn}
\sum_{\sigma_a \in S_4 } \jabc[\sigma_a][\sigma_b][\sigma_c][\perp] &\langle \sigma_a \, \sigma_a | \equiv
\bra{\sigma_b \, \sigma_c}
\lf 
\overline{ \uc[u][4]}
- 
\overline{\uc[u][2]} \otimes \overline{ \uc[u][2]}
\ri \\
&\equiv
\bra{\sigma_b\,\sigma_c}
\lf 
P_\parallel^{(4)}
- 
P_\parallel^{(2)} \otimes P_\parallel^{(2)}
\ri 
\notag
\end{align}
where we have used the equivalence between the parallel projection and the Haar averaged stack of unitaries. 

The right-hand-side of Eq.~\ref{eq:perptridefn} gives the $\perp$ triangle weight as the difference of two terms. The first term is the ordinary triangle weight for $S_4$. The second term vanishes unless $\sigma_a\in S_2\times S_2$. Depending on whether $\sigma_a \in S_2 \otimes S_2$, the large $q$ expansion yields
\begin{equation}
\label{eq:perp_tri_expansion}
\jabc[\sigma_a][\sigma_b][\sigma_c][\perp] \left\lbrace
  \begin{aligned}
    & \sim q^{- |\sigma_a^{-1} \sigma_b| - |\sigma_a^{-1} \sigma_c| } & \, \sigma_a \notin S_2 \otimes S_2 \\
    & \lesssim q^{- |\sigma_a^{-1} \sigma_b| - |\sigma_a^{-1} \sigma_c| - 2 }  & \, \sigma_a \in S_2 \otimes S_2 \\
  \end{aligned} \right. 
\end{equation}
For the first case, the $\perp$ triangle weight is exactly equal to an ordinary triangle
\begin{align}\label{eq:perpequalsnoperp}
\jabc[\sigma_a][\sigma_b][\sigma_c][\perp]  & = \jabc[\sigma_a][\sigma_b][\sigma_c],
&
\sigma_a & \notin S_2\times S_2.
\end{align}
We again interpret the corresponding expansion (the first line) in Eq.~\ref{eq:perp_tri_expansion} as a cost of  ${q}^{-1}$ for each outgoing elementary domain wall. For the case of $\sigma_a \in S_2 \otimes S_2$, we can show using the large $q$ expansion techniques in Ref.~\onlinecite{zhou2019emergent} that the term of order $q^{- |\sigma_a^{-1} \sigma_b| - |\sigma_a^{-1} \sigma_c| } $ cancels, which therefore incurs an additional cost of $q^{-2}$. In particular, if $\sigma_b , \sigma_c$ are also in $S_2 \times S_2$, then the $\perp$ triangle vanishes exactly:
\begin{equation}
\label{eq:perp_s2_s2_s2}
\jabc[\sigma_a][\sigma_b][\sigma_c][\perp] = 0, \quad \sigma_a, \sigma_b, \sigma_c \in S_2 \times S_2.
\end{equation}

With this in hand, we can estimate the cost of a $\perp$ cluster in Fig.~\ref{fig:supp_perp_cluster}. Because of the vanishing weight in Eq.~\ref{eq:perp_s2_s2_s2}, each triangle inside $\perp$ region must host at least one spin that is not in $S_2 \times S_2$. Therefore in each time slice that contains  a  $\perp$ there are at least four elementary domain walls. By the estimates above, each costs at least $\frac{1}{q}$ per time step, regardless of whether it is inside the $\perp$ region or on the boundary between the $\perp$ and $\pm$ regions. A $\perp$ cluster that persists for $t$ steps is therefore  suppressed by a factor of $q^{-4t}$ or smaller. This completes the argument in Sec.~\ref{sec:tractable_limit}. 

The cost of $q^{-4t}$ is achieved by the following configuration. We set the spins in the $\perp$ cluster to $(13)(24)$. The triangle weights  inside the $\perp$ cluster are then $1$,  by  Eq.~\ref{eq:perpequalsnoperp}. There are two elementary domain walls making up the composite domain wall $(13)(24)$ on the left boundary of the $\perp$ cluster and two elementary domain walls making up $(14)(23)$ on the right boundary. They enter the $\perp$ cluster through the triangle 
\begin{equation}
\jabc[(13)(24)][\I][(12)(34)][\perp]
\end{equation}
which is equal to the ordinary triangle and costs $q^{-4}$. Meanwhile on the boundary, the four domain walls also cost $q^{-4}$, according to Eq.~\ref{eq:plus_tri_expansion} and Eq.~\ref{eq:minus_tri_expansion}. In total, the weight is $q^{-4t}$ for the squared average of the $\perp$ cluster.



\subsection{Minimal $\perp$ cluster and operator purity}
\label{app:perp_op_purity}

In Eq.~\ref{eq:hexbox}, we listed four non-trivial configurations for an isolated $\perp$ surrounded by $+$ and $-$. Here we compute their weights for a fixed unitary $u$ at the location of the $\perp$. We will see they are related to the operator purity of $u$. 

It is more transparent if we represent each $+$ and $-$ block graphically as in  Eq.~\ref{eq:ptensor}:
\begin{equation}
\hexbox[+][-][+][\perp][-][\sigma_1][\sigma_2]  \equiv \text{shaded part of} \lf \, \hexboxdecomp \, \ri.
\end{equation}
The weight on the left-hand-side is equal to the c-number given by contracting the colorful parts of the tensor on the right-hand-side. The 4-leg $\perp$ tensor  (red) is $u^{(4)}P_{\perp}$. We can rewrite it as $u^{(4)} - P_{\parallel} = u^{(4)} - P_{+} - P_{-}$. It is contracted with states $\langle +|$ and $\langle -|$ on the top (blue) and the kets at the bottom from the green tensors:
\begin{equation}
\label{eq:l-shape}
\begin{aligned}
\lshapetensor[+][+][l][green!50] &= \frac{K q}{q^2 - 1} \left[  |+\rangle  - \frac{1}{q^3}  |-\rangle  \right] \\
\lshapetensor[+][-][l][green!50] &= \frac{K}{q^2 - 1} \left[  |+\rangle - \frac{1}{q} |-\rangle  \right] \\
\lshapetensor[-][-][r][green!50] &= \frac{K}{q^2 - 1} \left[ |-\rangle - \frac{1}{q^3} |+\rangle \right] \\
\lshapetensor[-][+][r][green!50] &= \frac{K}{q^2 - 1} \left[  |+\rangle - \frac{1}{q} |-\rangle  \right] \\
\end{aligned}
\end{equation}
Combining these, we have 
\begin{equation}
\begin{aligned}
&\hexbox[+][-][+][\perp][-][+][-] = -2K^2 +  K^2 \left( \frac{q}{q^2 - 1} \right)^2 \times \\
&\left( \regboxteq[$u^{(4)}$][+][-][-][+]  
- \frac{1}{q^3}\regboxteq[$u^{(4)}$][+][-][+][+]
- \frac{1}{q^3}\regboxteq[$u^{(4)}$][+][-][-][-] 
+ \frac{1}{q^6}\regboxteq[$u^{(4)}$][+][-][+][-]
\right)
\end{aligned}
\end{equation}
where the  $\pm$ symbols around the $u^{(4)}$ block in the second line indicate contraction with $\bra{\pm}$ (at the top) or $\ket{\pm}$ (at the bottom). Similarly,
\begin{equation}
\begin{aligned}
&\hexbox[+][-][+][\perp][-][+][+] = \hexbox[+][-][+][\perp][-][-][-] =  -K^2  + K^2 \frac{q}{(q^2 - 1)^2}  \times \\
&\left( \regboxteq[$u^{(4)}$][+][-][+][+]
- \frac{1}{q} \regboxteq[$u^{(4)}$][+][-][-][+]
- \frac{1}{q^3} \regboxteq[$u^{(4)}$][+][-][+][-]
+ \frac{1}{q^4} \regboxteq[$u^{(4)}$][+][-][-][-] 
\right) 
\end{aligned}
\end{equation}
\begin{equation}
\begin{aligned}
&\hexbox[+][-][+][\perp][-][-][+] = K^2\frac{1}{(q^2 - 1)^2} \times \\
&\left( \regboxteq[$u^{(4)}$][+][-][+][-]
- \frac{1}{q} \regboxteq[$u^{(4)}$][+][-][+][+]
- \frac{1}{q} \regboxteq[$u^{(4)}$][+][-][-][-] 
+ \frac{1}{q^2} \regboxteq[$u^{(4)}$][+][-][-][+] \right)
\end{aligned}
\end{equation}

The blocks on the second line, representing different contractions of  $u^{(4)}$,
are the operator purities for different partitions. Two of them are actually independent of the unitary gate $u$:
\begin{equation}
\label{eq:perp_pp}
\regboxteq[$u^{(4)}$][+][-][+][+] = \regboxteq[$u^{(4)}$][+][-][-][-] = q^3
\end{equation}
This is because $|+\rangle|+\rangle $ and $|-\rangle|-\rangle $ are invariant under the action of $u^{(4)}$. The other two terms are proportional to the operator purity of vertical and diagonal partitions: 
\begin{equation}
\label{eq:zopsqu}
\regboxteq[$u^{(4)}$][+][-][-][+] = q^4 \mathcal{P}_{\opsqu[1]} \qquad  \regboxteq[$u^{(4)}$][+][-][+][-] = q^4 \mathcal{P}_{\opsqu[2]}
\end{equation}
Using these operator purities and simplifying, we obtain the expressions in Eq.~\ref{eq:124weights}.


\subsection{The size of the fluctuations}
\label{app:size_fluct}

In Sec.~\ref{sec:tractable_limit}, we listed four local configurations where an isolated $\perp$ lies at the path of the domain wall. The average weight of each configuration is zero in the Haar ensemble. In this appendix we estimate their fluctuations, i.e. their mean-square values: 
\begin{equation}
\label{eq:hexbox_squ}
\begin{aligned}
\overline{\underbrace{\hexbox[+][-][+][\perp][-][+][-][s]}_{\circled{1'}}} \quad 
\overline{\underbrace{\hexbox[+][-][+][\perp][-][+][+][s]}_{\circled{2'}}}
 = \overline{\underbrace{\hexbox[+][-][+][\perp][-][+][+][s]}_{\circled{3'}}} \quad
\overline{\underbrace{\hexbox[+][-][+][\perp][-][-][+][s]}_{\circled{4'}}} 
\end{aligned}
\end{equation}
$\circled{2'} = \circled{3'}$ by the symmetry under spatial reflection and $+/-$ exchange. We thus evaluate the other three. 

The averages of these diagrams generate the  $S_4$ model discussed in App.~\ref{app:supp_perp}. They can be expressed in terms of $\perp$ and $\pm$ triangles:
\begin{equation}
\joneperp[][\I][(12)(34)][\perp][\I][\sigma_b][(12)(34)][\sigma_c] 
\end{equation}
where $\sigma_b$ and $\sigma_c$ label the types of the triangles and are  determined by the spins $\pm$ in the bottom two blocks of the configurations in Eq.~\ref{eq:hexbox_squ}. 
Because of the doubling of layers (due to squaring), $+$ corresponds to (e.g.) ${\sigma_b=\mathbb{1} \in S_4}$, and $-$ corresponds to ${\sigma_b = (12)(34)}$. 
We can write these weights as
\begin{equation}
  w_{\perp} ( \sigma_b, \sigma_c) = \sum_{\sigma_a} w_{\perp}( \sigma_b, \sigma_c; \sigma_a )
\end{equation}
where the sum is over the  spin  $\sigma_a$ at the center vertex and
\begin{align}
&w_{\perp}( \sigma_b, \sigma_c; \sigma_a )\\
&= \sum_{\sigma_a} \jabc[\sigma_a][\I][(12)(34)][\perp] \times  \jabc[][\I][\sigma_a][\sigma_b]  \times \jabc[][\sigma_a][(12)(34)][\sigma_c]
\end{align}
The lower spins of the second and triangles in this product are fixed to be $\sigma_b$ and $\sigma_c$ respectively, so are not written explicitly.
According to Eqs.~\ref{eq:plus_tri_expansion} and \ref{eq:minus_tri_expansion}, the $\sigma_b$ and $\sigma_c$ triangles have the leading order expansion
\begin{align}
\jabc[][\I][\sigma_a][\sigma_b]  &\lesssim q^{ - | \sigma_b| - |\sigma^{-1}_b \sigma_a| } \le q^{- |\sigma_a |}  \\
\jabc[][\sigma_a][(12)(34)][\sigma_c] &\lesssim q^{  - |\sigma_c (12)(34)| - |\sigma^{-1}_c \sigma_a| }  \le q^{- |\sigma_a (12)(34) |} 
\end{align}
For $\sigma_a \in S_2 \times S_2$, we have $\sigma_a = \sigma_a^{-1}$, and
\begin{align}
\jabc[][\I][\sigma_a][\sigma_b] \times \jabc[][\sigma_a][(12)(34)][\sigma_c]  &\le q^{ - |\sigma_a (12)(34) | - | \sigma^{-1} _a| }   \notag  \\
 & \le q^{-2}  
\end{align}
\begin{equation}
\label{eq:I_12_34_perp}
\jabc[\sigma_a][\I][(12)(34)][\perp]  \le q^{-8} 
\end{equation}
where Eq.~\ref{eq:I_12_34_perp} is based on the large $q$ expansion in Eq.(60) of Ref.~\cite{zhou2019emergent}. 

Hence
\begin{equation}
w_{\perp}( \sigma_b, \sigma_c; \sigma_a ) \le  q^{-10} \quad \text{ for } \sigma_a \in S_2 \times S_2.
\end{equation}

On the other hand, if $\sigma_a \notin S_2 \times S_2$, the $\perp$ triangle is equal to an ordinary triangle (Eq.~\ref{eq:perpequalsnoperp}). There will be at least four out-going domain walls in the $\perp$ triangle, and we have
\begin{align}
\jabc[][\I][\sigma_a][\sigma_b] \jabc[][\sigma_a][(12)(34)][\sigma_c] & \le q^{ - |\sigma_a (12)(34) | - | \sigma _a| } \\
&\le q^{-4}   
\end{align}
and
\begin{equation}
\jabc[\sigma_a][\I][(12)(34)][\perp] \sim q^{ - |\sigma_a (12)(34) | - | \sigma _a| } \le q^{-4}
\end{equation}
So
\begin{equation}
w_{\perp}( \sigma_b, \sigma_c; \sigma_a ) \le  q^{-8} \quad \text{ for } \sigma_a \notin S_2 \times S_2 
\end{equation}
These inequalities can be saturated {\it only} when
\begin{align}
\sigma_b & = \I, & 
\sigma_c & = (12)(34), & 
\sigma_a & = (13)(24) \text{ or } (14)(23).
\end{align}

Therefore,  the largest terms in Eq.~\ref{eq:hexbox_squ} are:
\begin{align}
w_{\perp}  ( \I, (12)(34) ) \sim &  w_{\perp} ( \I, (12)(34); (13)(24) )  \\
 & + w_{\perp} ( \I, (12)(34); (14)(23) ) \\
  \sim & 2 q^{ -8}.
\end{align}
In other words, $\circled{1'}$ dominates and scales as 
\begin{equation}
\circled{1'} \sim 2 q^{-8} 
\end{equation}

Note that we may also write
\begin{align}
\overline{\circled{1}^n}  = &\overline{\left( Z_u - \overline{Z_u } \right)^n}.
\end{align}
where we have used ${P_\perp=1-P_{\parallel}}$ to split $\circled{1}$ into  $Z_u- \overline{Z_u}$, and 
\begin{equation}
Z_u = \hexbox[+][-][+][u][-][+][-]. 
\end{equation}
We may then use the result for $\overline{Z_u^n}$ in Sec.~VI~A of \cite{zhou2019emergent}.


\section{Further details of numerical results}
\label{app:num_res}

\subsection{Protocol to compute $Z(x,t)$ and $W(x,t)$}
\label{subsec:protocol_WZ}

In the main text we defined the partition function $Z(x,t)$ for a domain wall with fixed upper and lower endpoints (fixed by contracting with $+$ and $-$ states at the top boundary and $(++)^*$ and $(--)^*$ states at the bottom, see Eq.~\ref{eq:Z_x_y_t}).

A na\"ive way to evaluate $Z(x,t)$ numerically is by contracting $U^{(2)}$ with these boundary states  as per its definition. However, it is numerically inefficient to store the full four-copy unitary $U^{(2)}$. 

It is much easier to compute modified partition functions whose bottom boundary conditions only contain $\ket{\pm}$. We have discussed a subset of these modified partition functions, which we called $Z_{\rm op}(x,t)$, in the context of dual-unitary circuits in Sec.~\ref{eq:Wbyrecursion}.
We can view the unitary circuit representing the single-copy time evolution operator $U(t)$ as a state on $2L$ spins, and view all the legs with the $+$ boundary condition as a subsystem. The modified partition function $Z_{\rm op}$ is then proportional to the operator purity for this partition. By rewriting the states at the bottom boundary via 
\begin{equation}
\begin{aligned}
  |(++)^* \rangle  &= \frac{1}{q^4 - 1} \lf | ++ \rangle  - \frac{1}{q^2} | -- \rangle  \ri,  \\
  |(--)^* \rangle  &= \frac{1}{q^4 - 1} \lf | -- \rangle  - \frac{1}{q^2} | ++ \rangle  \ri, 
\end{aligned}
\end{equation}
we find that $Z(x,t)$ is a linear combination of these operator purities. Our numerical scheme then computes these operator purities and obtains $Z(x,t)$ as the linear combination. We can then recursively compute $W(x,t)$ through Eq.~\ref{eq:Wbyrecursion}. This scheme enables us to compute up to $t_{\rm max} = 8$ for a circuit. 

For general Floquet evolutions that do not have a circuit representation, discussed in Sec.~\ref{sec:beyond_circuit}, the \textit{exact} computation of $Z(x,t)$ requires in principle infinitely many operator purities of the Floquet operator on an infinite chain. 
As argued in the text, in practice we may truncate the Floquet operator to a chain of  $L$ sites.
To minimize finite size effects, we choose the entry and exit locations of the membrane (at the top and bottom respectively) to be symmetric about the center of the chain. 
To compute the partition function $Z(x,t)$, in the one-site scheme we contract on the top with $ \langle + + \cdots - -|$, where last $+$ is at $x_1$,
and we contract on the bottom with $| +^* +^* \cdots -^* -^* \rangle $, where the last $+^*$ is at site $x_2$
(i.e. the domain wall enters at the link to the right of $x_1$ and leaves at the link to the right of $x_2$).
For the two-site scheme we use two-site boundary states,  $\bra{\pm\pm}$ at the top and $\ket{(\pm\pm)^*}$ at the bottom, except that at the spatial boundaries we use one-site states  when commensuration effects make this necessary. For both schemes we take ${x_1 = \lceil \frac{L - x}{2}  \rceil}$, and ${x_2 = \lceil \frac{L + x}{2}  \rceil}$.

The computational cost is dominated by the largest $L$ we study.
In addition to the cost of computing each purity, we must sum over of order $2^{L/2}$ or $2^L$ (for the two-and one-site insertions respectively) different purities.
We use the same $L$ for each $x$ and $t$ (in principle one could reduce $L$ for the smaller values of $t$). Since we use different top boundary conditions for different $x$, there is an additional polynomial factor in the cost proportional to the number of distinct $x$ and $t$ values.

\subsection{The random gate}
\label{subsec:u_rand_gate}

The matrix elements of the random gate used in Fig.~\ref{fig:rand_gate} are approximately  
\begin{align}
\label{eq:u_rand}
  u &=
  \begin{bmatrix}
    -0.10054 & 0.01426 & -0.53043 & 0.35559 \\
    0.48842 & 0.52408 & -0.44873 & -0.19577 \\
    -0.22473 & 0.50727 & 0.11189 & 0.26009 \\
    0.77248 & 0.02169 & 0.26725 & 0.12918
  \end{bmatrix}\\
  &+i\begin{bmatrix}
    -0.04094 & 0.27291 & -0.05859 & 0.70873 \\
    -0.28822 & 0.22707 & -0.15763 & -0.29600 \\
    0.11952 & -0.46156 & -0.61862 & 0.04465 \\
    0.07122 & -0.35814 & 0.14956 & 0.39873
  \end{bmatrix}
\end{align}
This gate is less entangling than the Haar average but nevertheless the scheme for extracting $\mathcal{E}(v)$ seems to work well.

\subsection{Protocol to estimate $v_B$}
\label{subsec:protocol_vb}

\begin{figure}[t]
\centering
\subfigure[]{
\label{fig:pauli_weight}
\includegraphics[width = 0.9 \linewidth]{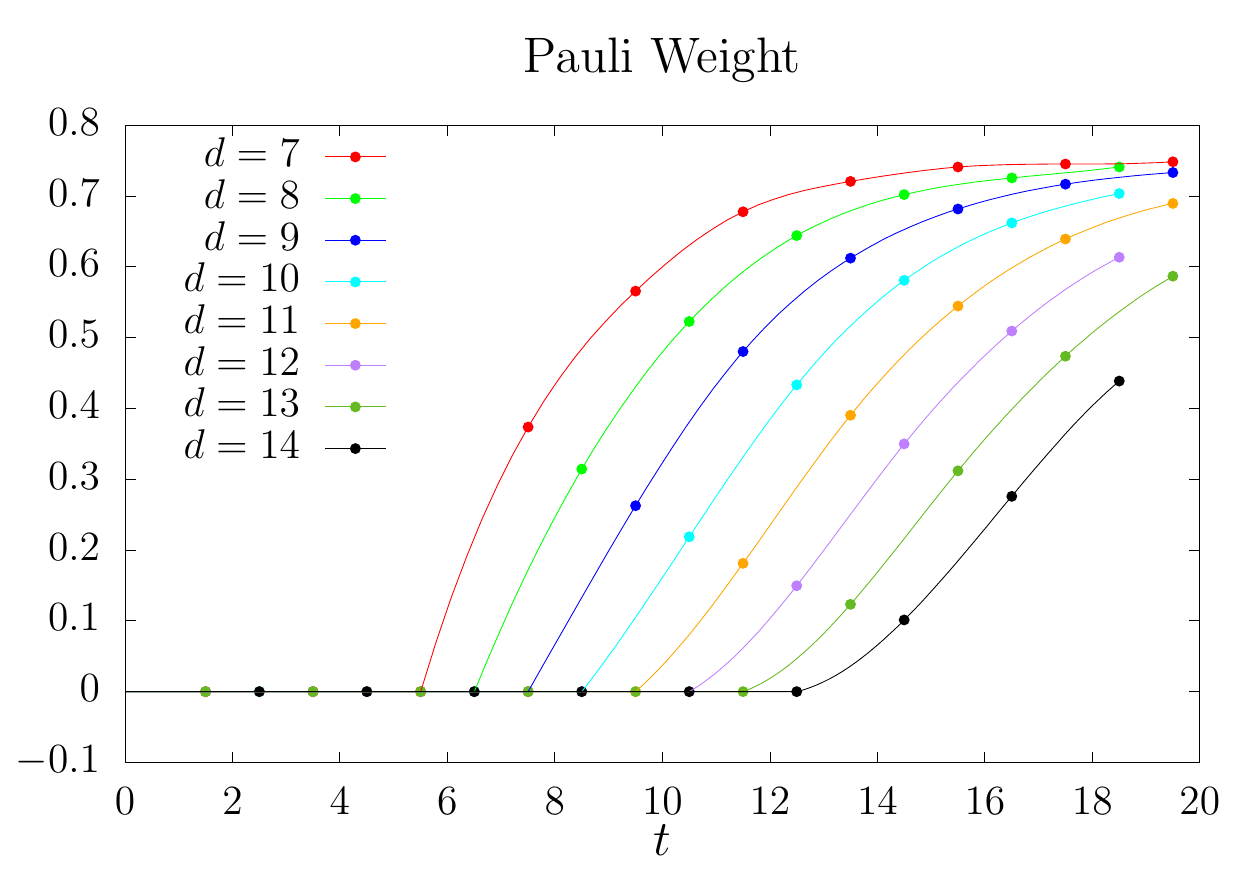}
}
\subfigure[]{
\label{fig:vb_fit}
\includegraphics[width = 0.9 \linewidth]{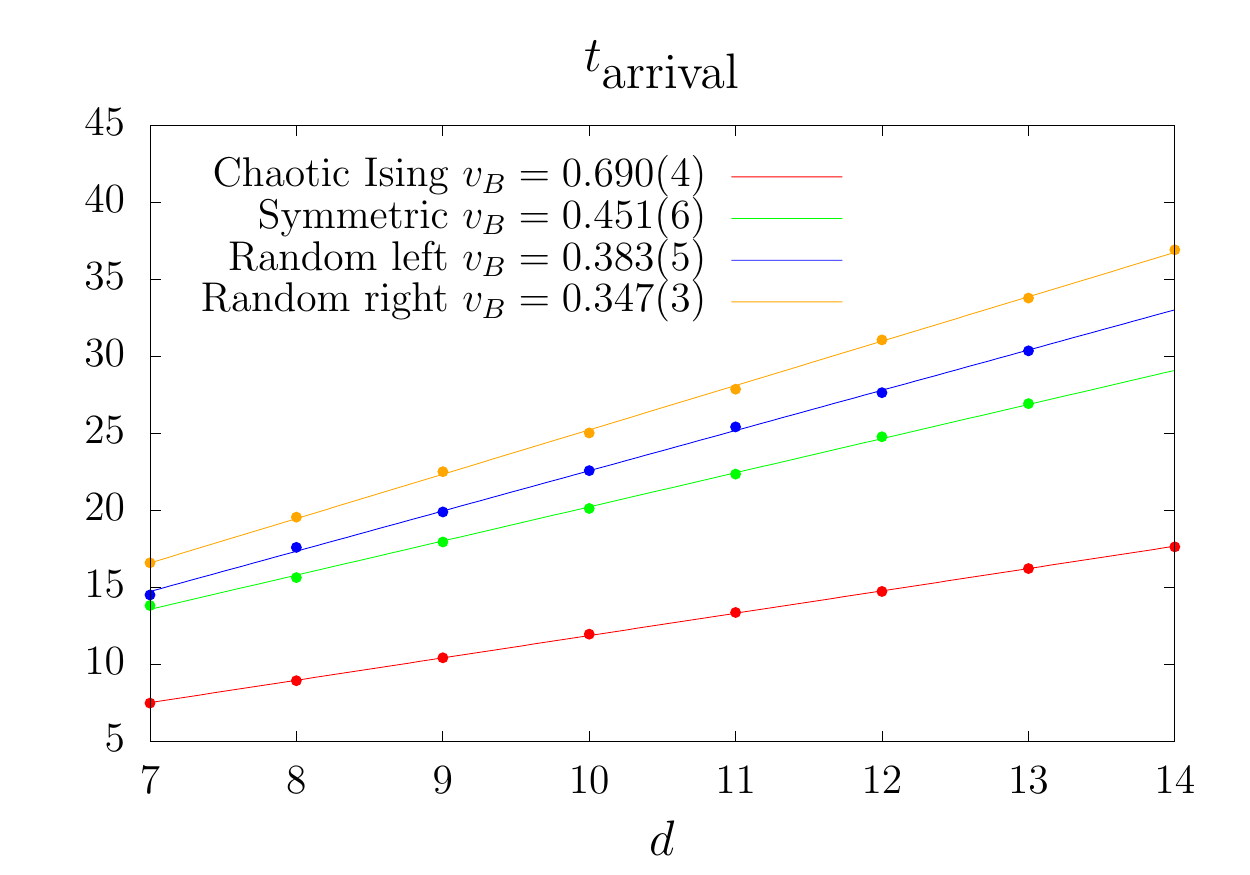}
}
\caption{(a) Pauli weight for the kicked Ising model with parameters in Eq.~\ref{eq:isinggeneric}. 
(b) Estimation of $v_B$ by fitting $t_{\rm arrival} $ versus $d$.
``Chaotic Ising'' is the kicked Ising model with the parameters in Eq.~\ref{eq:isinggeneric}, while
``Symmetric'' and ``Random'' refer to the  generic parity-symmetric gate with $x=0.8$ (Eq.~\ref{eq:symmetricgate}) and the fixed random parity-asymmetric gate (Sec.~\ref{subsec:sym_rand} and App.~\ref{subsec:u_rand_gate}).
For the asymmetric gate
we fit separately for the left and right butterfly velocities. For the dual unitary model (not shown) $v_B=1$ exactly.}
\end{figure}

We use the Pauli weight \cite{roberts2015localized, mezei2017entanglement, jonay_coarse-grained_2018} to estimate $v_B$ in the Floquet circuit. The Pauli weight is defined to be the fraction of the operator weight at a site on non-identity operators. To minimize the number of distinct length scales affecting the data, we place $\sigma^x$ at the leftmost site $0$ and probe its Pauli weight at the rightmost site $d = L-1$:
\begin{equation}
\begin{aligned}
W_{\rm Pauli} (t, d ) &= - \frac{1}{8}\sum_{\alpha = x, y,z} \tr( [\sigma^x_0(t ), \sigma^\alpha_d ]^2 ) / 2^L  \\
&= \frac{3}{4}  - \frac{1}{4} \sum_{\alpha = x, y,z} \tr( \sigma^x_0(t ) \sigma^\alpha_d \sigma^x_0(t ) \sigma^\alpha_d  ) / 2^L.
\end{aligned}
\end{equation}

According to the operator spreading picture, the Pauli strings in the time-evolved operator $\sigma_0^x(t)$ spread to the right with mean velocity $v_B$. Before the Pauli strings reach the right boundary, i.e. when $v_B t \ll d $, the Pauli weight is close to zero. At long times  when $v_B t \gg d$, $\sigma_0^x(t)$ can be regarded as a random operator for the purpose of calculating the Pauli weight, which is then $\frac{3}{4}$. 

Fig.~\ref{fig:pauli_weight} shows the Pauli weight for the kicked Ising model with the parameters in Eq.~\ref{eq:isinggeneric}. Due to the circuit structure, the Pauli weights at consecutive integer time steps are equal. Hence we place the data at half integer points, and interpolate the curve (with cubic spline) when the data is greater than zero. Following Ref.~\cite{jonay_coarse-grained_2018}, we define $t_{\rm arrival}$ as the time when the interpolated curve is at $\frac{3}{8}$. Fitting $d$ with $t_{\rm arrival}$ gives us an estimation of $v_B$. Fig.~\ref{fig:vb_fit} shows the results of fitting for the generic symmetric gate and the fixed but randomly chosen gate (Sec.~\ref{subsec:sym_rand}) 
and for the kicked Ising model (Eq.~\ref{eq:isinggeneric}).

\subsection{The approximate entanglement velocity}

\begin{figure}[htb]
\centering
\includegraphics[width=\columnwidth]{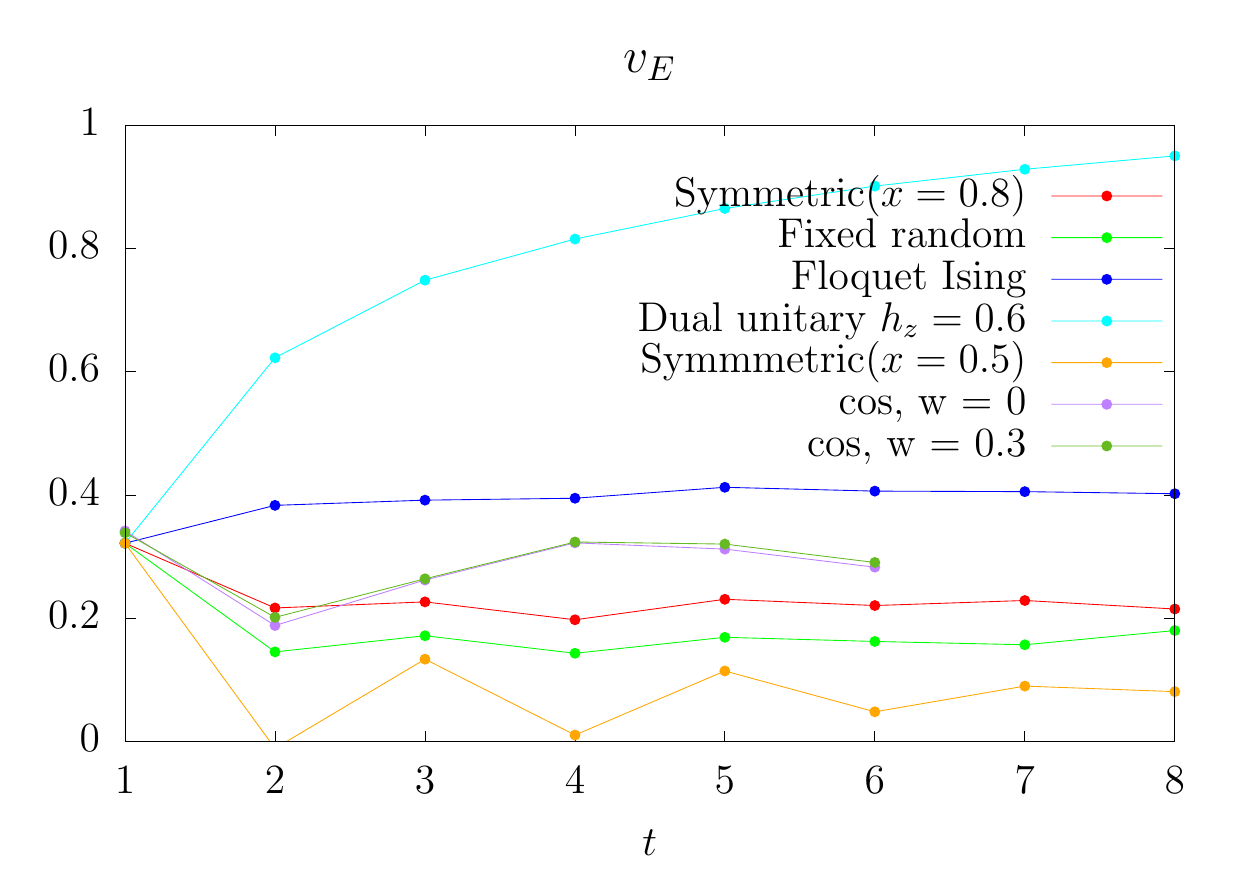}
\caption{The approximate entanglement velocity $v_E$ computed by truncating to times $\leq t$, for  $t$ up to $8$. 
The first four curves are for the four circuits studied in Sec.~\ref{sec:floq_app} (the first three are also the subject of Fig.~\ref{fig:vb_fit} for $v_B$).
The fifth shows the generic symmetric gate (Eq.~\ref{eq:symmetricgate}) at a weaker coupling strength.
The two curves with $w = 0$ and $w = 0.3$ are the entanglement velocities for the Floquet Hamiltonian with sinusoidal field in Eq.~\ref{eq:cosinehx}. Due to the absence of circuit structure, their starting values at $t = 1$ are not identical to that obtained from the random average; while the others are.}
\label{fig:vE}
\end{figure}

The second R\'enyi entropy after  a quench corresponds to a free boundary condition at the bottom for the domain wall. Therefore the system will select out $v_E = \min_v \mathcal{E}(v)$ as the entanglement speed for $S_2$. 
We extract the approximate entanglement speed from our numerical calculation of $\mathcal{E}(v)$, and show its dependence on the truncation time $t$ in in Fig.~\ref{fig:vE}. 

All the curves start off for $t=1$ at the purity velocity of the random circuit. Differences between the gates then drive the curves to their respective entanglement velocities, either above or below the value at $t = 1$. 

Notably, the curve for the dual-unitary gate shows a clear trend towards the analytical result $v_E = 1$. 
Curves for the Floquet Ising model, the generic reflection symmetric gate and the fixed random gate behave reasonably well.
Fluctuations are still visible for $t_{\rm max}$ up to $8$, and it is hard to determine the functional form for convergence directly from this plot.
However in the next subsection we will give evidence for exponential suppression of the irreducible step weights in these models at large $t$.

We also include in the plot a continuous time model with a static Hamiltonian (the limit $w=0$ of the Floquet model with a sinusoidal field, Eq.~\ref{eq:cosinehx}). 
This model would require a separate theoretical analysis and we include it only for comparison.
Due to continuous time translation symmetry it has a diffusive conserved energy density. This is expected to suppress $v_E$ to zero for the second R\'enyi entropy \cite{rakovszky2019sub,huang2019dynamics,zhou2019diffusive}, and is likely to introduce parametrically larger (in $t$) finite time effects.\footnote{We expect that $\mathcal{E}(v)$ for $v\neq 0$ will still be nontrivial. The suppression of $v_E$ all the way to zero is due to rare Feynman trajectories  in which a large spatial region is depleted of energy and cannot entangle. Since the dynamics of the energy density is diffusive, these empty regions can persist for a long time along a ray at $v=0$, suppressing the cost of a membrane with $v=0$, but not along a ray at nonzero $v$.}

\subsection{The suppression of irreducible weights}
\label{subsec:W_decay}

In Sec.~\ref{sec:expsup}, we conjectured that $\perp$ clusters inside a thick domain wall are exponentially suppressed at large $t$, leading in turn to an exponential suppression of the irreducible step weights at large $t$. We showed this explicitly for a typical fixed realization of a random circuit at large $q$: there, long steps of size $t$ are suppressed by $\mathcal{O}(q^{-t})$ compared to the leading order contributions to the domain wall free energy on this scale. 

Here we examine the nature of the decay in step weights at large $t$ numerically for translation-invariant spin-1/2 models.
Quantitatively, we can compare $W(x,t)$ with $\exp( - s_{\rm eq} v_E t )$, where the former represents the irreducible step weight for a single step of size $t$, while the latter represents the asymptotic scaling of the  partition function $Z(0,t)$ for a domain wall of duration $t\gg 1$. Their ratio should decay exponentially according to our conjecture.

\begin{figure}[h]
\centering
\includegraphics[width=\columnwidth]{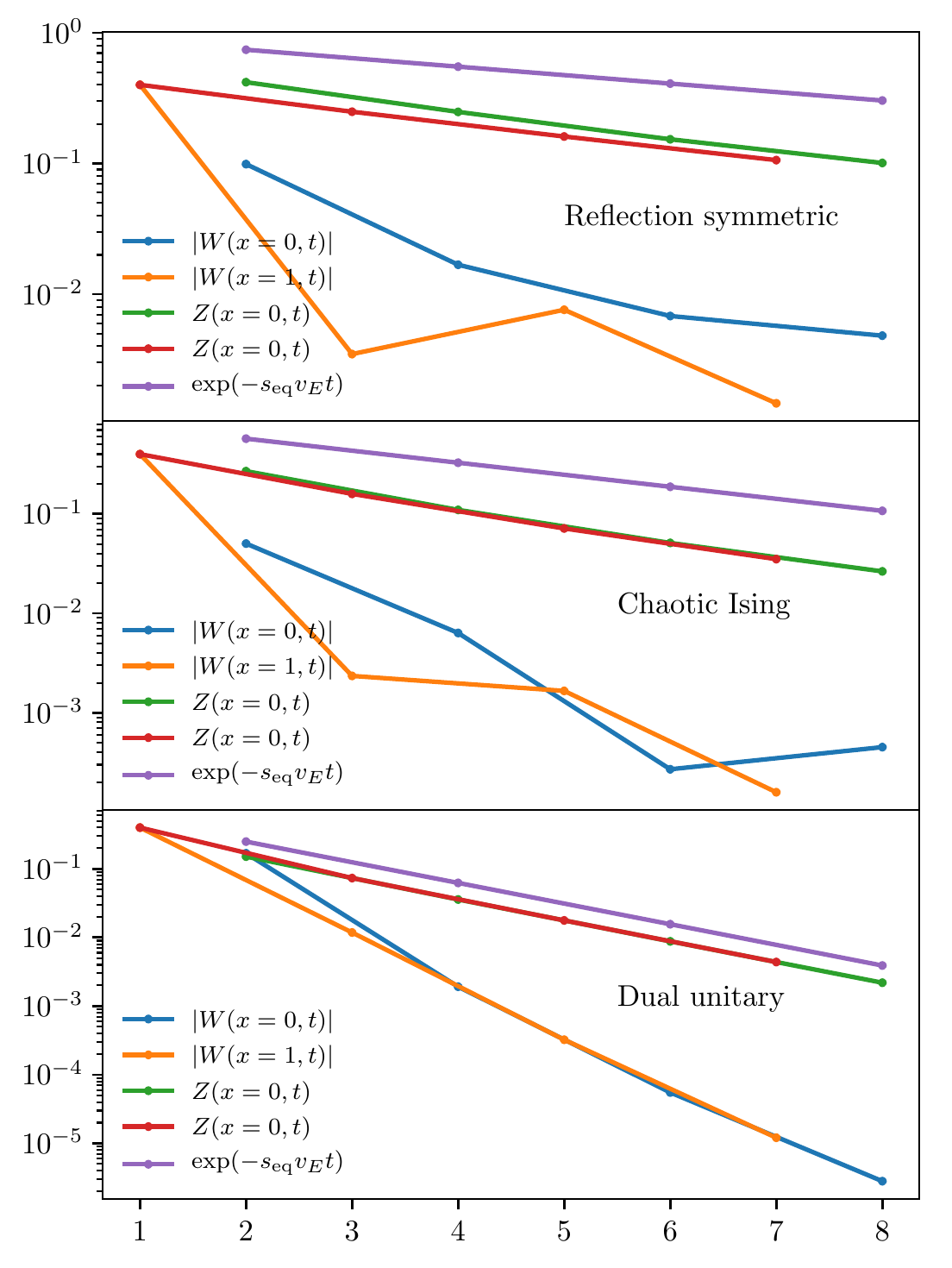}
\caption{ The decay of $|W|$  and $Z$ at $x = 0, 1$ and $\exp( - s_{\rm eq} v_E t )$ for (Top) reflection symmetric gate in Eq.~\ref{eq:symmetricgate}, (Center) Floquet Ising model in Eq.~\ref{eq:isinggeneric} and (Bottom) dual unitary gate in Eq.~\ref{eq:dualunitaryspecific}. 
}
\label{fig:decay}
\end{figure}

In Fig.~\ref{fig:decay}, we plot $W(x,t)$ along with both $Z(x,t)$ and $\exp( - s_{\rm eq} v_E t )$ as comparison. The best case is the dual unitary circuit, where a faster exponential decay of $W(x,t)$ w.r.t. both $Z(x,t)$ and 
$\exp( - s_{\rm eq} v_E t )$ is clearly visible. The Floquet Ising model follows a similar trend. Though this is less clear for the reflection symmetric gate at $x = 0$, that is consistent with the fact that this gate is more weakly entangling than the others (Fig.~\ref{fig:vE})
and takes larger $t$ to reach convergence.

\subsection{Diffusive character of the domain wall}
\label{subsec:diffusivescaling}

\begin{figure}[h]
\centering
\includegraphics[width=0.92\columnwidth]{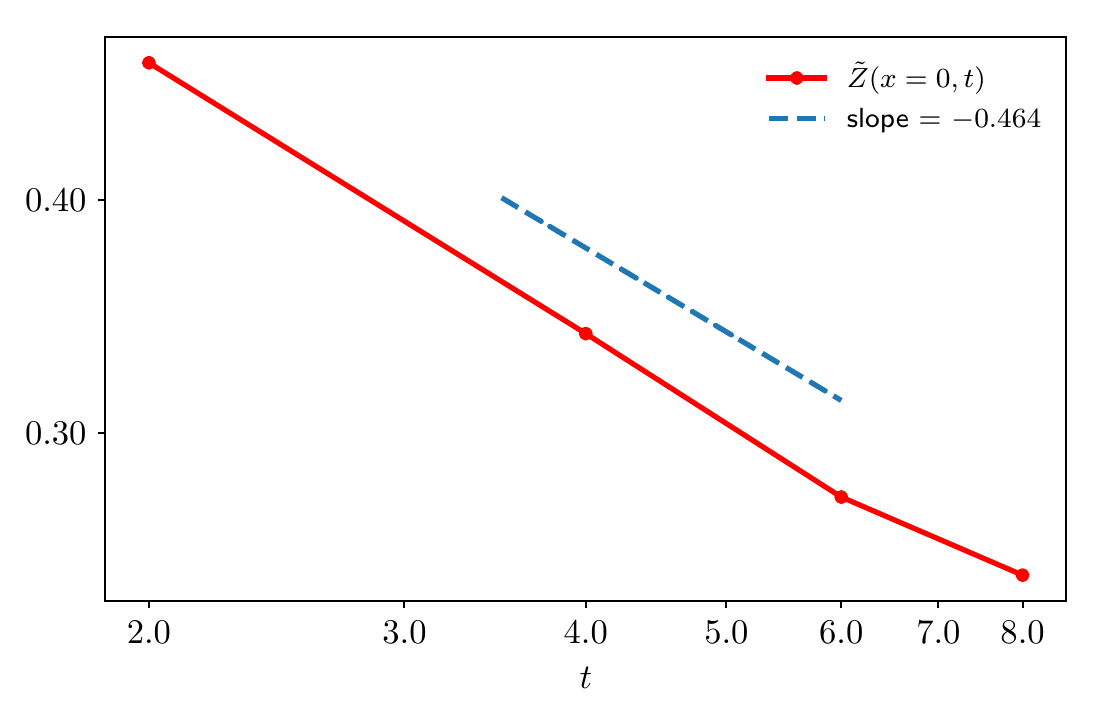}
\caption{The diffusive decay of $\tilde{Z}(x,t)$ at $x = 0$ for the chaotic Ising model in \eqref{eq:isinggeneric}. Both axes are logarithmic.}
\label{fig:Z_decay}
\end{figure}

In Sec.~\ref{sec:domainwalldualunitary} we argued that for generic choices of gate the wandering of the domain wall was diffusive at large scales. One consequence of this is the $t^{-1/2}$ scaling in Eq.~\ref{eq:Zdiffusion}: $\widetilde Z(0,t)\sim t^{-1/2}$. We stated in the caption to Fig.~\ref{fig:huse_Z_tilde} that the data was in good agreement with this scaling. This is demonstrated in Fig.~\ref{fig:Z_decay} which shows $\widetilde Z(0,t)$ versus $t$ on a log-log plot.

\section{Relation between root and line-tension}
\label{app:legendre}

We give details of the relation between $\mathcal{E}(v)$ and the root $b_0(s)$ described in Sec.~\ref{sec:generatingfunctions}. Let 
\begin{equation}\label{eq:littlez}
z(\sdensity, t ) \equiv \sum_x Z(x, t ) e^{ \sdensity x}. 
\end{equation}
This is the partition function with a fixed force on the endpoint rather than a fixed endpoint.
Asymptotically, it should scale as 
\begin{equation}
z(\sdensity, t ) \sim \exp( - s_{\rm eq} \Gamma( \sdensity )  t )
\end{equation}
with \cite{jonay_coarse-grained_2018}
\begin{equation}
\Gamma(\sdensity ) =    \min_{v } \left[ \mathcal{E}(v  ) - \f{v \sdensity}{s_{\rm eq}} \right],
\end{equation}
so that $-s_{\rm eq} \Gamma(s)$ and $s_{\rm eq} \mathcal{E}(v)$ form a standard Legendre transform pair.
On the other hand
\begin{equation}
\label{eq:z_omega}
\sum_t z(\sdensity, t) b^t  = \frac{\omega( \sdensity, b )}{1 - \omega( \sdensity, b )}.
\end{equation}
Assuming that the smallest root of $\omega( \sdensity, b)$ for a given $\sdensity$ is $b_0(\sdensity)$, then for $|b| \lesssim |b_0| $
\begin{equation}
\sum_t z(\sdensity, t ) b^t \sim \frac{1}{1 - {b}/{b_0}}.
\end{equation}
Matching the expansions in $b$,
\begin{equation}
s_{\rm eq} \Gamma(\sdensity ) =  \ln |b_0(\sdensity) |.
\end{equation}
Finally, inverting the Legendre transform yields the expression (\ref{eq:ev_legendre}) in the main text for $\mathcal{E}(v)$.

Generically $b_0(s)$ is a single root. But in the dual-unitary case, $b_0$ is a double root for $s=0$. In Sec.~\ref{sec:domainwalldualunitary}, we postulate that $Z(x,t)$ for the dual-unitary model is approximately constant in space, and so ${z(s = 0, t ) \sim t e^{ - s_{\rm eq } t}}$. The factor of $t$ in the front indicates a double root on the right hand side of Eq.~\ref{eq:z_omega} in the limit of large $t_\text{max}$.


\section{The continuum equation for $\widetilde{Z}$}
\label{app:eq_Z_tilde}

For completeness, let us give a more general derivation of the continuum equation which makes contact with the equation in Ref.~\onlinecite{jonay_coarse-grained_2018} for the dynamical evolution of the state entanglement. 
Writing $Z=\exp(-S)$, 
we assume that the solution is such that higher derivatives of $S(x,t)$ are small at late times (to be checked self-consistently), and expand in derivatives.

So far, we have considered a domain wall which is tethered at both the top and the bottom. To compute the entanglement $S(x,t)$ of a state across an entanglement cut at position $x$, we should instead take the domain wall to be tethered at position $x$ at the top, and free at the bottom; the initial condition ${Z(x,0)=\exp(-S(x,0))}$ is set by the entanglement in the initial state. The recursion relation still applies. 

To save clutter, let us write $(y,\tau)$ instead of $(\Delta x, \Delta t)$ in the recursion relation. In terms of $S(x,t)$, it reads:
\begin{align}
e^{-S(x,t)} & =\sum_{y,\tau} W(y,\tau) e^{-S(x-y, t-\tau)}. 
\end{align} 
Expanding in the exponent gives
\begin{align}\label{eq:expandS}
1 & =\sum_{y,\tau} W(y,\tau) \exp \lf 
y S' + \tau \dot S -\f{y^2}{2} S'' - \f{\tau^2}{2} \ddot S - y \tau \dot S'
\ri.
\end{align}
Generically, $S$ is of order $t$,  $S'$ and $\dot S$ are of order 1, and the higher derivatives are smaller.
Therefore the leading order equation is simply
\begin{align}
1 & =\sum_{y,\tau} W(y,\tau) \exp \lf 
y S' + \tau \dot S
\ri.
\end{align}
Using the definitions in Sec.~\ref{sec:generatingfunctions}, this is solved by
\begin{align}\label{eq:leadingorderSevo}
\dot S = s_\text{eq} \, \Gamma(S'),
\end{align}
so that $\Gamma$ is the growth rate, which depends on the derivative $S'$ of the entanglement as a function of the spatial location.
We are interested in the subleading corrections to this equation. 
Let us define an $s$-dependent ``average'' of an arbitrary function $F$ of the step:
\begin{equation}
\< F(y,\tau)  \>_s \equiv 
\sum_{y,\tau} W(y,\tau) e^{ 
 s y +  s_\text{eq} \Gamma(s) \tau
}  F(y,\tau).
\end{equation}
From the relations for the generation functions, 
\begin{equation}\label{eq:averageof1}
\< 1\>_s = 1.
\end{equation}
We also define the associated velocity ${v(s) = {\< y \>_s}/{\< \tau \>_s}}$. By differentiating (\ref{eq:averageof1}) with respect to $s$, we have
\begin{equation}\label{eq:vGamma}
v(s) = - \seq \f{\dd \Gamma(s)}{\dd s}. 
\end{equation}
This is the coarse grained velocity of the membrane since it also solves the Legendre transform (Eq.~\eqref{eq:ev_legendre}) relating the growth rate $\Gamma[s]$ and the line tension $\mathcal{E}(v)$. 

We may then write (\ref{eq:expandS}) as 
\begin{align}
\<
e^{\tau \left[ \dot S - \seq \Gamma(S') \right] }
e^{- \f{y^2}{2} S'' - \f{\tau^2}{2} \ddot S - y \tau \dot S'+\ldots}
\>_{S'},
\end{align}
where the $\ldots$ are higher derivatives. The exponent is now written in terms of quantities that are small at late times, so may be expanded,
\begin{align}
{
\dot S = \seq \Gamma(S') + \f{\<y^2\>_{S'} S'' + \<\tau^2\>_{S'} \ddot S + 2 \< y \tau\>_{S'} \dot S'}{2\<\tau\>_{S'}}.
}
\end{align}
We may use the leading-order solution to replace the time derivatives on the right hand side with space derivatives (cf.  Eqs.~\ref{eq:leadingorderSevo},~\ref{eq:vGamma}): ${\dot S' \simeq - v(S') S''}$ and ${\ddot S \simeq v(S')^2 S''}$.
Then
\begin{align}
\dot S = 
\seq \Gamma(S') +
\f{\< (y - v(S') \tau)^2 \>}{2\<\tau\>} S''.
\end{align}
By differentiating (\ref{eq:averageof1}) twice, 
\begin{equation}
\f{\< (y - v(s) \tau)^2 \>}{\<\tau\>} = - \f{\dd^2 \Gamma(s)}{\dd s^2},
\end{equation}
so 
\begin{align}\label{eq:S_dot_eq_Gamma}
\dot S = 
\seq \lf
\Gamma(S') - \f{1}{2} \Gamma''(S') S''
\ri.
\end{align}
(The primes on $\Gamma$ indicate derivatives with respect to its argument.)

We may also write everything in terms of $\mathcal{E}(v)$. From the Legendre transform relationship between $\mathcal{E}(v)$ and $\Gamma(s)$, and Eq.~\ref{eq:vGamma},
\begin{equation}
  \mathcal{E}(v) = \Gamma[s] + \frac{v(s) s}{s_{\rm eq}}, 
\end{equation}
we differentiate twice and obtain 
\begin{align}
\mathcal{E}'(v) & = \frac{s}{s_{\rm eq}} 
&
\mathcal{E}''(v) & = \frac{1}{s_{\rm eq} v'(s)} = - \frac{1}{s^2_{\rm eq} \Gamma''[s]}.
\end{align}
This along with Eq.~\ref{eq:S_dot_eq_Gamma} gives a diffusive equation
\begin{equation}
  \dot{S} = s_{\rm eq} \mathcal{E}(v)  + \frac{1}{2 s_{\rm eq} \mathcal{E}''(v)} S''
\end{equation}
with diffusion constant ${D = {1}/({2 s_{\rm eq} \mathcal{E}(v)})}$, consistent with Eq.~\ref{eq:diffusivity}.


\section{Triangle weights for two commuting domain walls}
\label{app:perp_tri_weight}

For reference and use in App.~\ref{app:size_fluct}, 
here we list exact expressions for triangle weights $J(\I, (12)(34); \sigma_a)$. This corresponds to two incoming ``commutating'' elementary domain walls at the top of the triangle. (An elementary domain wall is a transposition.)  We also list their order relative to the weight of two separated domain walls. We use the notation from Ref.~\onlinecite{zhou2019emergent}, drawing one line for each elementary domain wall crossing an edge of the triangle. 

There are in total 6 classes. Within each class the configurations are related by the combination of reflection symmetry about the center axis, conjugation by $(12)$, $(34)$ or both.

We categorize them in terms of $\sigma_a$: 
\begin{enumerate}
\item $\I$, $(12)(34)$. 
\begin{equation}
\begin{aligned}
\jabc[(12)(34)][\I][(12)(34)] &= \ddw[l] =  \frac{q^8 - q^6 -12 q^4 + 14 q^2 + 22}{(q^4 - 9 )(q^4 - 4 ) (q^2 + 1)} \\
&= \left[ \sdw[l] \right]^2 \left( 1 + \frac{2}{q^6} + \mathcal{O}(\frac{1}{q^{12}})\right) 
\end{aligned}
\end{equation}
\item $(12)$,$(34)$
\begin{equation}
\begin{aligned}
\jabc[(12)][\I][(12)(34)] &= \ddw[lr] = \frac{(q^8- 12q^4+12)(q^2 - 1)}{(q^4 - 9)(q^4-4)(q^2 + 1) q^2} \\
&= \left[ \sdw[l] \right]^2 \left( 1 - \frac{12}{q^8} + \mathcal{O}(\frac{1}{q^{12}}) \right)
\end{aligned}
\end{equation}
\item $(13)(24)$, $(14)(23)$

\begin{equation}
\begin{aligned}
\jabc[(13)(24)][\I][(12)(34)] &= \ddw[specvert] = 
  \frac{q^2 - 1}{(q^4 - 9)(q^2 + 1)}  \\
  &=\left[ \sdw[l] \right]^2 \left( \frac{1}{q^2 } + \mathcal{O}(\frac{1}{q^6}) \right)
\end{aligned}
\end{equation}
\item $(13)$, $(24)$, $(14)$, $(23)$, $(1234)$, $(1423)$, $(1243)$, $(1342)$
\begin{equation}
\begin{aligned}
\jabc[(1234)][\I][(12)(34)] &= \ddw[lp] = 
  - \frac{(q^4+6)(q^2 - 1)}{(q^4 - 9)(q^4 - 4)(q^2 + 1) q^2} \\
  &= -\left[ \sdw[l] \right]^2 \left( \frac{1}{q^4 } + \mathcal{O}(\frac{1}{q^8})\right)
\end{aligned}
\end{equation}
\item $(123)$, $(124)$, $(132)$, $(142)$, $(234)$, $(134)$, $(243)$, $(143)$
\begin{equation}
\begin{aligned}
\jabc[(123)][\I][(12)(34)] &= \ddw[lrp] = 
  \frac{5(q^2-1)}{(q^4 - 9)(q^4 - 4)(q^2 + 1)}  \\
  &= \left[ \sdw[l] \right]^2 \left( \frac{5}{q^6 } + \mathcal{O}(\frac{1}{q^{10}})\right)
\end{aligned}
\end{equation}
\item $(1324)$, $(1423)$
\begin{equation}
\begin{aligned}
\jabc[(1324)][\I][(12)(34)] &= \ddw[lrpp] = 
  - \frac{3(q^2 - 1)}{(q^4 - 9)(q^2 + 1) q^2 } \\
  &= -\left[ \sdw[l] \right]^2 \left( \frac{3}{q^4 } + \mathcal{O}(\frac{1}{q^8})\right)
\end{aligned}
\end{equation}
\end{enumerate}


\bibliographystyle{unsrt} 
\bibliography{questionmarkrefs}

\end{document}